\documentclass[amsmath,amssymb,aps,twocolumn,pra,freq,floatfix, superscriptaddress]{revtex4-2}
\usepackage{graphicx}
\usepackage{dcolumn}
\usepackage{bm}
\usepackage{hyperref}
\usepackage{color}
\usepackage{subfigure}
\usepackage{flexisym}
\usepackage{float}
\usepackage{placeins}

\begin{document}

\title{High Field Diamond Magnetometry Towards Tokamak Diagnostics}

\author{S. M. Graham}
\altaffiliation{Stuart.M.G.Graham@warwick.ac.uk}
\affiliation{Department of Physics University of Warwick Gibbet Hill Road Coventry CV4 7AL United Kingdom}
\affiliation{Diamond Science and Technology Centre for Doctoral Training University of Warwick Coventry CV4 7AL United  Kingdom}%

\author{C. J. Stephen}
\affiliation{Department of Physics University of Warwick Gibbet Hill Road Coventry CV4 7AL United Kingdom}

 \author{A. J. Newman}
 \affiliation{Department of Physics University of Warwick Gibbet Hill Road Coventry CV4 7AL United Kingdom}
 \affiliation{Diamond Science and Technology Centre for Doctoral Training University of Warwick Coventry CV4 7AL United Kingdom}%

\author{A. M. Edmonds}
\affiliation{Element Six Innovation Fermi  Avenue Harwell  Oxford Didcot  OX11 0QR  Oxfordshire  United  Kingdom}%

\author{M. L. Markham}
\affiliation{Element Six Innovation Fermi  Avenue Harwell  Oxford Didcot  OX11 0QR  Oxfordshire  United  Kingdom}%

\author{G. W. Morley}
\altaffiliation{gavin.morley@warwick.ac.uk}
\affiliation{Department of Physics University of Warwick Gibbet Hill Road Coventry CV4 7AL United Kingdom}%
\affiliation{Diamond Science and Technology Centre for Doctoral Training University of Warwick Coventry CV4 7AL United Kingdom}%


%


\begin{abstract}

Nitrogen vacancy centres (NVC) in diamond have been widely used for near-dc magnetometry. The intrinsic properties of diamonds make them potential candidates for tokamak fusion power diagnostics, where radiation-hard magnetometers will be essential for efficient control. An NVC magnetometer placed in a tokamak will need to operate within a $\geq$ 1 T magnetic field. In this work, we demonstrate fibre-coupled ensemble NVC optically detected magnetic resonance (ODMR) and magnetometry measurements at magnetic fields up to 1.2 T. Sensitivities of approximately 240 to 600 nT/$\sqrt{\textrm{Hz}}$ and 110 nT/$\sqrt{\textrm{Hz}}$ are achieved in a (10-150) Hz frequency range, for non-degenerate and near-$\langle$111$\rangle$ field alignments respectively.

\end{abstract}

\maketitle

\section{Introduction}
 
Diamond magnetometers have been proposed for use in a variety of applications in industry \cite{newman2024tensor, kubota2023wide, zhang2021battery, zhou2021imaging, hatano2021simultaneous, chatzidrosos2019eddy} and physics research \cite{afach2021search, waxman2014diamond, bouchard2011detection}, as well as in biology and medical applications such as breast cancer surgery \cite{steinert2013magnetic, glenn2015single, barry2016optical, davis2018mapping, newman2025endoscopic, barry2016optical}. Diamond's robust nature makes it suitable for use in extremes of temperature (up to 600 K) \cite{plakhotnik2014all, toyli2013fluorescence, toyli2012measurement, liu2019coherent}, pressure (up to 60 GPa) \cite{doherty2014electronic, ivady2014pressure, hsieh2019imaging} and radiation \cite{shimaoka2022recent}. Fusion power could be a major contributor to the global energy supply by the 2030s or 2040s \cite{kikuchi2010review}. Tokamaks require magnetometers to detect the shape and position of the plasma. These measurements can then be used as feedback to control the plasma. This is considered an important factor in achieving net gain energy generation \cite{international2023iaea2}. Current tokamaks use in-vessel inductive coils and loops for magnetic sensing. However, these rely on integration and errors grow for long plasma durations \cite{biel2019diagnostics, entler2019prospects}.
This makes them unsuitable for future DEMO tokamaks, with plasma durations $\geq$ 1000 s, or potential commercial steady state devices \cite{hodapp1995magnetic, international2023iaea2}. Proposed steady-state magnetic field sensors include Hall probes, as well as fibre devices using magneto-optical effects \cite{vduran2017development, entler2019prospects, quercia2022long, wang2024development, aerssens2011faraday, moreau2011test, dandu2023distributed, sarancha2025fiber}.
However, high neutron fluences, up to four orders of magnitude higher than ITER, are anticipated for future DEMO and commercial tokamaks, as well as high levels of gamma radiation \cite{entler2019prospects}. In-vessel, plasma facing sensors could also be destroyed by runaway, relativistic electrons \cite{ataeiseresht2023runaway}. 
Additionally, sensors located in-vessel will be required to operate at temperatures up to 300$^{\circ}$C, from factors including radiation-induced heating \cite{entler2019prospects}. 
Traditional Si-based Hall probes are insufficiently radiation-hard, have temperature-dependent responses and suffer from noise given the small voltages involved \cite{international2023iaea2}. Although various metallic Hall probes have been tested, steady-state magnetometry remains an open challenge for future DEMO and commercial tokamaks beyond ITER \cite{kovarik2012status, quercia2022long}. 
Diamond magnetometers are a potential alternative or supplement to Hall probes, given their combination of robustness and sensitivity, as well as their dynamic and frequency range. Arrays of instrumented mm-scale diamond sensor heads could be integrated into a tokamak, with sensitive optics and electronics kept outside of the bioshield. Diamond magnetometers are also capable of highly accurate vector measurements and should be robust against electromagnetic interference and radiation-induced currents \cite{schloss2018simultaneous, biel2019diagnostics}. 

However, in addition to high radiation and temperatures, a diamond magnetometer in a fusion device would need to work in a high magnetic field $\geq$ 1 T \cite{biel2019diagnostics, creely2020overview}. This would mean operating in a magnetometry regime distinct from that of traditional ensemble NVC magnetometers, with the tokamak's combined poloidal and toroidal magnetic field acting as the bias field for each diamond sensor. For comparison, the bias fields of ensemble NVC magnetometers are usually on the order of 1 to 10 mT \cite{rondin2014magnetometry, barry2020sensitivity}. In typical use cases, the bias field would also often be aligned parallel to the symmetry axis of one of the four possible NVC orientations within the tetrahedral diamond lattice. This would not be easy to achieve in a fusion device.

Neglecting hyperfine, nuclear Zeeman, and quadrupole interactions, as well as transverse strain, the NVC spin-Hamiltonian in the NVC body frame is given by \cite{barry2020sensitivity}

\begin{equation}
\label{eq: SimplifiedNVHamiltonian}
H/h = DS_z^2 + \gamma_e(B_x'S_x+B_y'S_y+B_z'S_z),
\end{equation}

where D is the temperature-dependent zero field splitting (ZFS) parameter (D $\approx$ 2.87 GHz), $\gamma_e$ is the gyromagnetic ratio of the NVC ($\gamma_e$ $\approx$ 28 GHz/T) and $\textrm{S}_x$, $\textrm{S}_y$ and $\textrm{S}_z$ are S = 1 spin operators. The prime indicates the NVC body frame. The zero-field eigenstates of this spin-Hamiltonian are eigenstates of the $\textrm{S}_z$ operator 
and the quantisation axis is set by the NVC symmetry axis. This axis is the line between the nitrogen atom and the vacancy and is taken to be parallel to the z-axis in the NVC body frame. The zero-field eigenstates can be written as $\vert$$m_s$ = 0$\rangle$, $\vert$$m_s$ = -1$\rangle$ and $\vert$$m_s$ = +1$\rangle$, corresponding to the $m_s$ = 0, $m_s$ = -1 and $m_s$ = +1 spin sub-levels. There are three possible transitions, for each NVC orientation, between these three sub-levels, including the generally forbidden $\Delta$$m_s$ = $\pm$ 2 double quantum (DQ) transitions. 
When the external magnetic field is non zero and misaligned to the NVC symmetry axis, such that $\textrm{B}_x'$ = $\textrm{B}_y'$ $\not=$ 0, the eigenstates of the Hamiltonian become linear combinations of the zero-field eigenstates. 
This mixing occurs in both the ground and excited states, resulting in reduced spin-polarisation, fluorescence levels and contrast. All these factors combine to impair the sensitivity for misaligned NVC orientations \cite{rondin2014magnetometry, tetienne2012magnetic}. 
A 1 T bias magnetic field places the NVC ensemble in a regime above the ground-state level anti-crossing (GSLAC) at 102.4 mT. The spin-Hamiltonian is dominated by the electronic Zeeman term, and the quantisation (and thus sensitive) axis for each NVC orientation is no longer set by the NVC symmetry axis and depends largely on the bias field orientation \cite{rondin2014magnetometry, tetienne2012magnetic}. In this case, which we will refer to as the high field, as opposed to low field, regime the performance is strongly dependent upon the field alignment, with spin-polarisation disappearing entirely for a $\langle$100$\rangle$ alignment, being minimally impacted for a $\langle$111$\rangle$ alignment and with intermediate effects for other alignments. Ideally, for vector measurements the diamond would be sat such that eight resonances (two from each NVC orientation) are visible; we shall refer to such alignments as non-degenerate in this work. 

NVC electron paramagnetic resonance (EPR) and optically-detected magnetic resonance (ODMR) has been demonstrated at fields up to 10 T \cite{stepanov2015high, jeong2017understanding, fortman2020demonstration, magaletti2022quantum, kollarics2024terahertz, kollarics2023magneto, an2024micron}, typically employing a $\langle$111$\rangle$ alignment with single NVCs. 
In this work, fibre-coupled ensemble NVC magnetometry is demonstrated at fields up to 1.2 T, looking towards potential diagnostic applications in tokamaks. For a frequency range of (10-150) Hz, sensitivities of approximately 240 to 600 nT/$\sqrt{\textrm{Hz}}$ and 110 nT/$\sqrt{\textrm{Hz}}$ are achieved for non-degenerate and near-$\langle$111$\rangle$ alignments respectively.

\FloatBarrier
 
\section{Methods}

\begin{figure}[h!]
\centering
\includegraphics[width=\columnwidth]{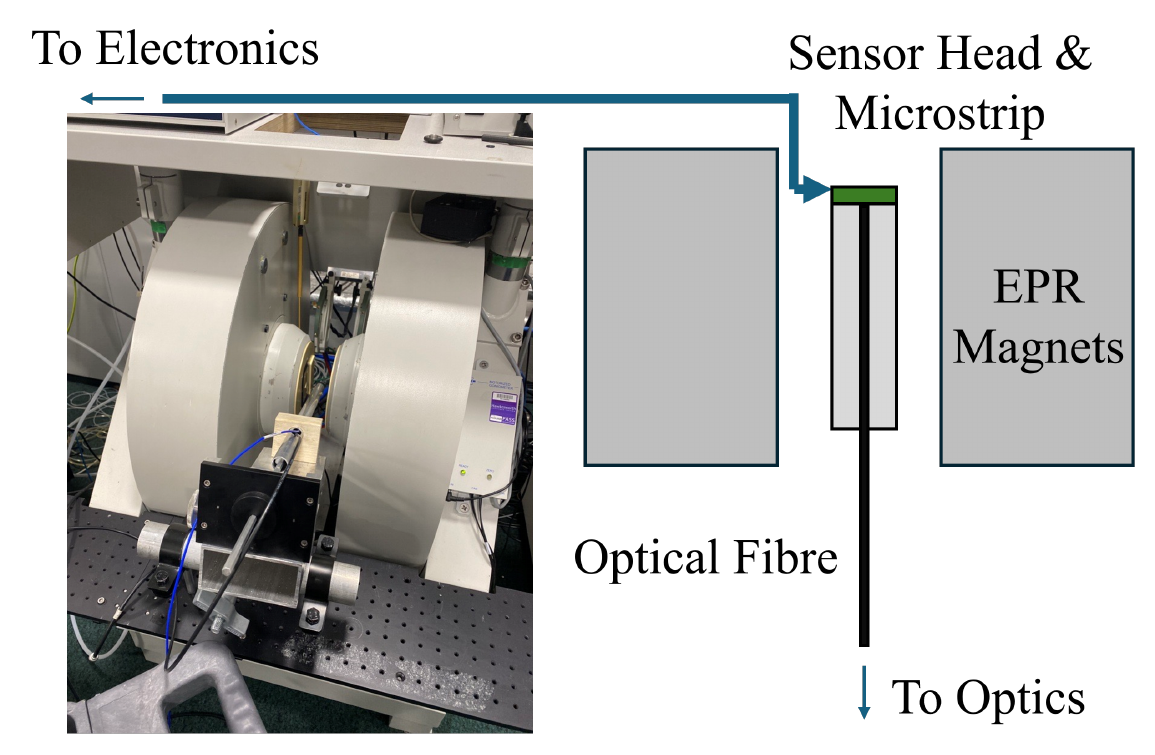} 
\caption{\small A schematic of the experimental setup showing the sensor head within the electron paramagnetic resonance (EPR) Magnets. The inset shows a photograph of the holder with its goniometer and the EPR magnets.}
\label{fig: ExperimentalSetupSchematics}
\end{figure}

Figure \ref{fig: ExperimentalSetupSchematics} shows a schematic of the experimental setup used in this work (see appendix D for opto-electronics box and sensor head).
A continuous-wave (cw) ODMR magnetometry scheme was employed, with a 532-nm Laser Quantum GEM laser being used to excite the NVC ensemble. The laser power was set to 1 W. Approximately 1\% of the laser power was collected with a beam sampler and directed to a Thorlab’s PD450A balanced detector for common-mode noise cancellation. A fibre-coupled sensor head was employed, with both the laser beam and the red fluorescence passing through the same 3-m multimode optical fibre, which had a diameter of 910 $\mu$m and an N.A. of 0.39.

An Element Six 1-$\textrm{mm}^3$, $\langle$100$\rangle$-oriented CVD-grown, $^{12}\textrm{C}$ purified diamond was used. It had around 13 ppm nitrogen and 3 ppm negatively charged NVC before and after electron irradiation and annealing  respectively \cite{edmonds2021characterisation}. Instead of a loop-antenna design, limited to frequencies up to approximately 3 GHz, a microstrip transmission line was used for microwave delivery.  The diamond was seated in the middle of an aperture in the transmission line and held in place by solder. The bare end of the optical fibre was directly coupled to the diamond via index matching gel. The microstrip was attached to an Al cylinder, through which the optical fibre was threaded and held in place by two plastic set screws. 

The Al cylinder sensor head was placed into a rod (see Fig. \ref{fig: ExperimentalSetupSchematics}) that was attached to a mount, with a goniometer, that allowed the sensor head to rotate 350$^{\circ}$ around the axis of the rod. The sensor head was placed into the bore of a Bruker ELEXSYS-580 cw-EPR spectrometer electromagnet. This could generate magnetic fields from 0 to $\pm$ 1.2 T (see appendix K). When aligning the sensor head within the EPR magnets, a low field of 1 to 4 mT was used, before increasing the field to 0.9 to 1.2 T. In this work, fields of approximately 0.2 T and higher are referred to as high-field and fields $<$ 100 mT as low-field.
This EPR spectrometer also has modulation coils within the bore that allowed test fields to be applied parallel to the bias field. 

For bias fields of 0.95 T, resonances can be found in the range 20 to 30 GHz (and higher including the DQ transitions). Accordingly, a KeySight E8257D  microwave source (100 kHz to 67 GHz) was employed with 2.92 mm microwave cables, suitable for frequencies up to 40 GHz. Sine-wave frequency modulation (FM) of the microwaves was employed to enable lock-in detection with a Zurich MFLI-500 kHz lock-in amplifier (LIA). A LIA output scaling of $\times$1000 was used for all measurements and the phase setting was adjusted to maximise the signal in the X-channel. The optimum microwave power and FM parameters at both low- and high-field (modulation frequency and depth) were determined using the methods outlined in Refs. \cite{el2017optimised, patel2020subnanotesla} and appendix J. For high-field measurements, a modulation depth of 4 MHz and a modulation frequency of 3.5 kHz were used, with the microwave power at the source set to +20 dBm. The optimum microwave and FM parameters were found to differ between low- and high-fields, in part because of differences in the ODMR hyperfine structure. 
This is detailed in appendices F and J.
For the ODMR spectra, a LIA low-pass filter (LPF) of 5 Hz was employed, while for the sensitivity measurements it was set to 150 Hz. This was the limiting factor on the frequency range in these measurements. The demodulated X-channel LIA signal was observed on a PicoScope 5442 oscilloscope.

\section{Results}

\begin{figure}[h!]
\centering
\includegraphics[width=\columnwidth, trim={1.5cm 1cm 1.5cm 1.5cm}]{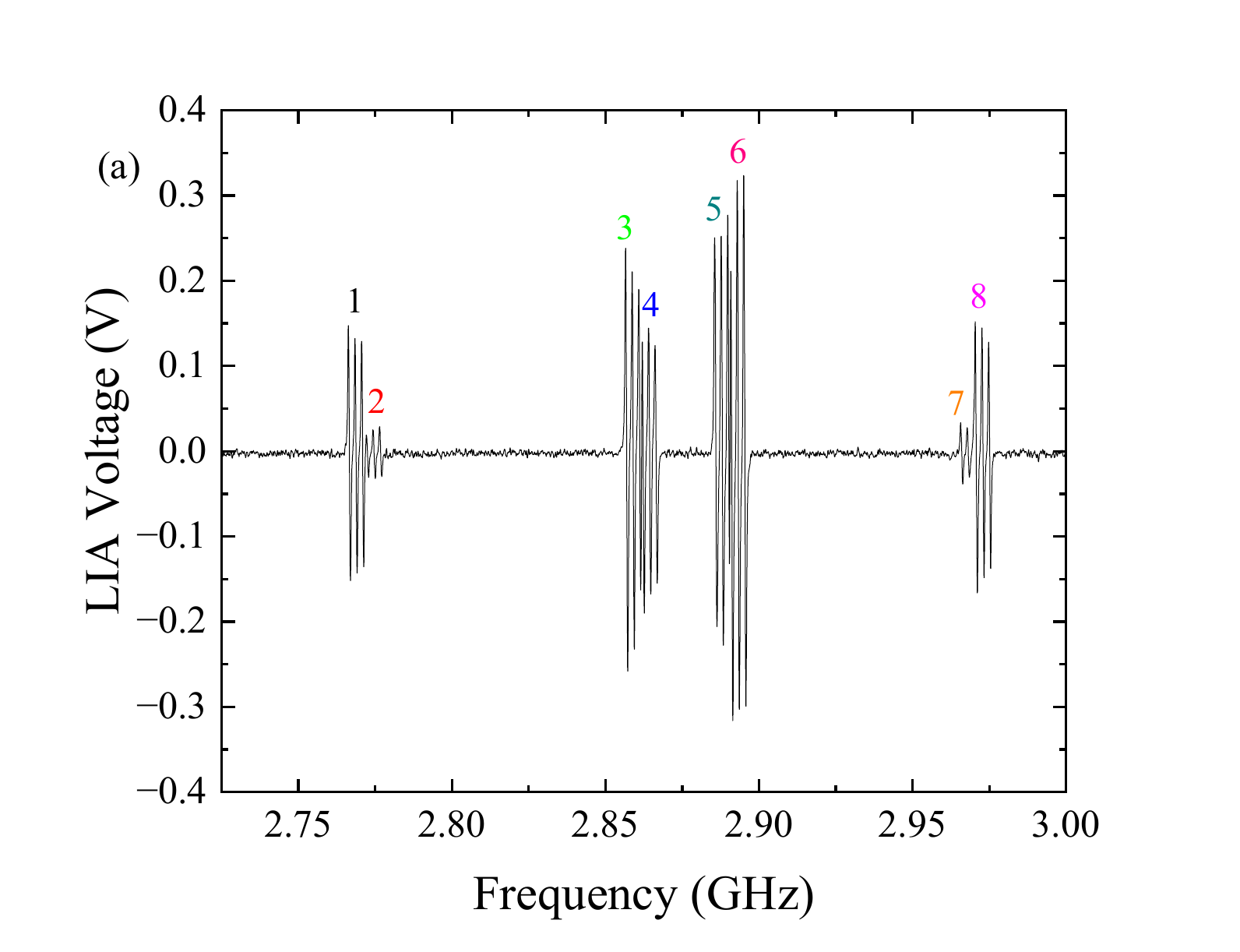}
\includegraphics[width=\columnwidth, trim={1.5cm 1cm 1.5cm 1.5cm}]{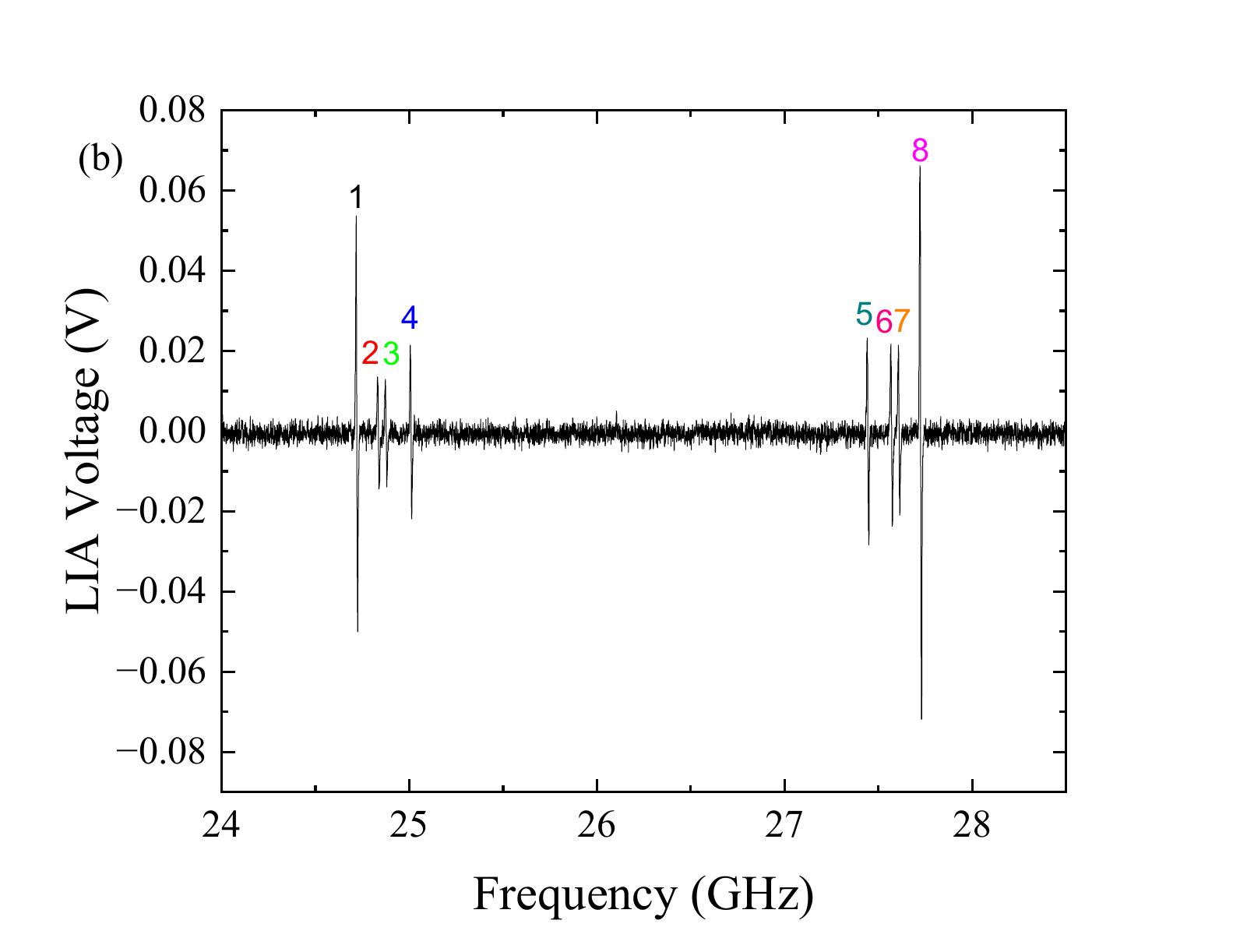}
\vspace{-5mm}
\caption{\small (a) Demodulated ODMR spectrum at a magnetic field strength of approximately 4 mT for a non-degenerate bias field alignment. A modulation depth of 400 kHz was used, such that the hyperfine resonances were visible. (b) Demodulated ODMR spectrum at a magnetic field strength of approximately 0.95 T for a non-degenerate bias field alignment. A modulation depth of 4 MHz was used. The resonances are labelled 1 to 8 for both spectra. 
}
\label{fig:NondegenerateODMR}
\end{figure}

For single-axis measurements, a $\langle$111$\rangle$ alignment is ideal. However, there is then only a significant signal for one of the four possible NVC orientations. There is also a large splitting between the resonances, which complicates the microwave engineering for some vector implementations. Vector capability is desirable and thus we took ODMR for a non-degenerate alignment. It would be difficult to align sensors precisely in a tokamak and thus this is more representative of the real use case than a perfect $\langle$111$\rangle$ alignment. These ODMR spectra are shown in Fig. \ref{fig:NondegenerateODMR}. Note that one of the NVC orientations has significantly weaker resonances; this is probably due to the microwave field ($B_1$) orientation. 

\begin{figure}[h!]
\centering
\includegraphics[width=\columnwidth, trim={1.5cm 1cm 1.5cm 1.5cm}]{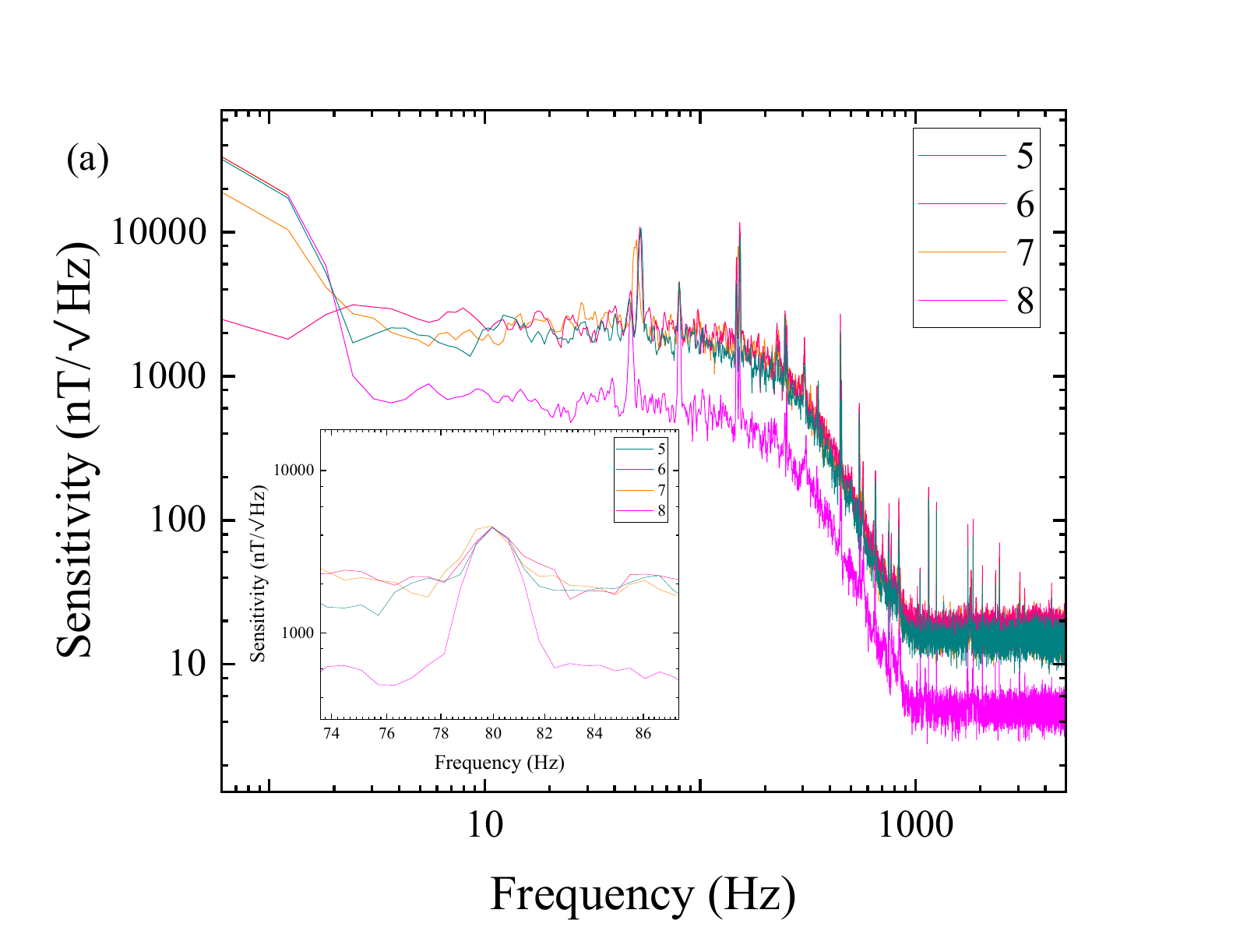}
\includegraphics[width=\columnwidth, trim={1.5cm 1cm 1.5cm 1.5cm}]{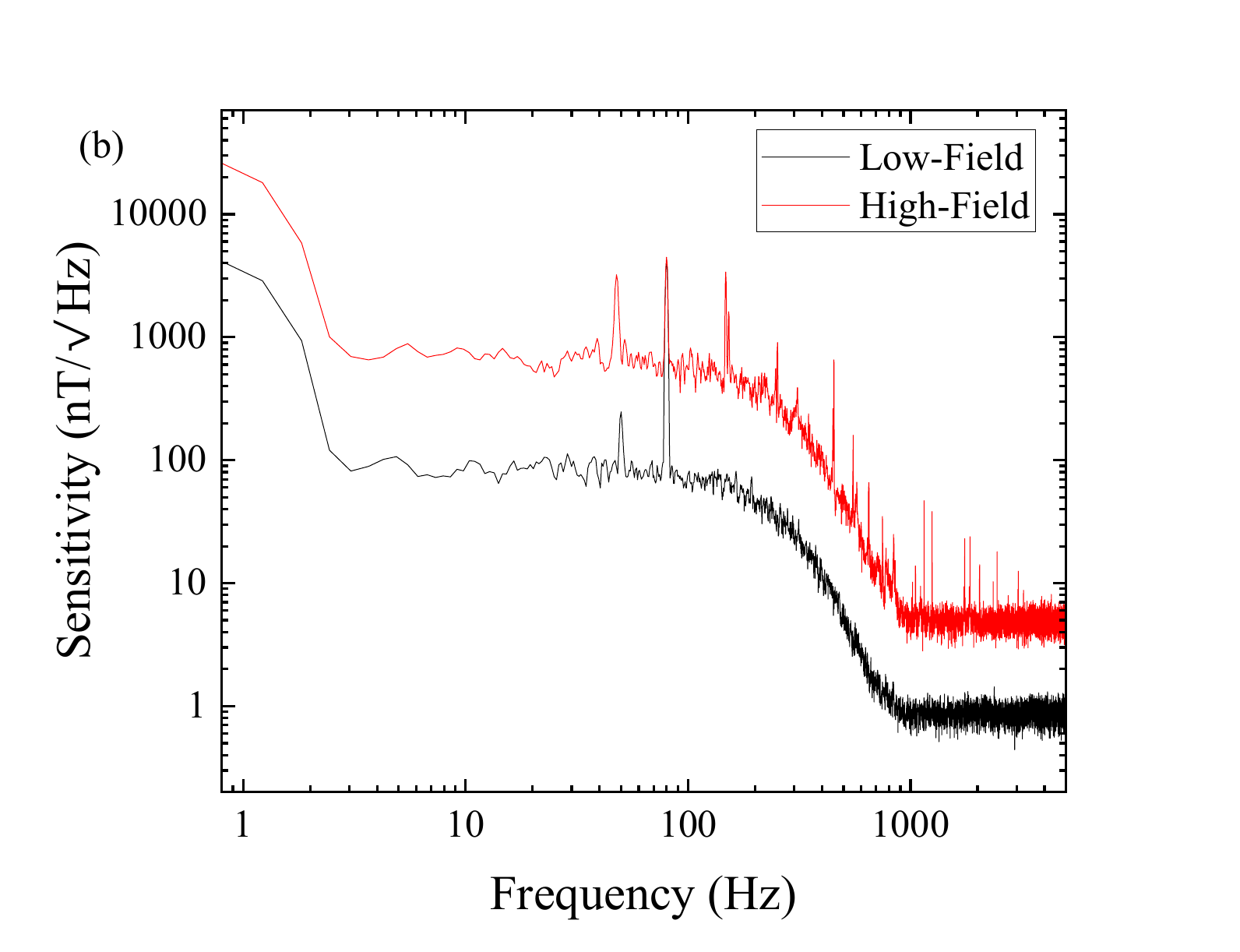}
\vspace{-5mm}
\caption{\small For a non-degenerate alignment, (a) sensitivity spectra taken on the resonances 5 to 8, labelled in Fig. \ref{fig:NondegenerateODMR}, at a magnetic field strength of approximately 0.95 T. The inset shows the 80 Hz test field. (b) Sensitivity spectra taken at low and high-field (approximately 4 mT and 0.95 T respectively) for resonance 8. A LIA LPF of 150 Hz was used.}
\label{fig:NondegenerateSensitivitySpectra}
\end{figure}

\begin{figure}[h!]
\centering
\includegraphics[width=\columnwidth, trim={1.5cm 1cm 1.5cm 1.5cm}]{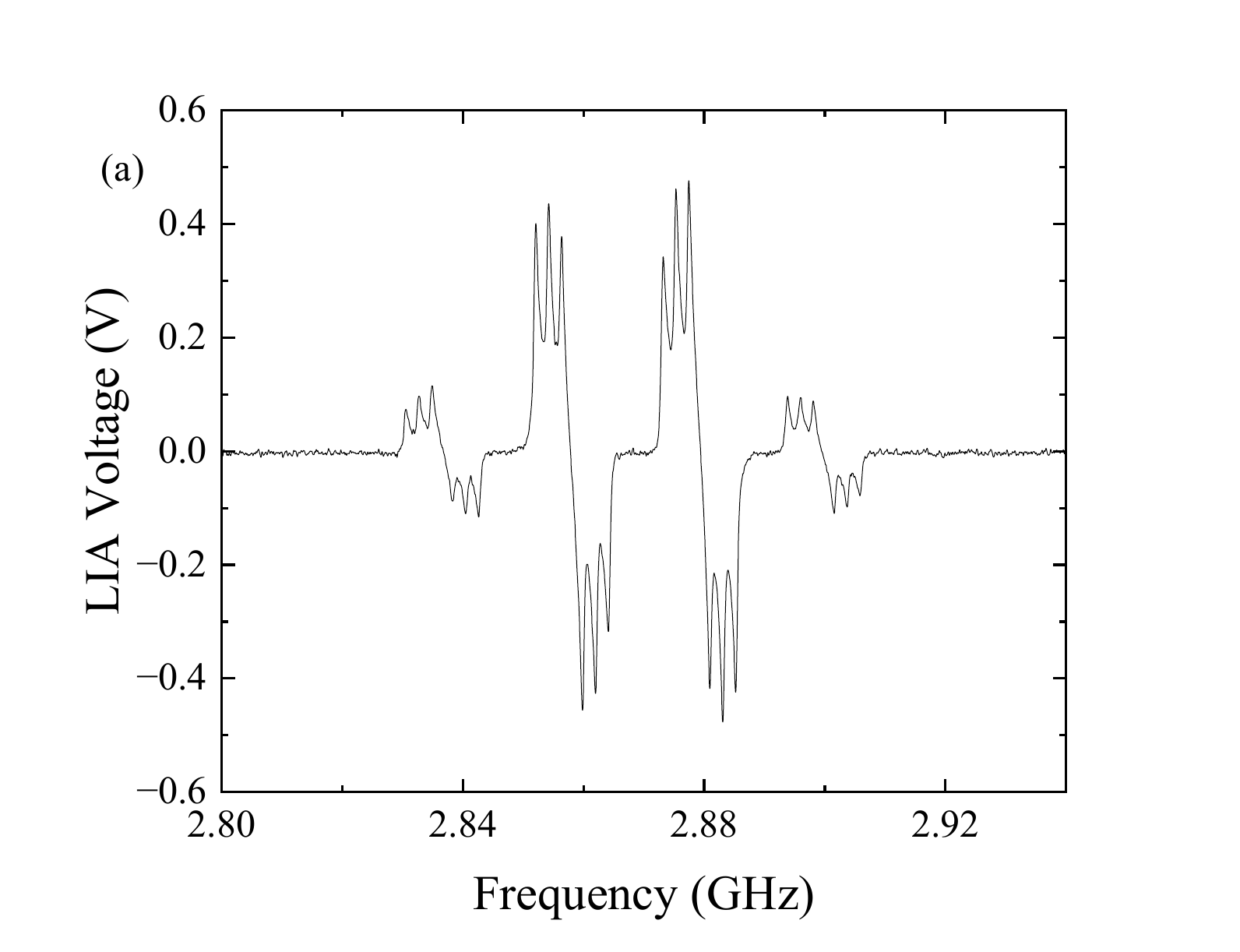}
\includegraphics[width=\columnwidth, trim={1.5cm 1cm 1.5cm 1.5cm}]{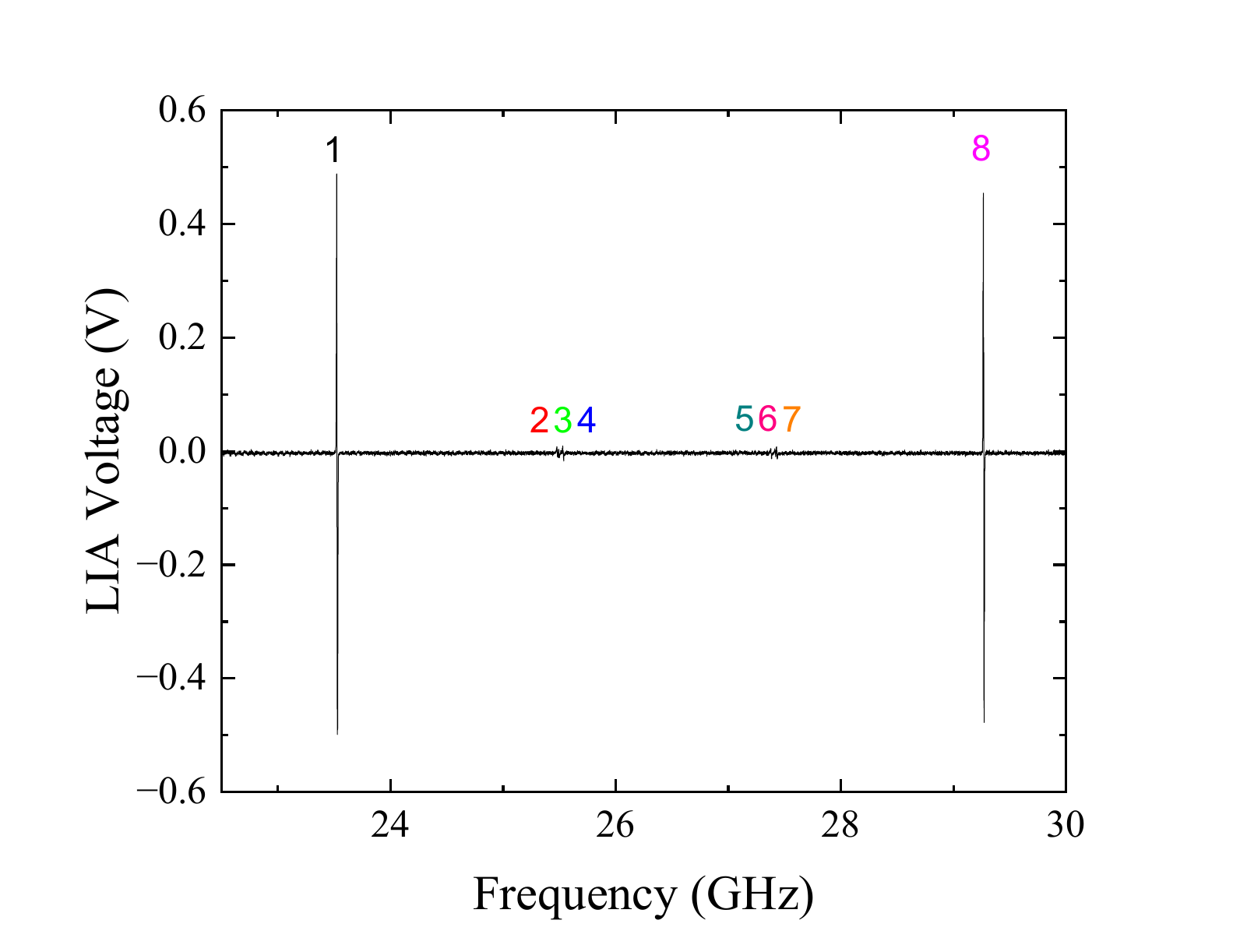}
\vspace{-5mm}
\caption{\small (a) Demodulated ODMR spectrum at a magnetic field strength of approximately 1 mT for a near-$\langle$111$\rangle$ bias field alignment. (b) Demodulated ODMR spectrum at a magnetic field strength of approximately 0.95 T for a near-$\langle$111$\rangle$ bias field alignment. The resonances are labelled 1 to 8. The splitting between resonances 1 and 8 was approximately 5.7 GHz. A modulation depth of 4 MHz was used for both measurements.}
\label{fig:Near111ODMR}
\end{figure}

Single-axis sensitivity measurements were taken by fixing the microwave frequency to the resonances 5 to 8 in Fig. \ref{fig:NondegenerateODMR}. Figure \ref{fig:NondegenerateSensitivitySpectra}a shows these sensitivity measurements at high-field (0.95 T). For all of these measurements, an 80 Hz test field was applied parallel to the bias field. Twenty 1-s voltage-traces were taken, with the average of their power-spectral-density (PSD) spectra being square-rooted to obtain the amplitude-spectral-density (ASD) \cite{barry2024sensitive}. This spectrum was converted from V/$\sqrt{\textrm{Hz}}$ to nT/$\sqrt{\textrm{Hz}}$ by dividing by a calibration constant in V/nT. This calibration constant was calculated by multiplying the zero-crossing slope (in V/MHz) of the resonance by the gyromagnetic ratio of the NVC (in MHz/nT) \cite{graham2025road}. For resonance 8, a mean-sensitivity of (600 $\pm$ 100) nT/$\sqrt{\textrm{Hz}}$ was determined for a frequency range of (10-150) Hz. The sinusoidal noise peaks were masked when calculating this mean. This is around $\times$6 lower than the low-field sensitivity, though it is apparent (see appendix H) that other non-degenerate alignments can provide higher sensitivities even at 1.2 T. As expected, the amplitude of the 80 Hz test field was approximately the same for each resonance. At high field, the sensitive axis of each orientation is largely set by the bias field, with only a small contribution from each NVC symmetry axis. 
These sensitive axes are almost parallel at sufficiently high field, and so while vector information is present in the ODMR, it is less pronounced than in the low-field regime. The vector capabilities of the magnetometer at high field are further explored in appendix B. Figure \ref{fig:NondegenerateSensitivitySpectra}b shows the difference in sensitivity for this resonance at high and low field. See appendices C and E for further sensitivity and ZCS measurements as a function of magnetic field. 

\begin{figure}[h!]
\centering
\includegraphics[width=\columnwidth, trim={1.5cm 1cm 1.5cm 1.5cm}]{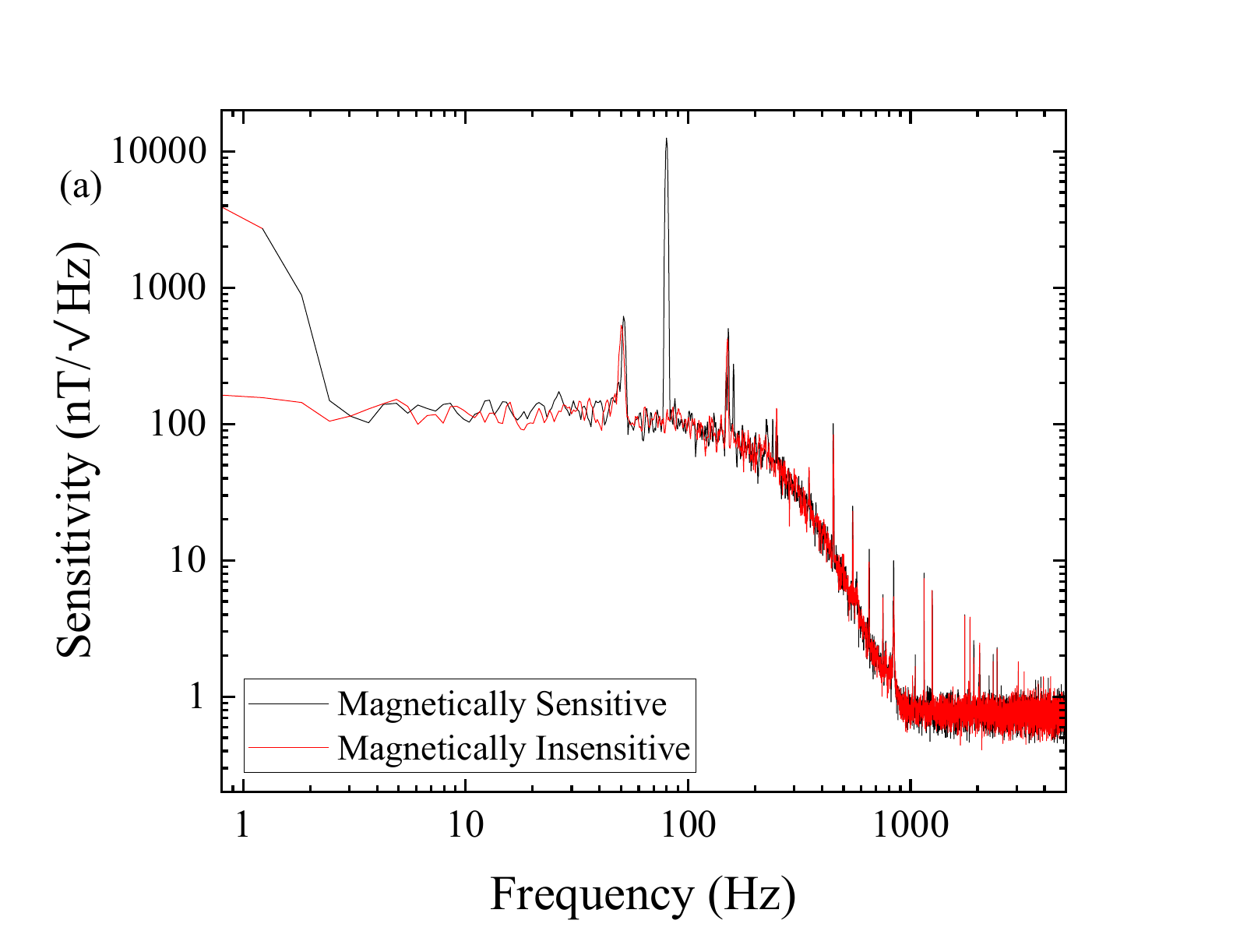}
\includegraphics[width=\columnwidth, trim={1.5cm 1cm 1.5cm 1.5cm}]{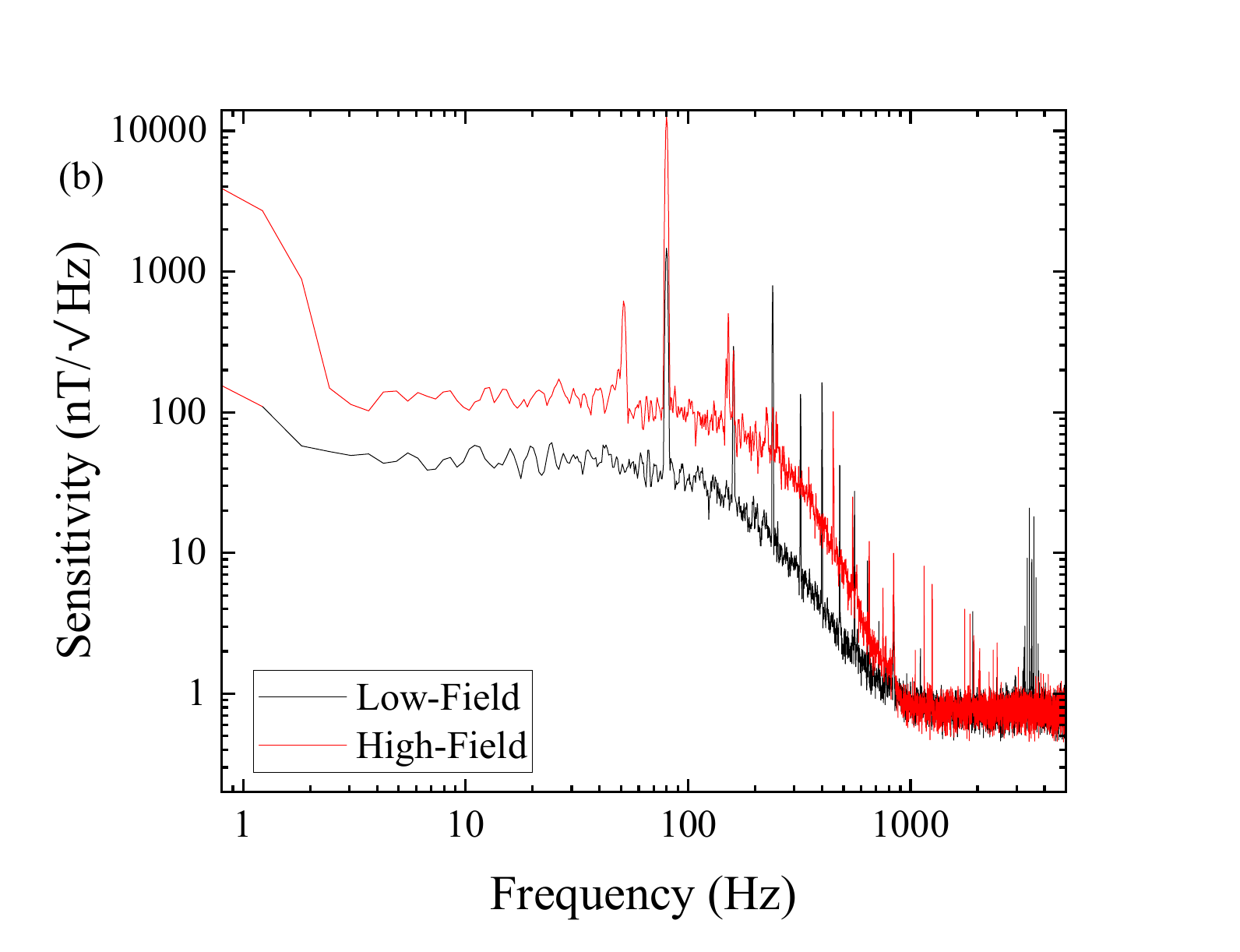}
\vspace{-5mm}
\caption{\small For a near-$\langle$111$\rangle$ alignment, (a) sensitivity spectra taken when magnetically sensitive (on-resonance) and magnetically insensitive (off-resonance) at a magnetic field strength of approximately 1 mT. (b) Sensitivity spectra taken at low and high-field (approximately 1 mT and 0.95 T respectively). A LIA LPF of 150 Hz was used. An 80 Hz test field was applied parallel to the bias field, the strength differed between low and high field.}
\label{fig:Near111SensitivitySpectra}
\end{figure}

Figures \ref{fig:Near111ODMR}a and \ref{fig:Near111ODMR}b show ODMR spectra taken with the sensor head in a near-$\langle$111$\rangle$ alignment, at low (1 to 4 mT) and high fields (0.95 T) respectively. Beyond the 1:3:3:1 structure of the ODMR spectrum at low field, we further confirmed that it was a near-$\langle$111$\rangle$ alignment by measuring the fluorescence as a function of magnetic field. This data is shown in appendix G.

The outermost resonances correspond to the NVC orientation with a symmetry axis approximately parallel to the bias field. The weaker inner resonances correspond to the other three NVC orientations, for which the transverse field components are substantial. The sensitivity measurement for resonance 8 in Fig. \ref{fig:Near111ODMR} is shown in Fig. \ref{fig:Near111SensitivitySpectra}a, alongside the noise level when off-resonance. A mean-sensitivity of (110 $\pm$ 20) nT/$\sqrt{\textrm{Hz}}$ was determined for a frequency range of (10-150) Hz. The frequency range was limited by the LIA LPF. The mean of the noise floor was (100 $\pm$ 20) nT/$\sqrt{\textrm{Hz}}$ off-resonance, suggesting that the sensitivity was limited by laser noise. The low-field sensitivity was lower than the non-degenerate case due to changes in the fluorescence collection. Figure \ref{fig:Near111SensitivitySpectra}b shows the on-resonance sensitivity measurement compared to a measurement
at low field. The sensitivity is found to be $\times$3 to $\times$4 lower at high than low field, accounting for differing balanced detection optimisation. 
For the outer resonance no reduction in sensitivity would be expected, however, the alignment is not perfectly along one of the $\langle$111$\rangle$ axes.

\FloatBarrier

\section{Discussion and Conclusions}

Although the sensitivity is significantly reduced, it is possible to obtain ODMR with sub-microtesla sensitivities at high fields.
Given the strength of the magnetic field, at high field the relative magnetic field sensitivity can be written as approximately 0.1 to 0.6 ppm/$\sqrt{\textrm{Hz}}$, compared to approximately 7 to 9 ppm/$\sqrt{\textrm{Hz}}$ at low field, accounting for differences in noise levels. 
The sensitivity of the magnetometer at both high and low fields is currently limited by technical noise such as that from the laser. These high-field sensitivities are more than sufficient for tokamak diagnostics, in a mm-scale package, and this is obtained starting with a low-field sensitivity significantly below the state of the art \cite{entler2019prospects, barry2024sensitive, barry2016optical}.  
Further improvements in sensitivity could be obtained via increasing the green-to-red photon conversion efficiency ($\approx$ 0.01\%) and refining the microstrip design. At high field, obtaining an alignment closer to a $\langle$111$\rangle$ could yield significant further improvements in sensitivity. Alternative non-degenerate alignments could also have higher sensitivities. Embedded sensors would have to be placed to avoid the total magnetic field being parallel to the $\langle$100$\rangle$ alignments (see appendix I). 
Further development would be required for operation within a future DEMO or commercial tokamak. A high dynamic range and slew rate would be necessary, on account of both the large static and dynamic magnetic fields.  
This would require multiple resonance tracking as opposed to calibration-dependent single-axis measurements \cite{schloss2018simultaneous, graham2025road}. In principle, NVC magnetometers would provide a high frequency range of up to 100 kHz and can achieve slew rates of at least 0.723 T/s \cite{barry2020sensitivity, quercia2022long, wang2023realization}. Magnetic field strengths $>$ 1 T would also require higher frequency microwaves, making coaxial cables less suitable. It is desirable to measure both the poloidal and radial components local to each sensor. Vector capabilities will be reduced at high field (see Appendix B). 

Operating at 300$^{\circ}$C would result in further reductions in sensitivity, assuming that cooling systems are not employed \cite{entler2019prospects, kubota2023wide, toyli2012measurement}. 
Long-term temperature stability would also be important (see Appendix B) \cite{fescenko2020diamond}. 
An array consisting of hundreds of diamonds would be required to cover the entirety of a tokamak, although it could be used in particular locations to supplement other magnetic sensors \cite{entler2019prospects}.  Although radiation levels are significantly lower behind the vacuum vessel, installation in-vessel is necessary to prevent shielding impacting the response time \cite{hodapp1995magnetic, biel2019diagnostics}. These sensors would need to be robust against electromagnetic and mechanical shocks
\cite{entler2019prospects}. The diamonds would need to be instrumented for remote operation with radiation-hard optical fibres or free-space optics, as well as radiation-hard microwave lines at distances of around 50 m. 
Depending on the sensor position, in situ maintenance or replacement would not be possible during the lifetime of the tokamak and a large number of feedthroughs are undesirable \cite{biel2019diagnostics}. Hollow-core fibres could be an option, though these are typically single-mode, single-wavelength and designed for wavelengths higher than 532 nm \cite{gu2023radiation, poletti2013hollow}. Two feedthroughs would be required for each sensor head, the diamonds themselves could be sub-mm in size, with thinner diamonds preferable for radiation hardness \cite{bauer1995radiation, passeri2021assessment}. 
There are also questions about the radiation hardness of diamond. Current tokamaks produce significantly lower fluences than is expected for future commercial devices, making testing difficult \cite{biel2019diagnostics}. Another concern is the effect of gamma radiation, which could photo-ionise negatively charged NVCs to the neutral charge state, which cannot be used for magnetometry \cite{barry2020sensitivity, razinkovas2021photoionization}. 

In conclusion, we present a fibre-coupled ensemble NVC magnetometer operational at fields up to 1.2 T with unshielded sensitivities of approximately 240 to 600 nT/$\sqrt{\textrm{Hz}}$ and 110 nT/$\sqrt{\textrm{Hz}}$ for non-degenerate and near-$\langle$111$\rangle$ alignments respectively. With further development, these could be used for fusion power diagnostics and other high-radiation applications.

\section{Acknowledgements}

The Ph.D. studentships of S. M. G. and A. J. N. are funded by the Defence Science and Technology Laboratory (DSTL) and the National Nuclear Laboratory (NNL) respectively. We thank Reza Mirfayzi at Tokamak Energy for many useful discussions about tokamak diagnostics. We thank Ben Green at the University of Warwick for useful conversations relating to forbidden transitions and the effect on the hyperfine structure. We thank the following people at UKAEA for useful discussions: David Croft, Norman Lam, David Ryan, John Steele, Eddie Pennington, Chantal Shand, Terry Thompson, Morten Lennholm, Neil Conway, Alexandros Plianos. This work is supported by the UK Hub NQIT (Networked Quantum Information Technologies) and the UK Hub in Quantum Computing and Simulation, part of the UK National Quantum Technologies Programme, with funding from UK Research and Innovation (UKRI) EPSRC Grants No. EP/M013243/1 and No. EP/T001062/1, respectively. This work is also supported by Innovate UK Grant No. 10003146, EPSRC Grant No. EP/V056778/1 and EPSRC Impact Acceleration Account (IAA) awards (Grants No. EP/R511808/1 and No. EP/X525844/1). This project has been supported by UK Atomic Energy Authority through the Fusion Industry Programme. The Fusion Industry Programme is stimulating the growth of the UK fusion ecosystem and preparing it for future global fusion powerplant market. More information about the Fusion Industry Programme can be found online: https://ccfe.ukaea.uk/programmes/fusion-industry-programme/

\FloatBarrier

\section*{Appendix A: NVC Physics and Simulations}

\begin{figure}[h!]
\centering
\includegraphics[width=\columnwidth]{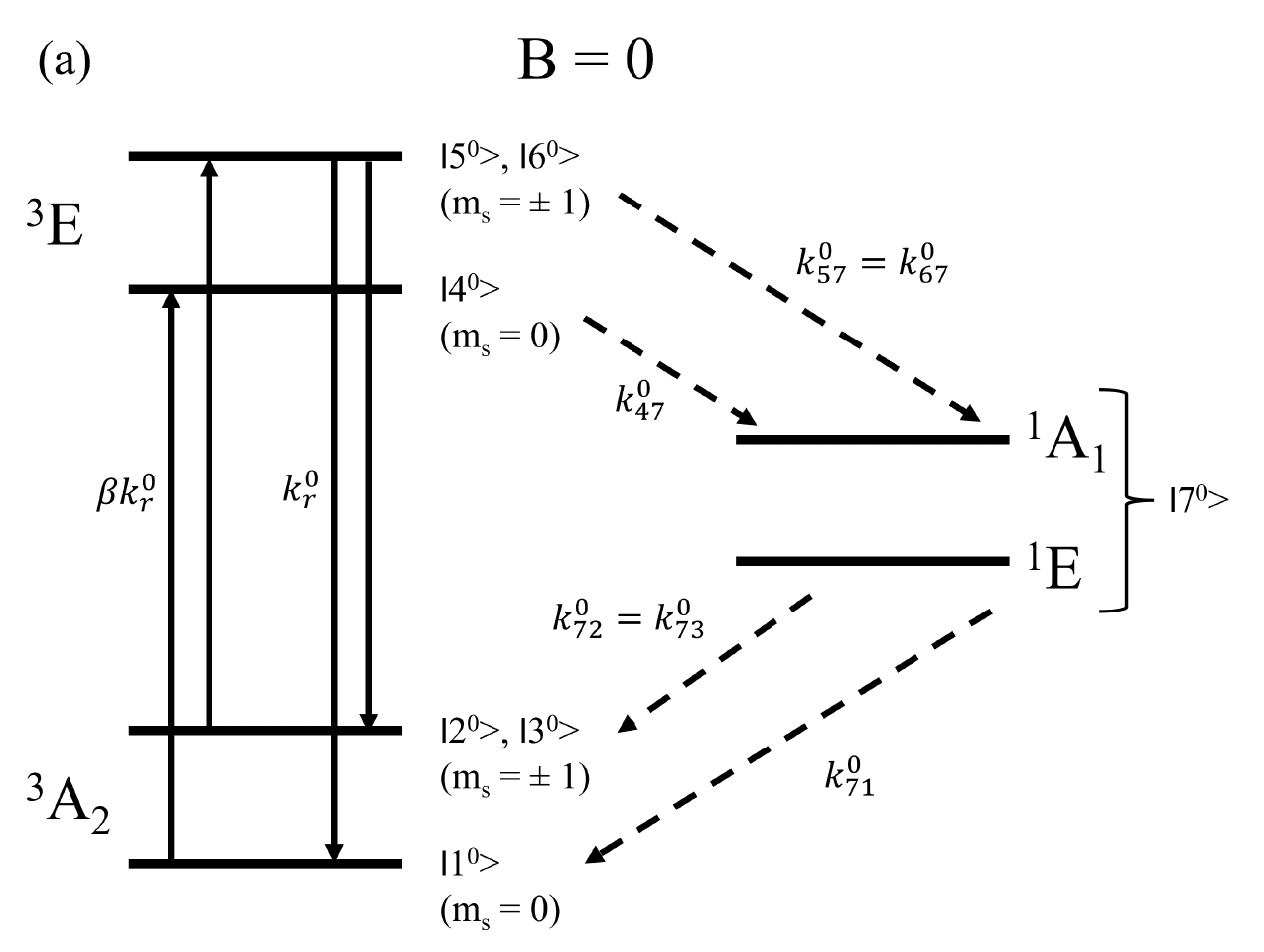}
\includegraphics[width=\columnwidth]{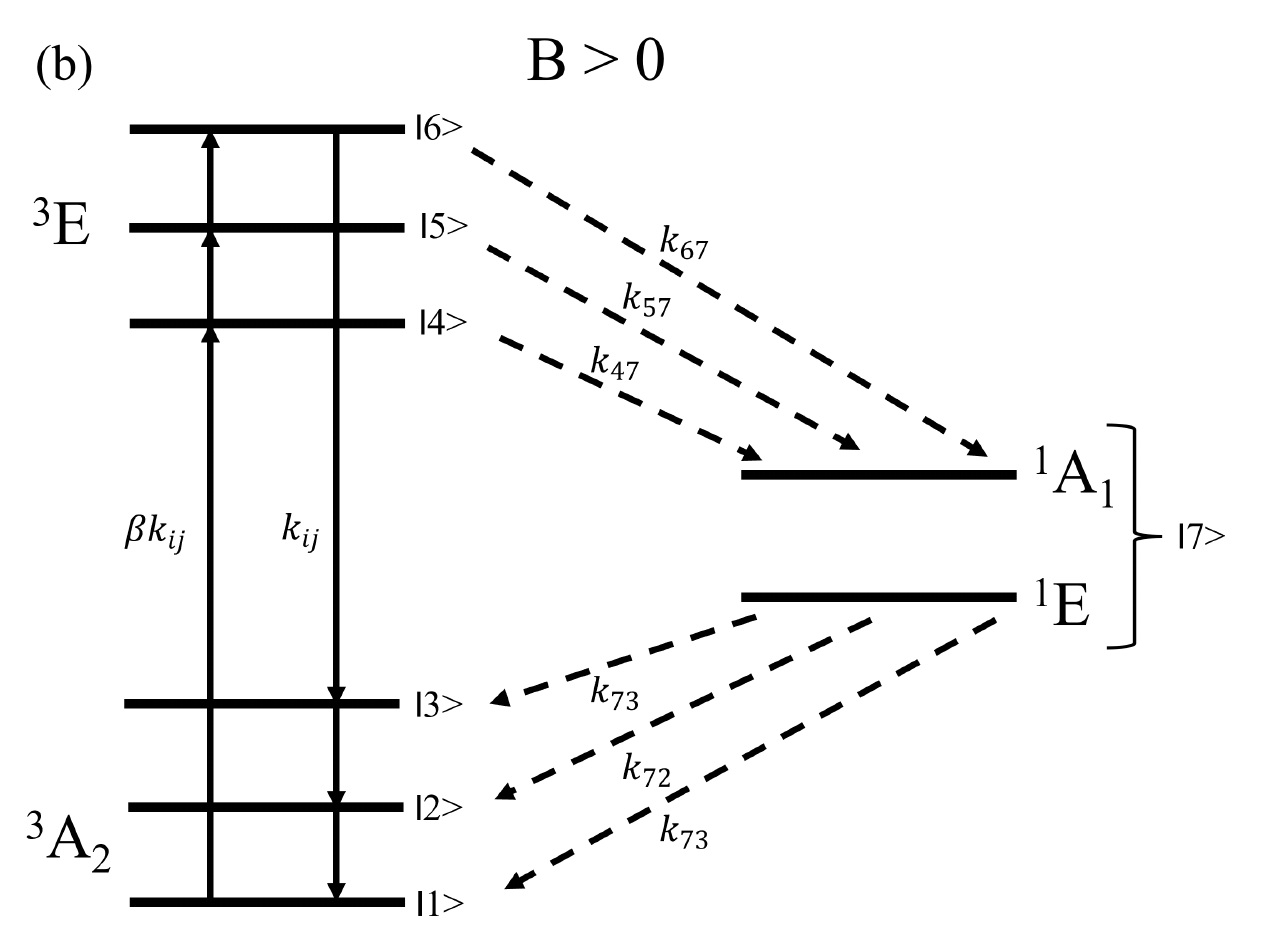}
\vspace{-5mm}
\caption{\small (a) and (b) Energy level structure of the NVC at zero-field and with a magnetic field applied respectively. The transition sub-levels and transition rates (at zero and non-zero field are indicated). $\beta$ is a proportionality constant. The B $>$ 0 energy level structure assumes a bias field $<$ 102.4 mT. Based on Ref. \cite{tetienne2012magnetic}.}
\label{fig: AppendixA-Physical+EnergyLevelStructureofNVC}
\end{figure}

The NVC is a point defect in diamond that consists of a nitrogen atom adjacent to a vacancy. It can exist in a positive, neutral, or negative charge state. The negative charge state is of interest for magnetometry and has an S = 1 spin state \cite{doherty2013nitrogen, barry2020sensitivity}. Figure \ref{fig: AppendixA-Physical+EnergyLevelStructureofNVC}a and \ref{fig: AppendixA-Physical+EnergyLevelStructureofNVC}b show the energy level structure of the negatively charged NVC at zero-field and with a magnetic field applied respectively. 

\begin{figure}[t]
\centering
\includegraphics[width=\columnwidth, trim={1.5cm 1cm 1.5cm 1.5cm}]{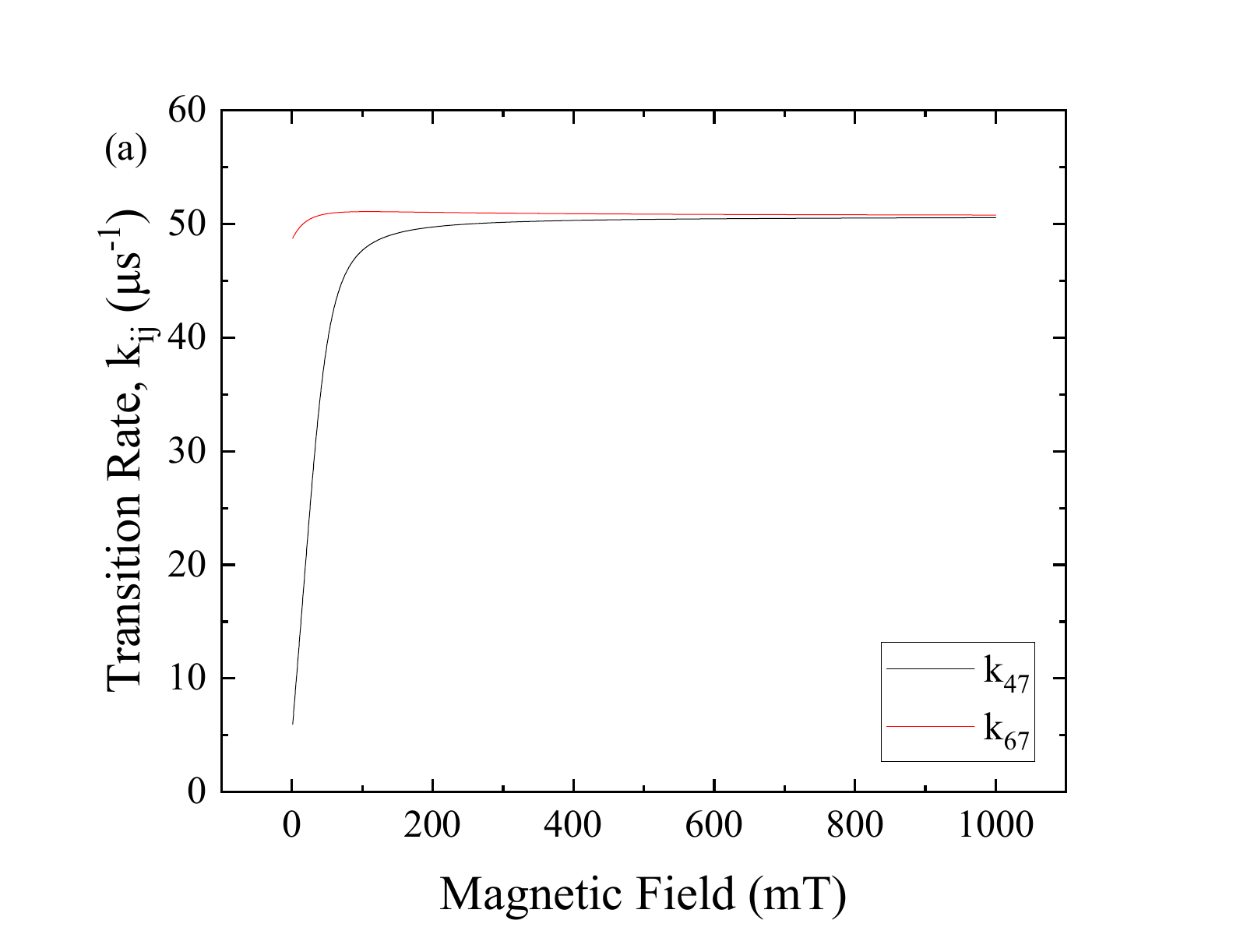}
\includegraphics[width=\columnwidth, trim={1.5cm 1cm 1.5cm 1.5cm}]{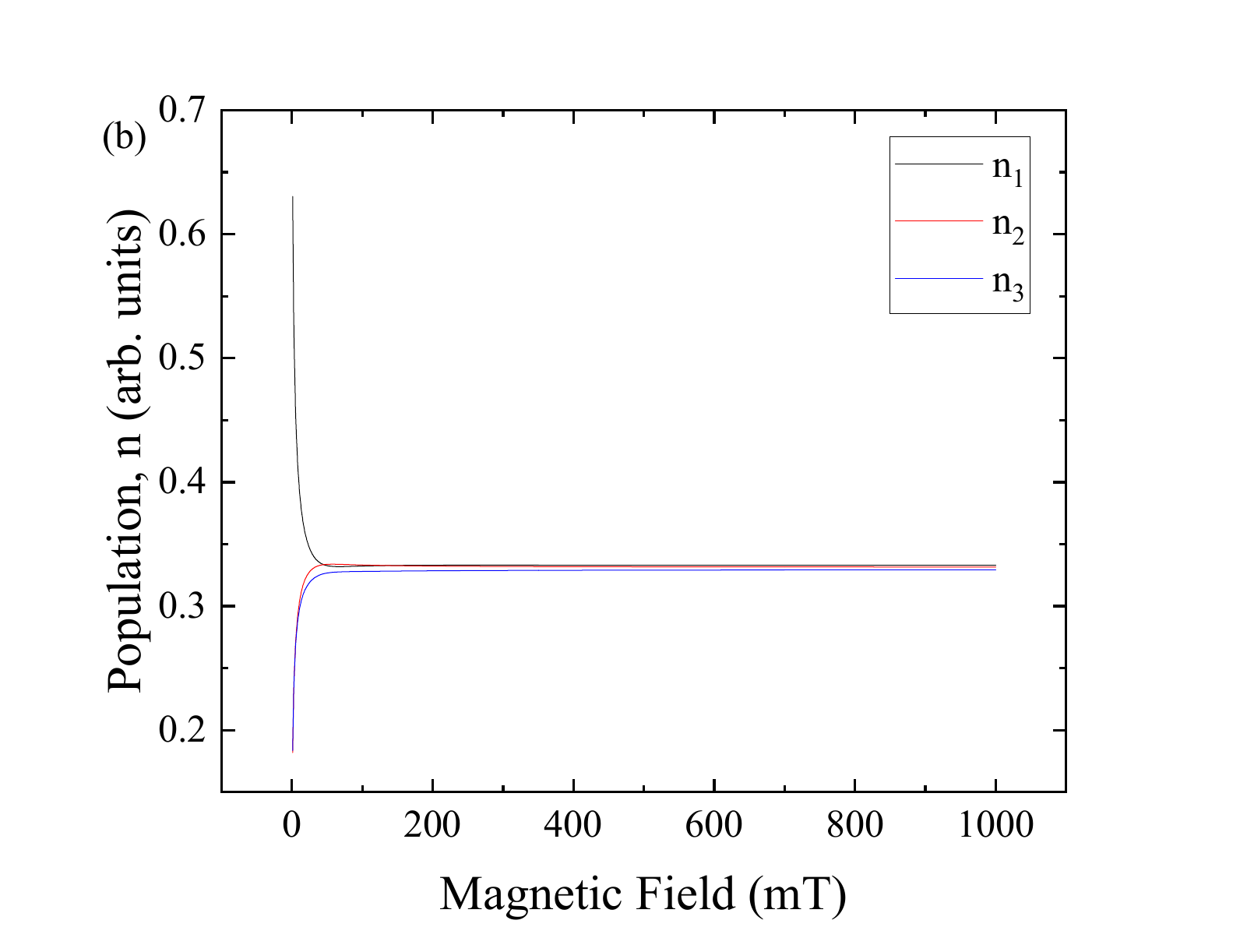}
\vspace{-5mm}
\caption{\small (a) The transition rates $k_{47}$ and $k_{67}$ calculated using the approach outlined in Ref. \cite{tetienne2012magnetic} as a function of magnetic field. (b) The corresponding relative populations of the states $\vert$1$\rangle$, $\vert$2$\rangle$, and $\vert$3$\rangle$ as a function of magnetic field. All for one of the four possible NVC orientations with a non-degenerate field alignment.}
\label{fig: AppendixA-Simulation-TransitionRates}
\end{figure}

There is a spin-triplet ground ($^3\textrm{A}_2$) and excited ($^3\textrm{E}$) state, as well as two singlet levels. For the energy-level structure, the zero-field ground state eigenstates are labelled $\vert$$1^0$$\rangle$, $\vert$$2^0$$\rangle$, and $\vert$$3^0$$\rangle$, and similarly for the excited state. The singlet states can be represented by one eigenstate, $\vert$$7^0$$\rangle$. The eigenstate $\vert$$1^0$$\rangle$ is represented as $\vert$$\psi$$\rangle$ = 0$\vert$$m_s$=-1$\rangle$ + 1$\vert$$m_s$=0$\rangle$ + 0$\vert$$m_s$=+1$\rangle$ = $\vert$$m_s$=0$\rangle$.
When the magnetic field, B $\not=$ 0, the ground-state eigenstates become $\vert$1$\rangle$, $\vert$2$\rangle$ and $\vert$3$\rangle$. As an example, for a bias field $\textrm{B}$ = (0.0335,0.5622,0.7650) T, the eigenstate $\vert$1$\rangle$ = (-0.0261 + 0.8737i)$\vert$$m_s$=-1$\rangle$ + (0.3388 - 0.3288i)$\vert$$m_s$=0$\rangle$ + (0-0.1141i)$\vert$$m_s$=+1$\rangle$, for one of the four possible NVC orientations. In this work, the lab frames x, y and z axes are defined by the crystallographic axes [100], [010] and [001], respectively. 
The zero-field and non zero-field transition rates, $k_{ij}$, are also labelled in the figure. Table \ref{table: AppendixA-TransitionRates} shows typical transition rates for zero field and a bias field, B = (0.0335,0.5622,0.7650). $k_{r}$ is the transition rate from the excited state to the ground state, this is spin-conserving at zero-field. From this table and Figs. \ref{fig: AppendixA-Simulation-TransitionRates}a and \ref{fig: AppendixA-Simulation-TransitionRates}b it can be seen why, in this picture, the spin-polarisation, contrast and fluorescence are reduced for large transverse fields, as $k_{47}$ approaches $k_{57}$ and $k_{67}$ for instance, while the populations in the ground-state equalise \cite{tetienne2012magnetic}.  

\begin{table}[h!]
\centering
\begin{tabular}{ |c|c|c|c|c|c| } 
 \hline
Transition Rate $k_{ij}$ & Zero-field ($\mu$$s^{-1}$) & High-field (ND) ($\mu$$s^{-1}$) \\ 
\hline
$k_{47}$ & 5.2 & 50.5\\
$k_{67}$ & 48.6 & 43.0\\
$k_{71}$ & 1.5 & 2.0\\
$k_{72}$ & 1.4 & 3.0\\
 \hline
\end{tabular}
\caption{Experimental transition rates for the NVC at zero-field, taken from Ref. \cite{tetienne2012magnetic}, 
as well as the modified transition rates at high-field, for the non-degenerate field alignment. The high-field transition rates were calculated considering the weighted contribution of each zero-field eigenstate to the new eigenstates, following the approach outlined in Refs. \cite{tetienne2012magnetic} and \cite{patel2024single}. High-field refers to a total magnetic field strength of 0.95 T and ND to a non-degenerate alignment. All for one of the four possible NVC orientations with a non-degenerate field alignment. 
}
\label{table: AppendixA-TransitionRates}
\end{table}

MATLAB simulations were used to help interpret the experimental data described in the main paper. Bias field alignments (in the lab frame $\textrm{B}$ = ($\textrm{B}_x$, $\textrm{B}_y$, $\textrm{B}_{z}$)).  were obtained using the fminsearch function to fit the low-field ODMR spectra with the full spin-Hamiltonian in the NVC body frame of each respective alignment 

\begin{multline}
\label{eq:NVGroundStateH}
H/h = DS_z^2 + \gamma_e(B_x'S_x+B_y'S_y+B_z'S_z) + P(I_z^2 - I(I+1)/3) \\ + A_\parallel I_z + A_\bot(S_xI_x + S_yI_y),
\end{multline}

where P is the quadrupole moment, I is the nuclear spin (I = 1 for a $^{14}\textrm{N}$ nuclei) and $\textrm{I}_x$, $\textrm{I}_y$, and $\textrm{I}_z$ are the nuclear spin operators. $A_\parallel$ and $A_\bot$ are the axial and transverse hyperfine constants, respectively \cite{barry2020sensitivity}. Rotation matrices were used to transfer between the lab and body frame. The spin-Hamiltonian could then be evaluated for a given magnetic field to obtain the eigenvalues (three for each NVC orientation, nine including hyperfine) and eigenvectors. The expected resonance frequencies at different field strengths up to 1.2 T and beyond could be determined. The relative strength of each resonance could also be calculated using the rate equations approach (see Ref. \cite{tetienne2012magnetic}), EasySpin simulations, and an approach outlined in Ref. \cite{jeong2017understanding}.
This latter approach assumed the relative strength of the resonances to be proportional to a factor $\kappa$, equal to

\begin{equation}
\label{eq:kappaFactor}
\kappa = \vert\langle\psi_{f}\vert\gamma_e(S_x+S_y)\vert\psi_{i}\rangle\vert^2(\Delta\langle\rho\rangle)(\Delta\langle S_z^2\rangle)
\end{equation}

where 

\begin{equation}
\label{eq:DeltaRho}
\Delta \langle \rho\rangle = \langle\psi_f\vert\rho\vert\psi_f\rangle - \langle \psi_i\vert\rho\vert\psi_i\rangle
\end{equation}

and 

\begin{equation}
\label{eq:DeltaSz2}
\Delta \langle S_z^2\rangle = \langle\psi_f\vert S_z^2\vert\psi_f\rangle - \langle \psi_i\vert S_z^2\vert\psi_i\rangle.
\end{equation}

$\rho$ is the density operator, which, assuming 100\% optical polarisation into the $m_s$ = 0 sub-level is given by,

\begin{equation}
\label{eq:Rho}
\rho = E - S_z^2
\end{equation}

where E is the identity operator. $\langle$$\psi_{f}$$\vert$($S_x$+$S_y$)$\vert$$\psi_{i}$$\rangle$ is the transition matrix element for the magnetic dipole transition between a given initial and final eigenstate, under the assumption of a $B_1$ field perpendicular to the NVC symmetry axis \cite{jeong2017understanding}. $\Delta$$\langle$$\rho$$\rangle$ relates to the population difference between the initial and final eigenstate and is a measure of the degree of spin-polarisation. $\Delta$$\langle$$S_z^2$$\rangle$ is proportional to the change in fluorescence, the terms being a measure of the $m_s$ = 0 (and thus bright or dark) character of each eigenstate. 
Simulations were also made using EasySpin \cite{stoll2006easyspin}.

\FloatBarrier

\section*{Appendix B: Vector Measurements}

Tokamak magnetic diagnostics require vector measurements \cite{entler2019prospects}. In the regime where $\gamma_e$$\textrm{B}$ $<<$ D, ensemble NVC magnetometers provide high-accuracy and low non-orthogonality vector measurements in a single diamond \cite{schloss2018simultaneous}. This exploits the presence of four distinct quantisation axes, set by the four fixed and unique $\langle$111$\rangle$ orientations of the NVC symmetry axis within the tetrahedral diamond lattice. From perturbation theory, in the low-field regime it can be shown that to second order the frequency of a given ($m_s$ = 0 to $m_s$ = +1 or $m_s$ = 0 to $m_s$ = -1) resonance is 

\begin{equation}
\label{eq:LowFieldPT}
f_{\pm} = D \pm \gamma_e \textrm{B}_{\textrm{NVC}} + \frac{3}{2}\frac{\gamma_e^2 \textrm{B}_{\perp}^2}{D},
\end{equation}

where $\textrm{B}_{\textrm{NVC}}$ is the magnetic field projection along a given NVC symmetry axis and $\textrm{B}_{\perp}$ the perpendicular field magnitude to this axis \cite{welter2022scanning, beaver2024optimizing}. This formula is expressed in the NVC body frame. For a small test-field, $\delta$B $<<$ $\textrm{B}$, to first order the frequency shift of each resonance, $\delta$f = $\gamma_e$$\delta \textrm{B}_{\textrm{NVC}}$. Accordingly, from measuring the shift in frequency for each resonance, neglecting higher order terms, the projection of the magnetic field shift along each NVC symmetry axis can be determined. This provides a set of simultaneous equations that can be solved to obtain the magnetic field shift components ($\delta \textrm{B}_x$, $\delta \textrm{B}_y$, $\delta \textrm{B}_z$) for an arbitrary lab Cartesian coordinate frame. 

\begin{figure*}[h!]
\centering
\includegraphics[width=\columnwidth]{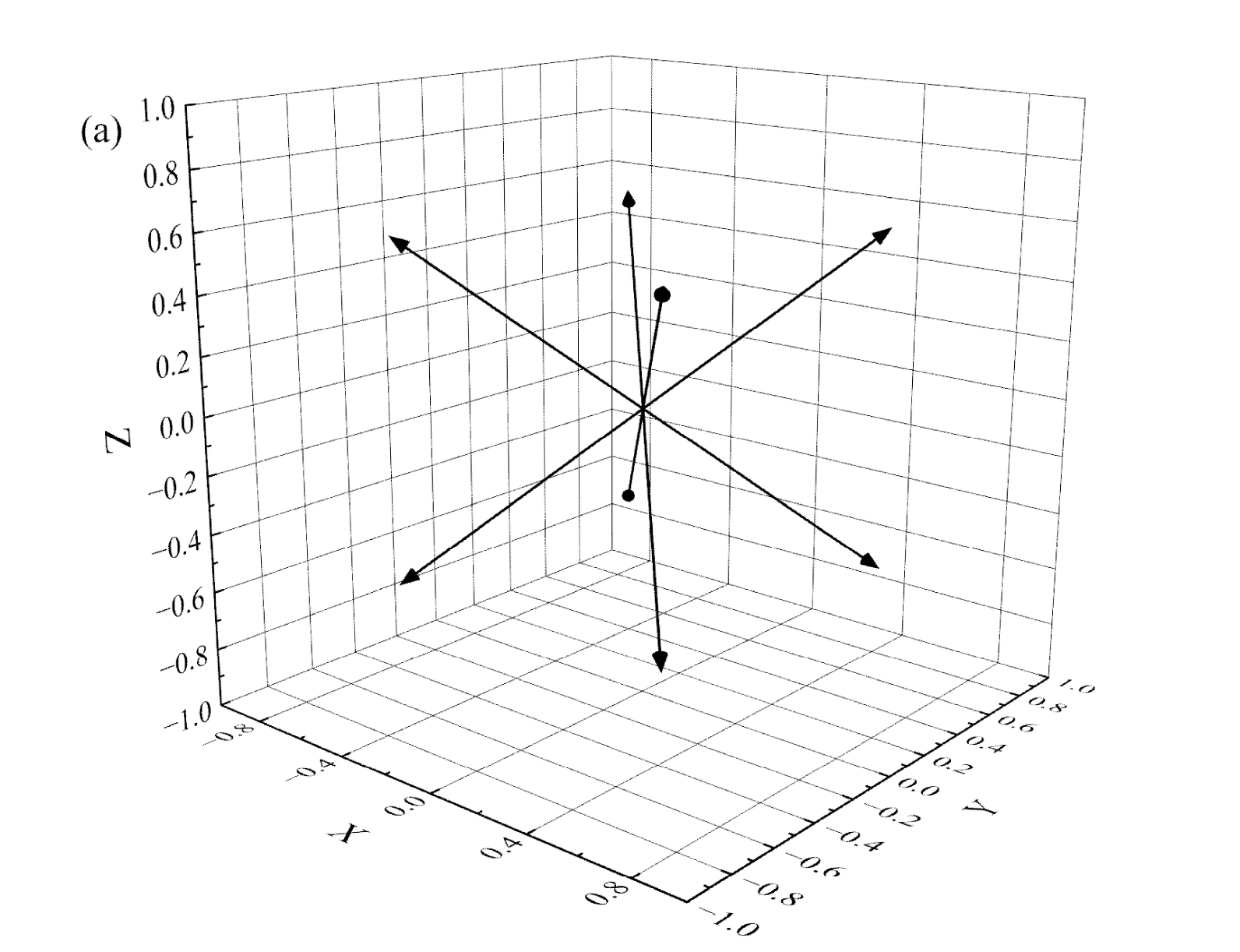}
\includegraphics[width=\columnwidth]{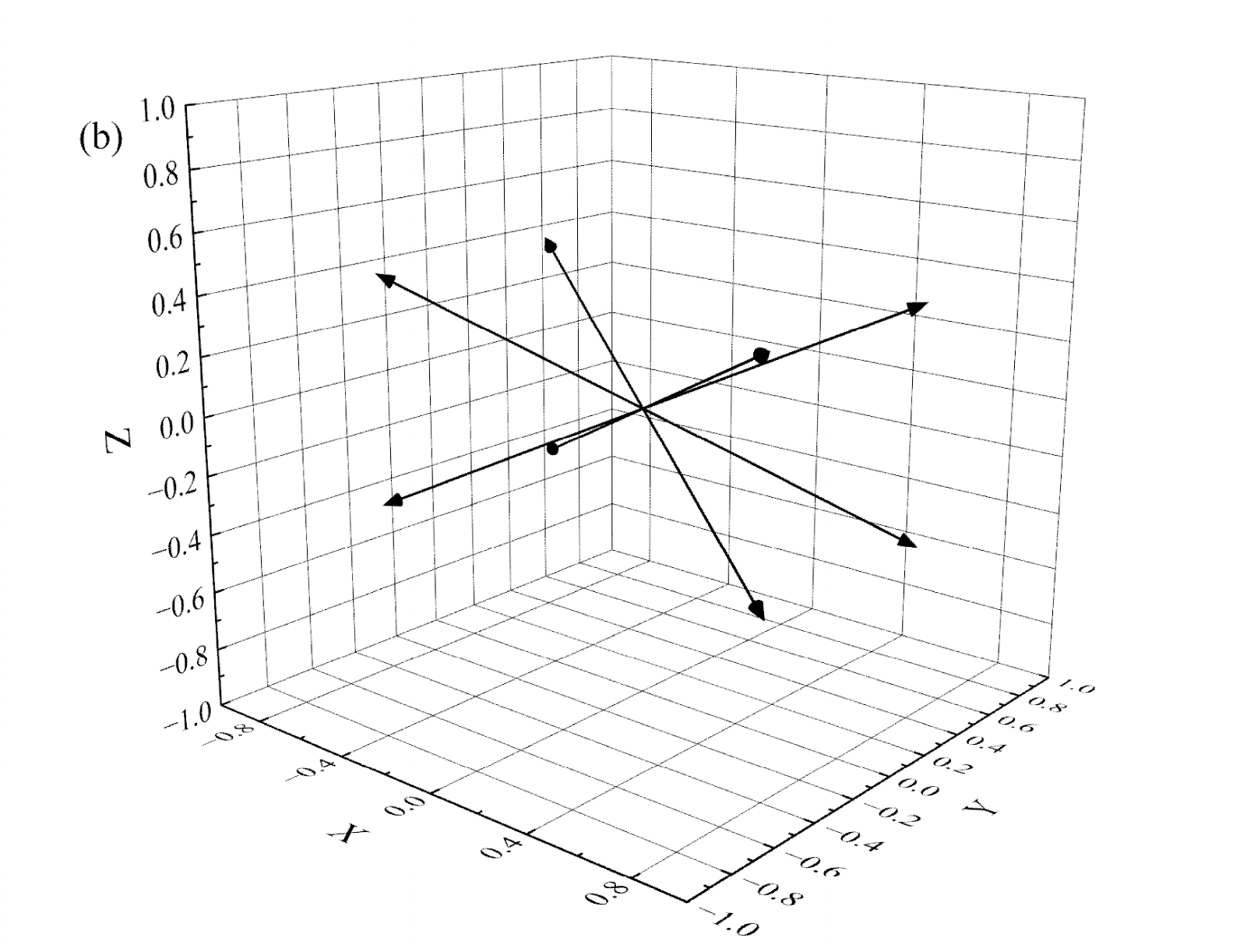}
\includegraphics[width=\columnwidth]{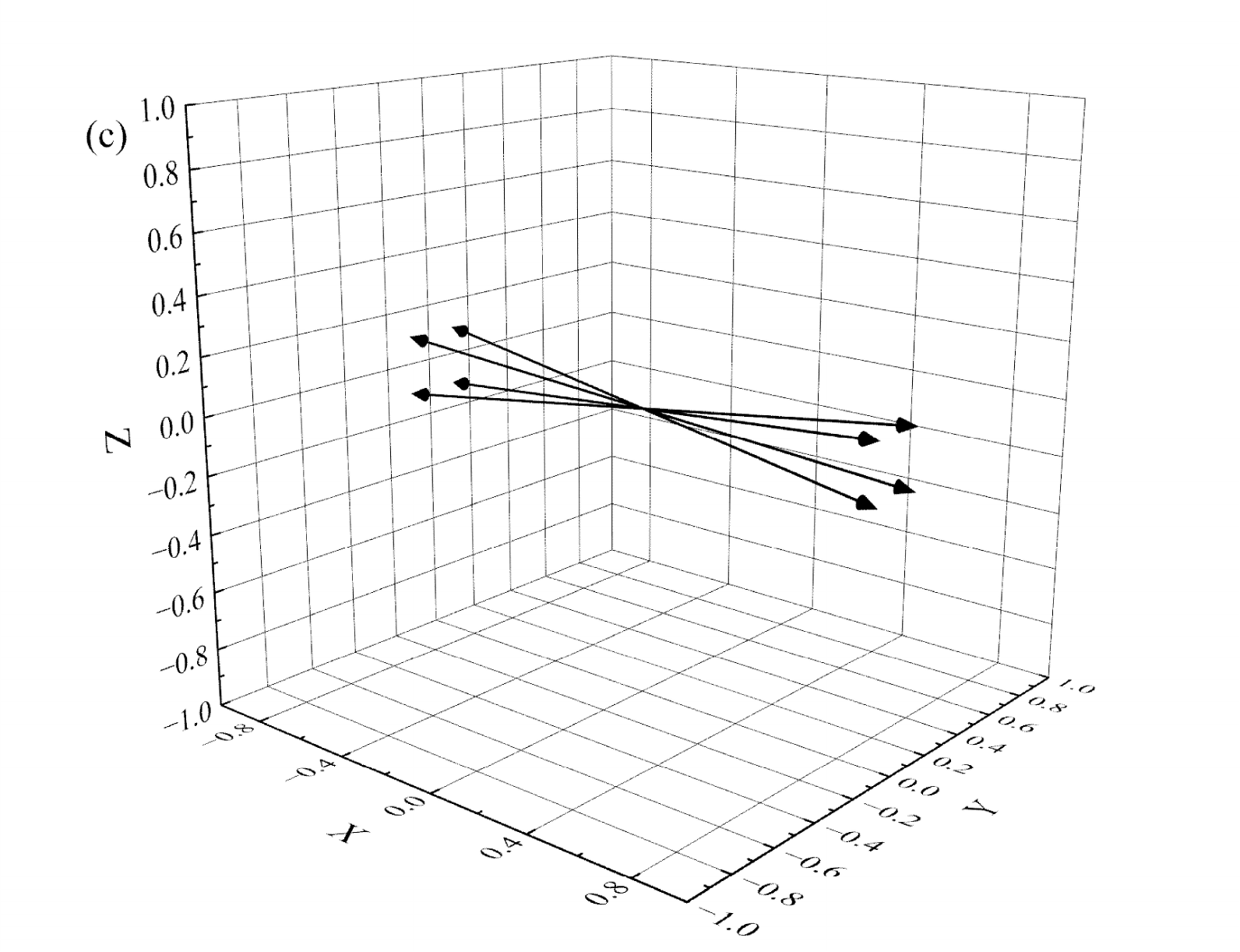}
\includegraphics[width=\columnwidth]{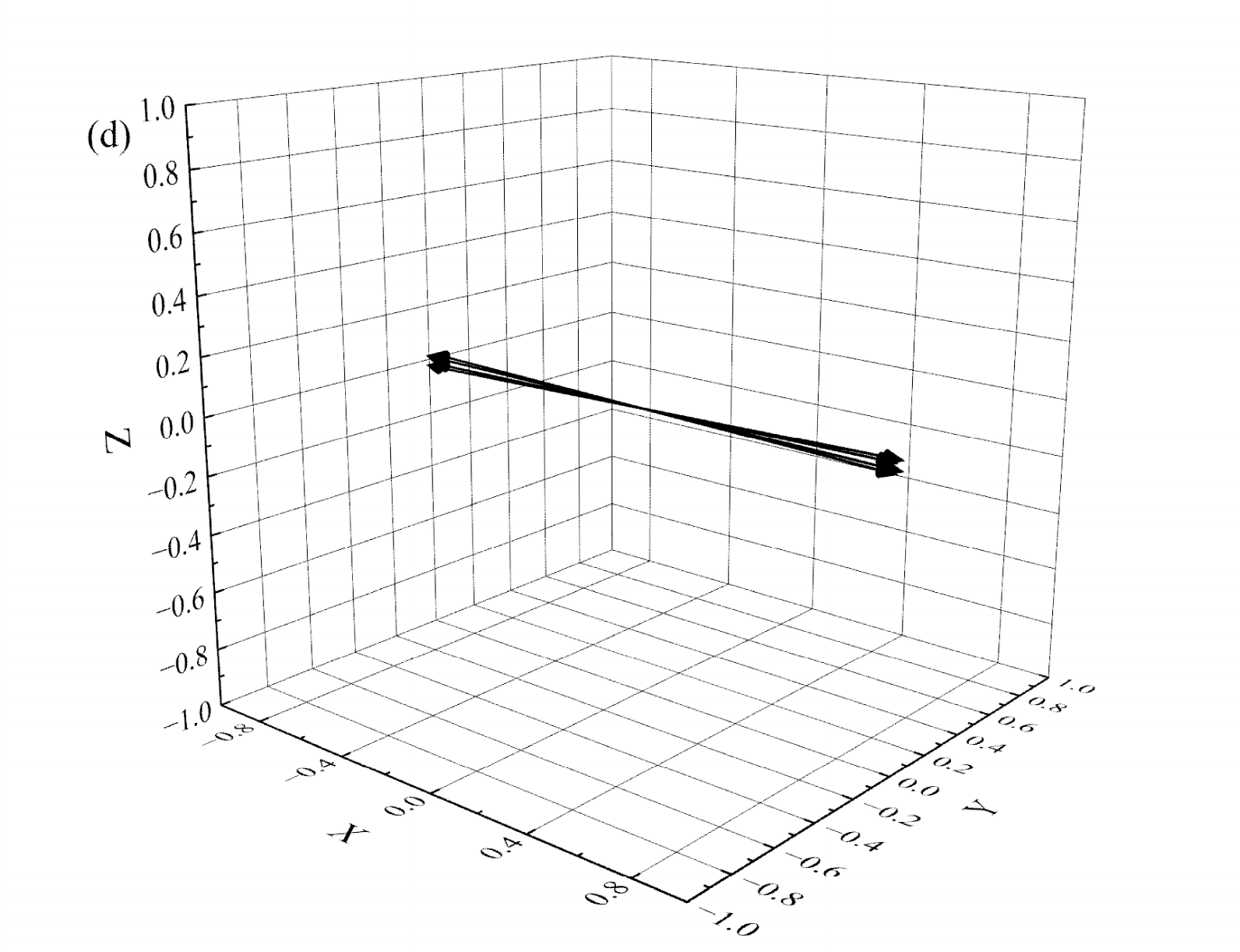}
\vspace{-5mm}
\caption{\small Matlab simulation showing the change in quantisation, and thus sensitive
axis, as a function of the magnetic field strength (1 mT, 200 mT, 1000 mT and 10000 mT) for a [100] bias field alignment.
A test field was applied at various angles and the frequency shift for the resonance measured in order to calculate the effective gyromagnetic ratio and thus the unit vector of the quantisation axis.}
\label{fig: AppendixB-ChangingQuantisationAxis}
\end{figure*}

However, in the high-field regime, D $<<$ $\gamma_e$$\textrm{B}$. Consequently, the quantisation axes are largely set by the bias field and Eq. \ref{eq:LowFieldPT} is no longer valid. This is shown in Fig. \ref{fig: AppendixB-ChangingQuantisationAxis} which shows the quantisation axis for each NVC orientation as a function of the magnetic field strength. 

\begin{figure}[h!]
\centering
\includegraphics[width=\columnwidth, trim={1.5cm 1cm 1.5cm 1.5cm}]{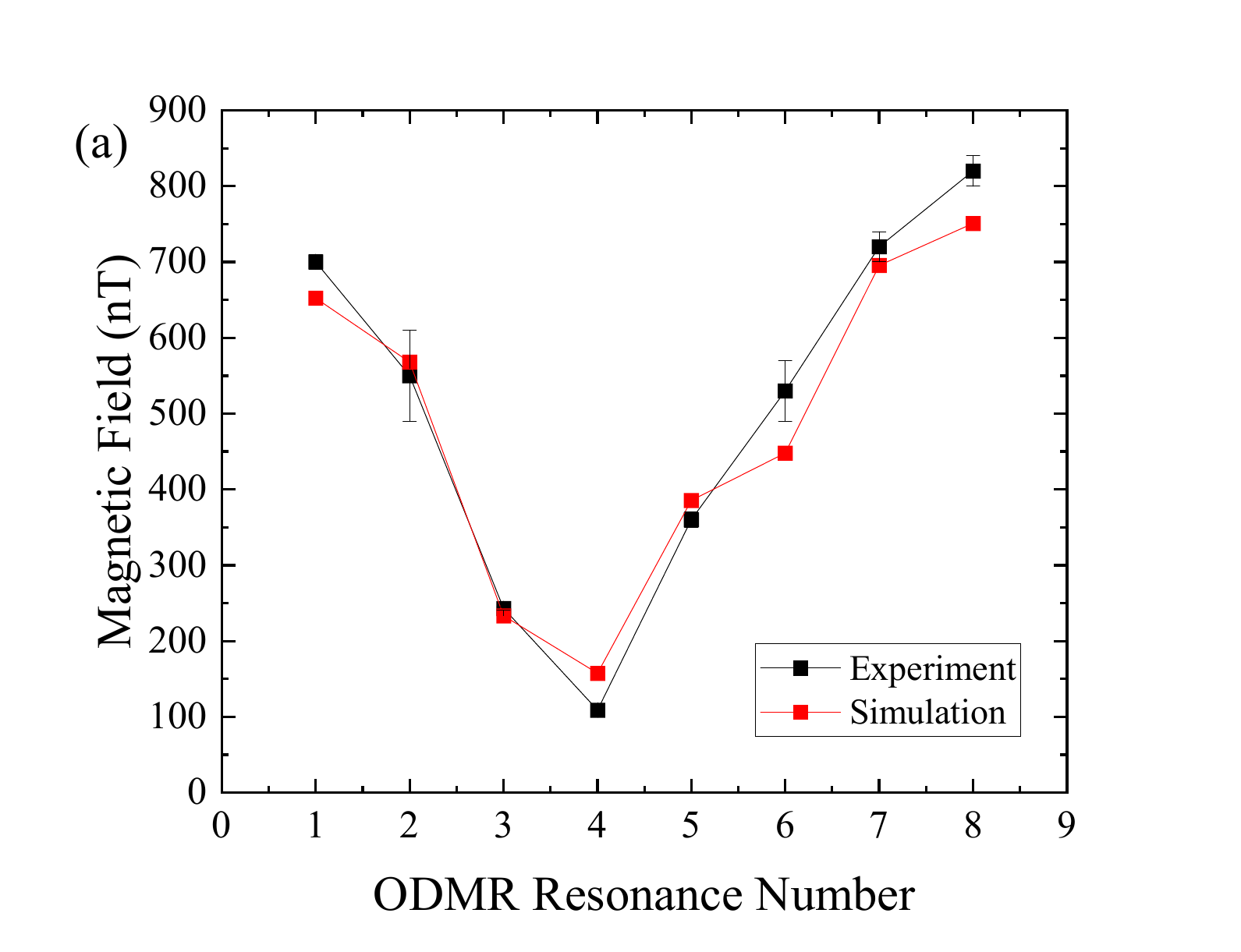}
\includegraphics[width=\columnwidth, trim={1.5cm 1cm 1.5cm 1.5cm}]{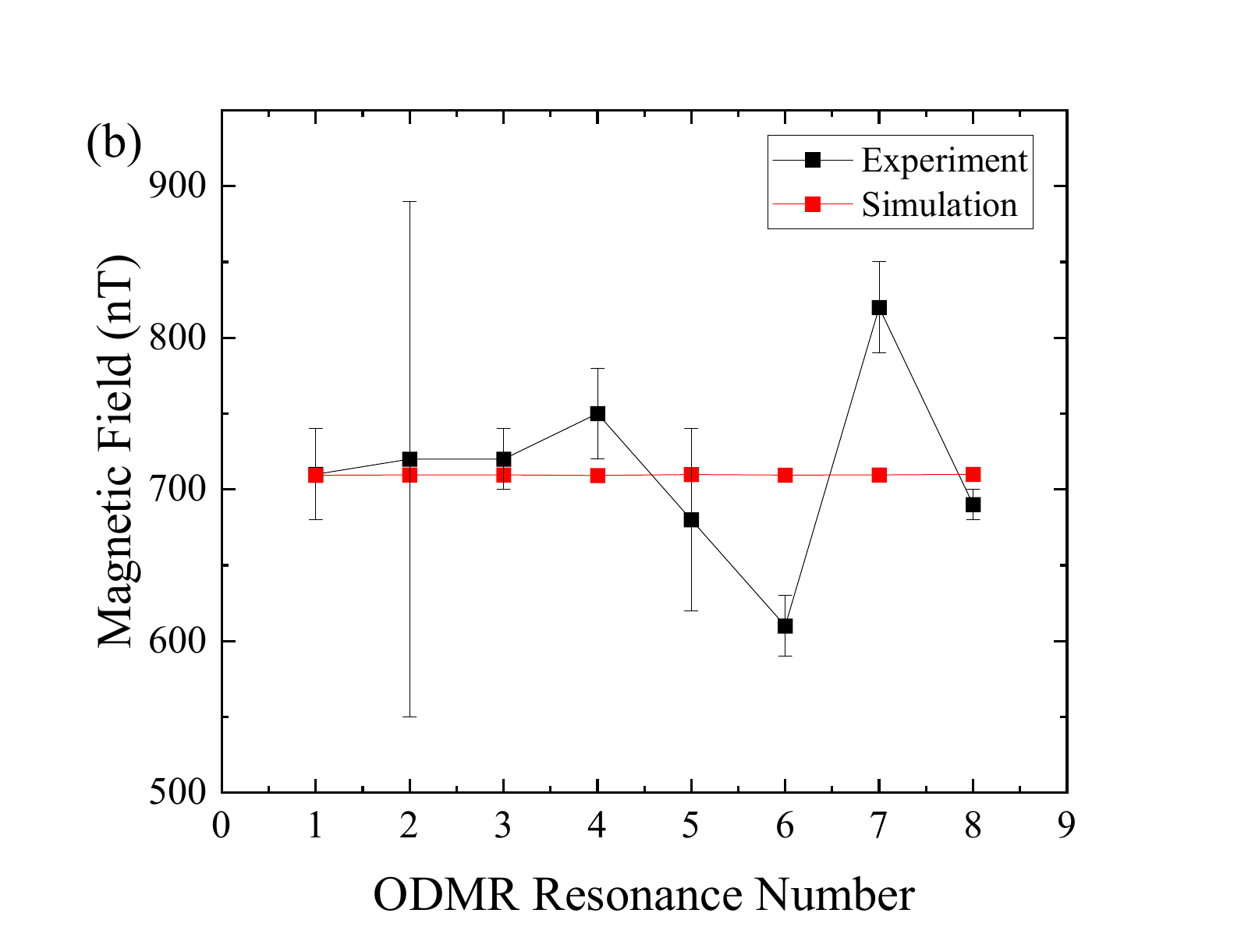}
\vspace{-5mm}
\caption{\small (a) The amplitude of a 65 Hz test field signal measured for each resonance in the low field regime (approximately 4 mT), (b) the amplitude of the 65 Hz test field signal measured for each resonance in the high-field regime (approximately 1000 mT). For both sets of measurements the expected results based on the solutions to the spin Hamiltonian are shown. The discrepancies observed were most likely due to uncertainty in the measurement of test-field amplitude.}
\label{fig: AppendixB-LowvsHighFieldProjection}
\end{figure}

Using pertubation theory at high-field, treating the ZFS as the perturbation as opposed to the Zeeman term, it can be shown that instead the resonance frequencies are given by

\begin{equation}
\label{eq:HighFieldPT}
f_{\pm} = \gamma_eB \pm \frac{D}{2}(3cos^2(\theta)-1),
\end{equation}

to first order, where $\theta$ is the angle between the bias field and the respective NVC symmetry axis. 
Figures \ref{fig: AppendixB-LowvsHighFieldProjection}a and \ref{fig: AppendixB-LowvsHighFieldProjection}b show the magnetic field shift due to a test field applied parallel to the bias field, calculated by dividing each $\delta$f by 28.024 GHz/T. At low-field, the outermost resonances have the largest value, with the Eq. \ref{eq:LowFieldPT} first order approximation $\delta f$ $\approx$ $\gamma_e$$\delta\textrm{B}_{\textrm{NVC}}$ holding. In contrast, at high field approximately the total field  $\delta\textrm{B}$ is obtained for each resonance, reflecting a situation in which for a small $\delta$B, $\delta f$ $\approx$ $\gamma_e$$\delta\textrm{B}_{\parallel bias}$ \cite{kollarics2023magneto}. Strictly, it is proportional to the field parallel to the quantisation axis. A second order, non-linear response is also anticipated for fields perpendicular to the bias field. The A-matrix should account for these effects. This second order term would be suppressed as $\textrm{B}$ increased.

In the low-, intermediate- and high-field regimes, more accurate real-time results can be numerically obtained by linearising the full spin-Hamiltonian about a particular magnetic field strength, assuming $\delta\textrm{B}$ $<<$ $\textrm{B}$. This process results in a sensitivity A-matrix given by \cite{schloss2018simultaneous}

\begin{equation}
\label{eq:AMatrix}
\mathbf{A} = \begin{pmatrix}
\frac{\partial{f_1}}{\partial{(\delta B_{\textrm{x}}})} & \frac{\partial{f_1}}{\partial{(\delta B_{\textrm{y}}})} & \frac{\partial{f_1}}{\partial{(\delta B_{\textrm{z}}})}\\
\frac{\partial{f_2}}{\partial{(\delta B_{\textrm{x}}})} & \frac{\partial{f_2}}{\partial{(\delta B_{\textrm{y}}})} & \frac{\partial{f_2}}{\partial{(\delta B_{\textrm{z}}})}\\
\frac{\partial{f_3}}{\partial{(\delta B_{\textrm{x}}})} & \frac{\partial{f_3}}{\partial{(\delta B_{\textrm{y}}})} & \frac{\partial{f_3}}{\partial{(\delta B_{\textrm{z}}})}\\
\frac{\partial{f_4}}{\partial{(\delta B_{\textrm{x}}})} & \frac{\partial{f_4}}{\partial{(\delta B_{\textrm{y}}})} & \frac{\partial{f_4}}{\partial{(\delta B_{\textrm{z}}})}\\
\end{pmatrix}
.
\end{equation}

The magnetic field shifts are calculated from four measured frequency shifts using the equation 

\begin{equation}
\label{eq:MagFieldCalculation}
\begin{pmatrix}
\delta B_{\textrm{x}}\\
\delta B_{\textrm{y}}\\
\delta B_{\textrm{z}}\\
\end{pmatrix} = 
\mathbf{A}^{-1}
\begin{pmatrix}
\delta f_1\\
\delta f_2\\
\delta f_3\\
\delta f_4\\
\end{pmatrix}
,
\end{equation}

where $\mathbf{A}^{-1}$ is the Moore-Penrose pseudo-inverse of the A-matrix. The elements of the A-matrix above can be determined using the spin-Hamiltonian MATLAB code. Small simulated test fields are applied along the defined lab-frame x, y and z axes (see appendix A) and the shift in the resonance frequencies relative to the frequencies set by the initial bias field determined.

As an example, at high-field, the non-degenerate alignment would have an A-matrix for the upper four resonances,

\begin{equation}
\label{eq:AMatrixHighField}
\mathbf{A} \, (\textrm{GHz}/\textrm{T}) = \begin{pmatrix}
-3.0 & 17.3 & 22.2\\
0.3 & 17.4 & 22.0\\
1.5 & 17.1 & 22.1\\
4.8 & 17.2 & 21.9\\
\end{pmatrix}
.
\end{equation}

This compares to the A-matrix at low-field of, 

\begin{equation}
\label{eq:AMatrixLowField}
\mathbf{A} \, (\textrm{GHz}/\textrm{T}) = \begin{pmatrix}
-16.5 & -16.1 & -15.9\\
16.6 & -16.1 & -15.9\\
-16.2 & 16.7 & -15.6\\
16.2 & 16.7 & -15.6\\
\end{pmatrix}
.
\end{equation}

\begin{figure*}[t]
\hspace{-0.5cm}
\centering
\includegraphics[width=\columnwidth, trim={1.5cm 1cm 1.5cm 1.5cm}]{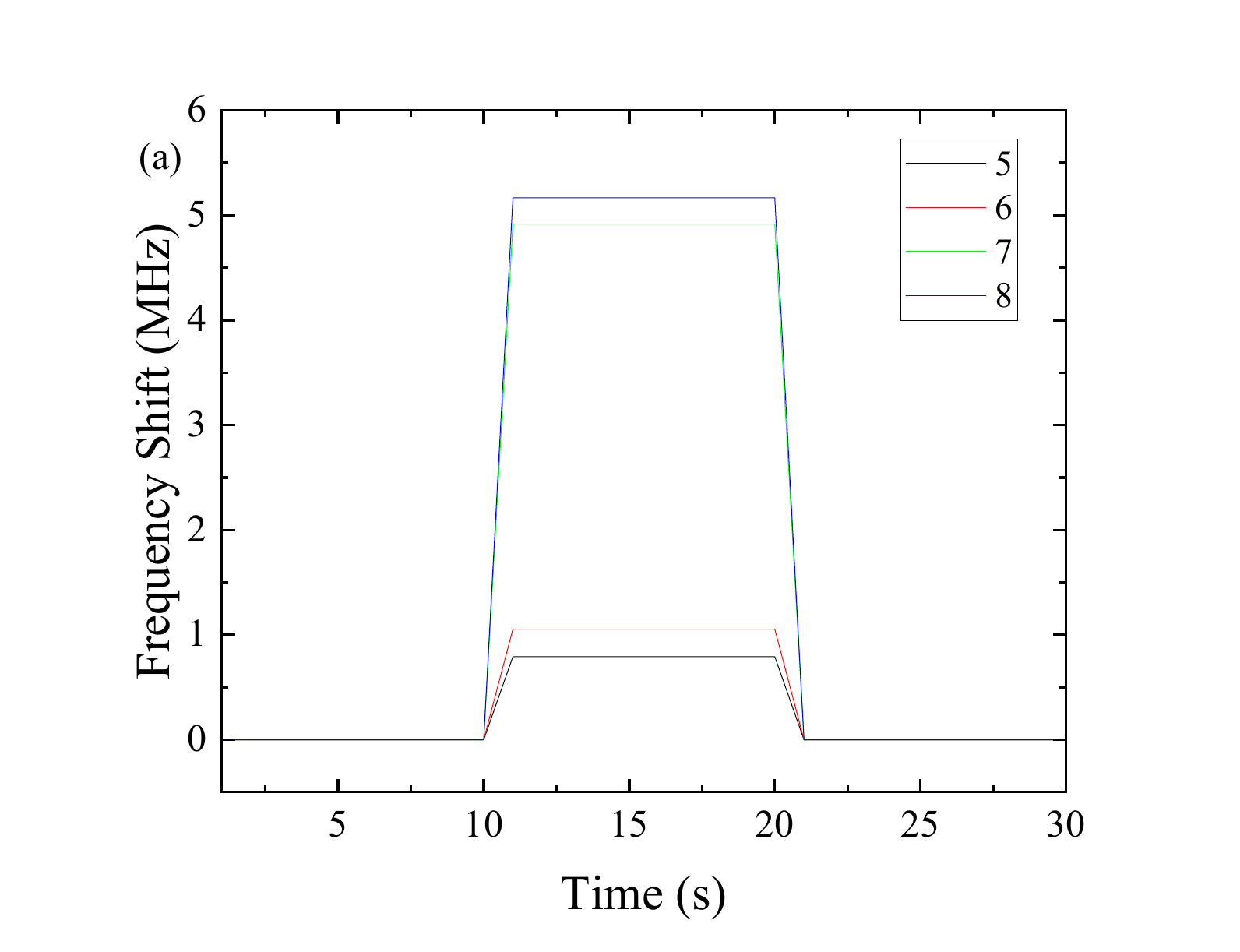} 
\includegraphics[width=\columnwidth, trim={1.5cm 1cm 1.5cm 1.5cm}]{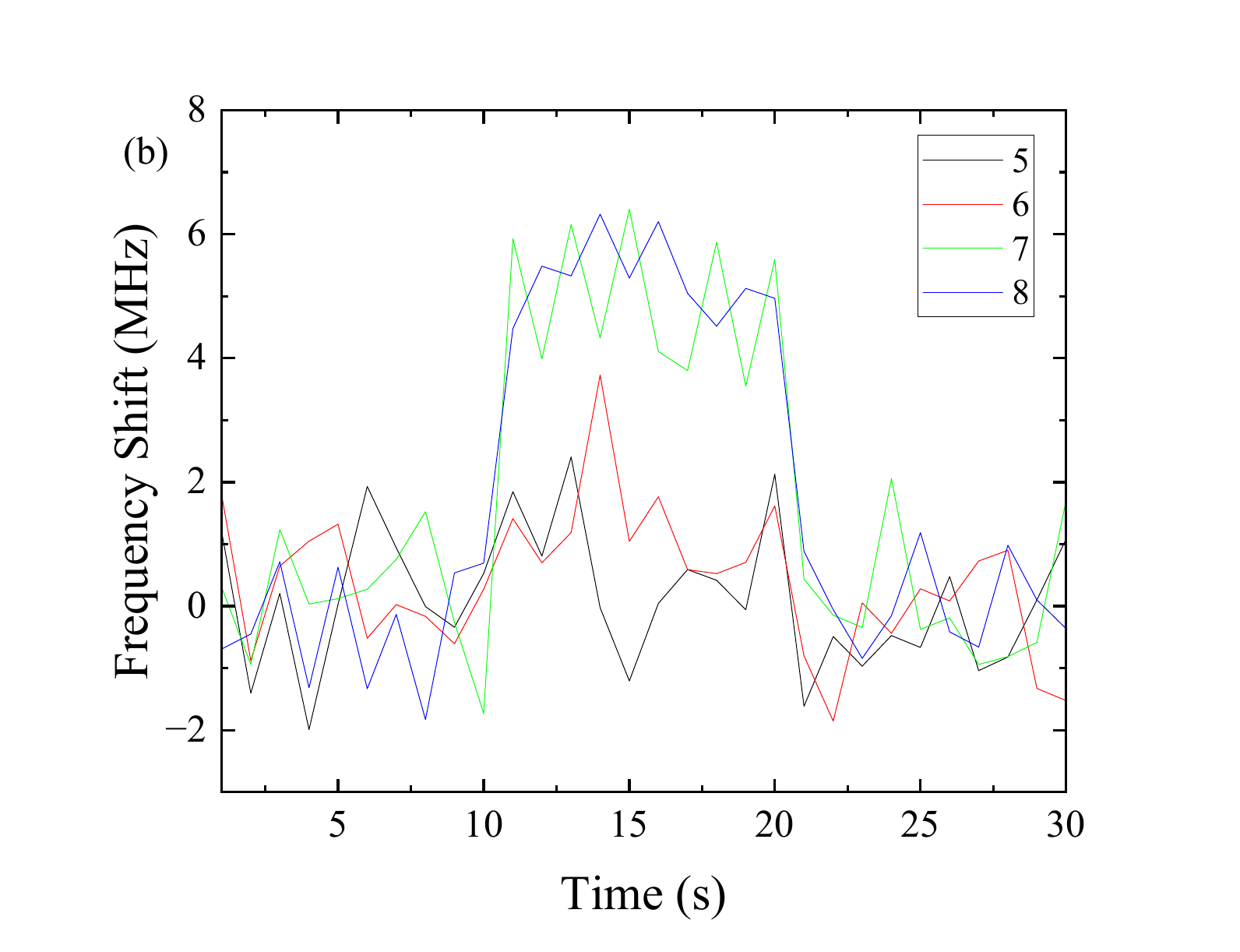} 
\includegraphics[width=\columnwidth, trim={1.5cm 1cm 1.5cm 1.5cm}]{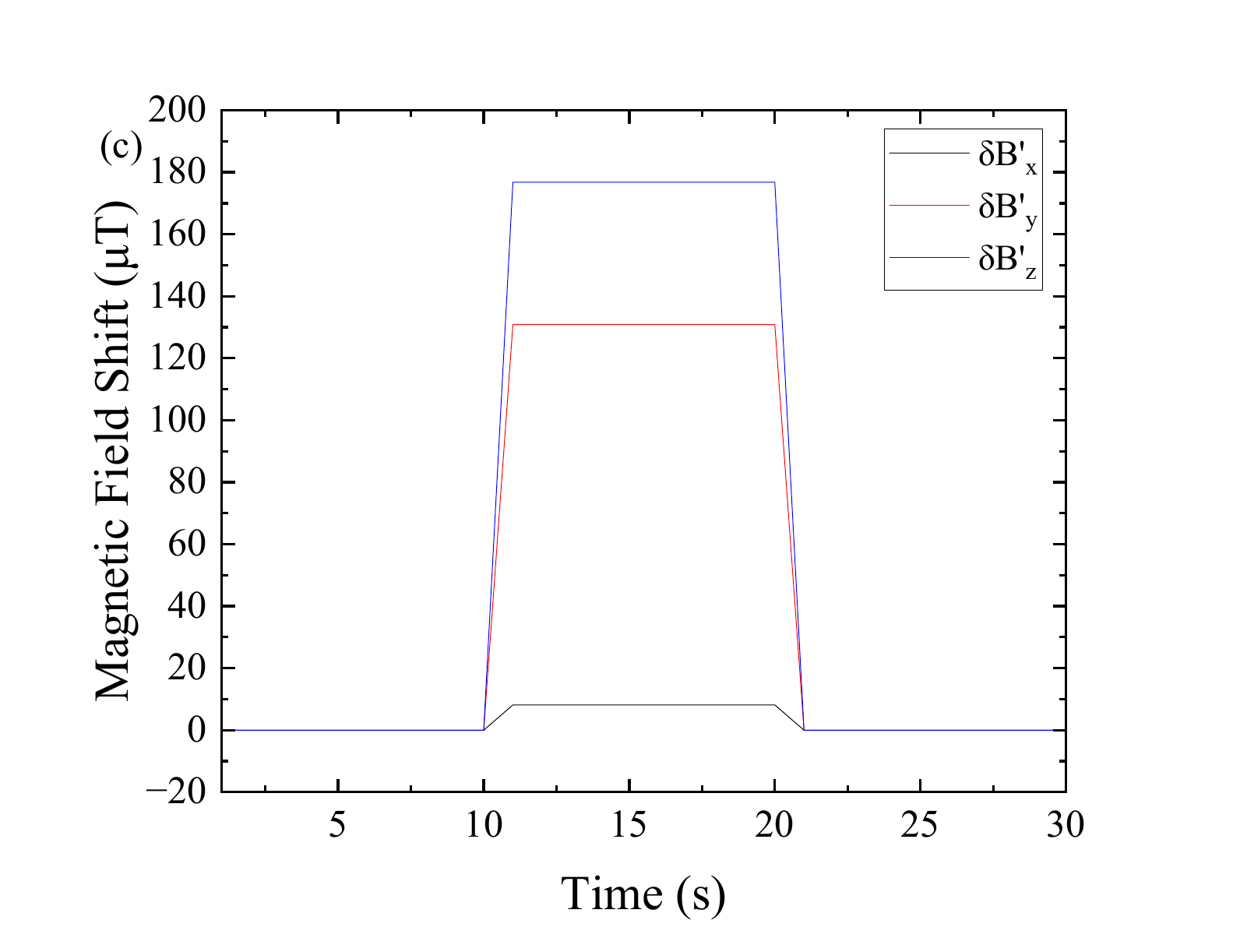} 
\includegraphics[width=\columnwidth, trim={1.5cm 1cm 1.5cm 1.5cm}]{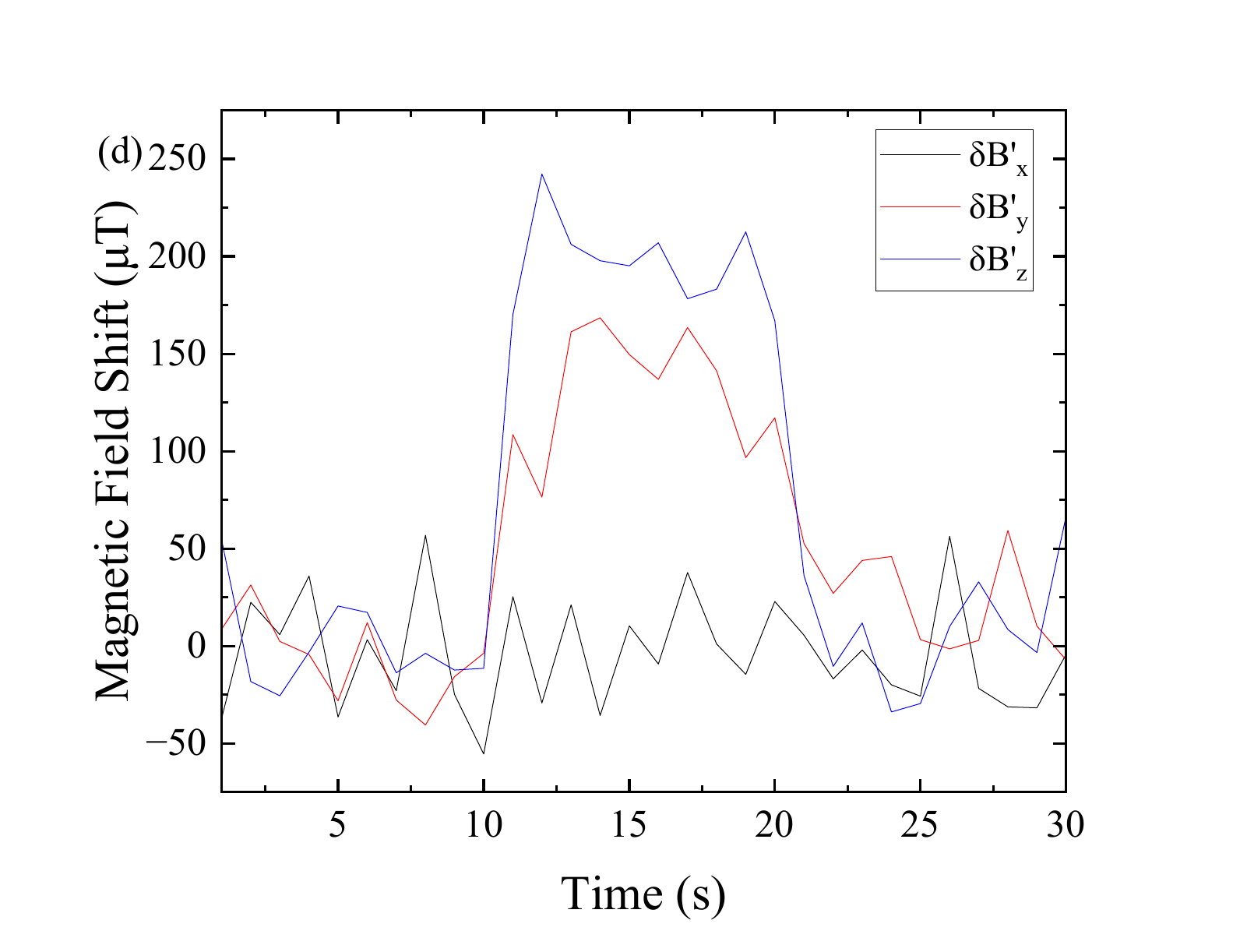} 
\caption{\small (a) Low-field: Frequency shifts for four upper resonances (see Fig. \ref{fig:NondegenerateODMR}) as a result of a simulated test field with no added Gaussian noise. (b) Frequency shifts for four resonances as a result of a simulated test field with added Gaussian noise (standard deviation of 30 $\mu$T). (c) Magnetic field shifts calculated using an A-matrix from the no-added noise frequency shifts. (d) Magnetic field shifts using the same A-matrix as (c) but from the added-noise frequency shifts. A bias field of 1 mT was used for the simulation with a non-degenerate alignment.}
\label{fig: AppendixB-SyntheticData-LowField}
\end{figure*}

\begin{figure*}[t]
\hspace{-0.5cm}
\centering
\includegraphics[width=\columnwidth, trim={1.5cm 1cm 1.5cm 1.5cm}]{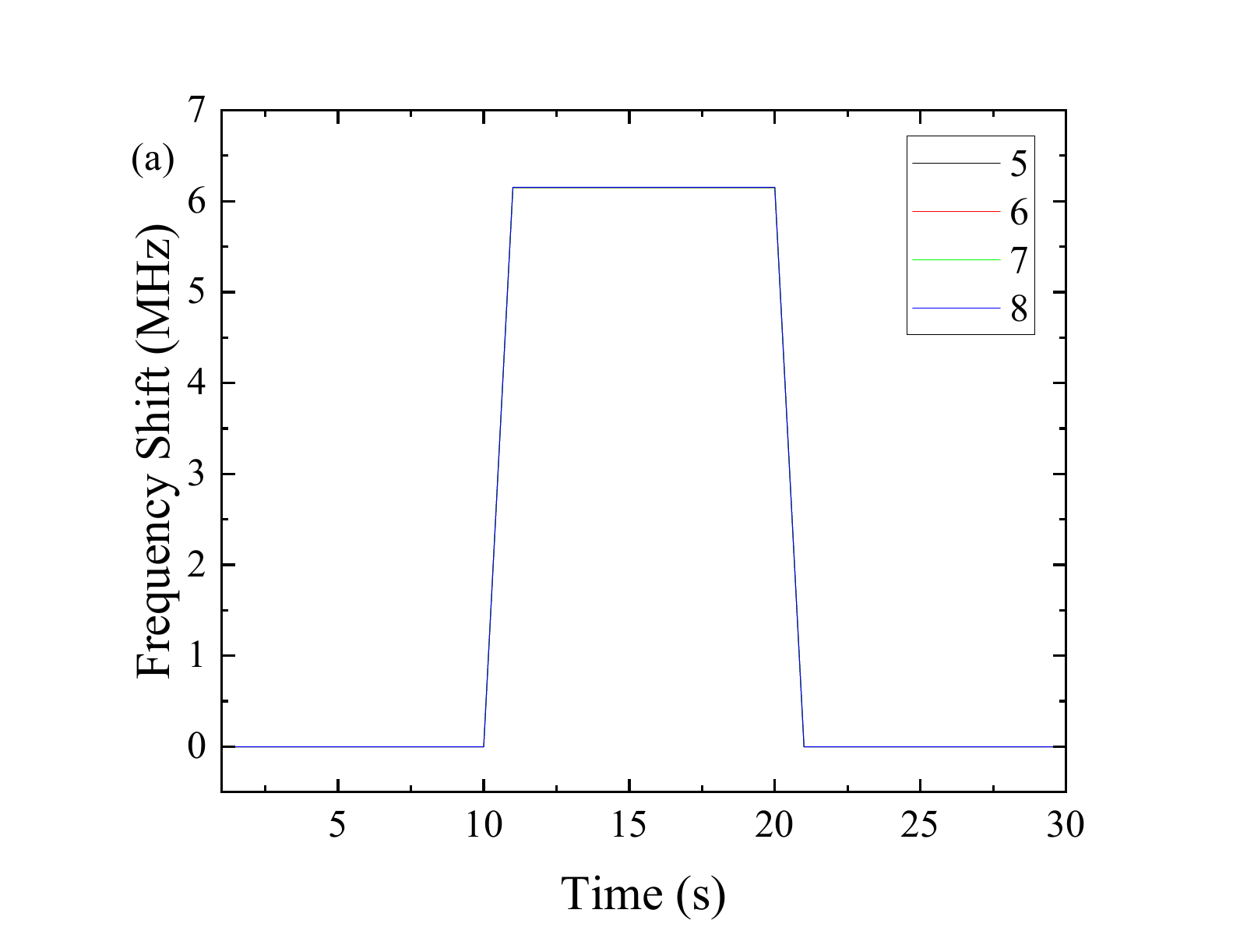} 
\includegraphics[width=\columnwidth, trim={1.5cm 1cm 1.5cm 1.5cm}]{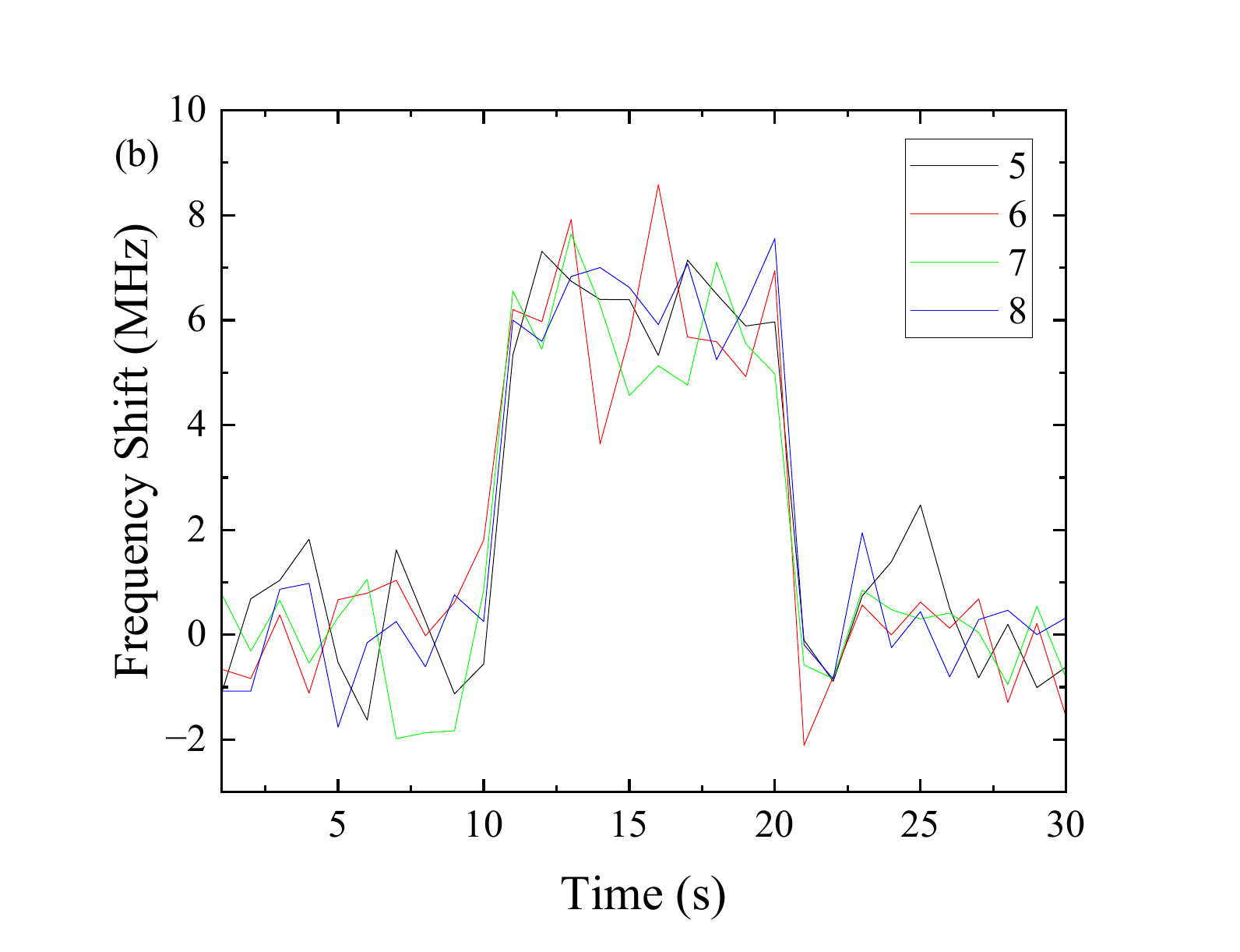} 
\includegraphics[width=\columnwidth, trim={1.5cm 1cm 1.5cm 1.5cm}]{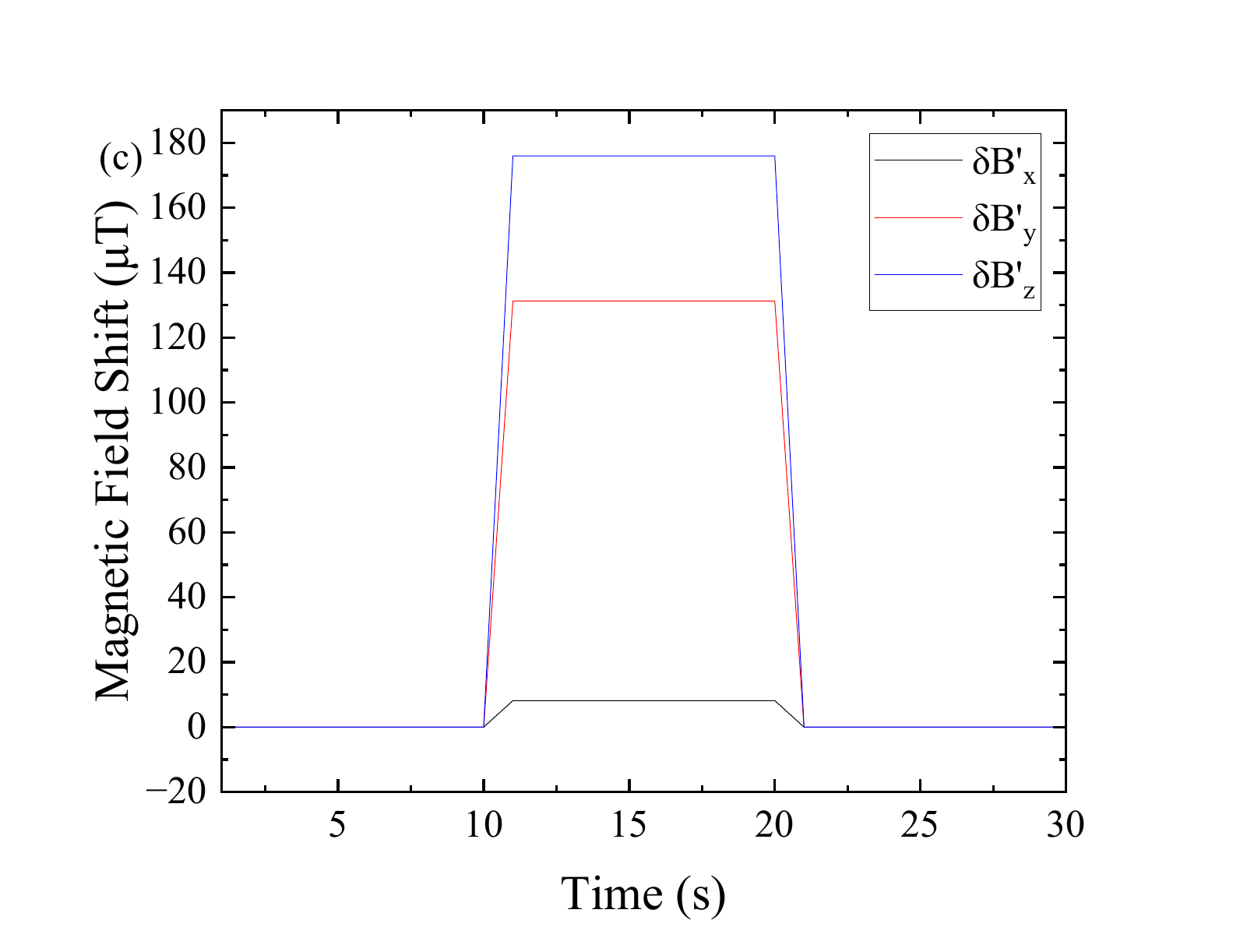} 
\includegraphics[width=\columnwidth, trim={1.5cm 1cm 1.5cm 1.5cm}]{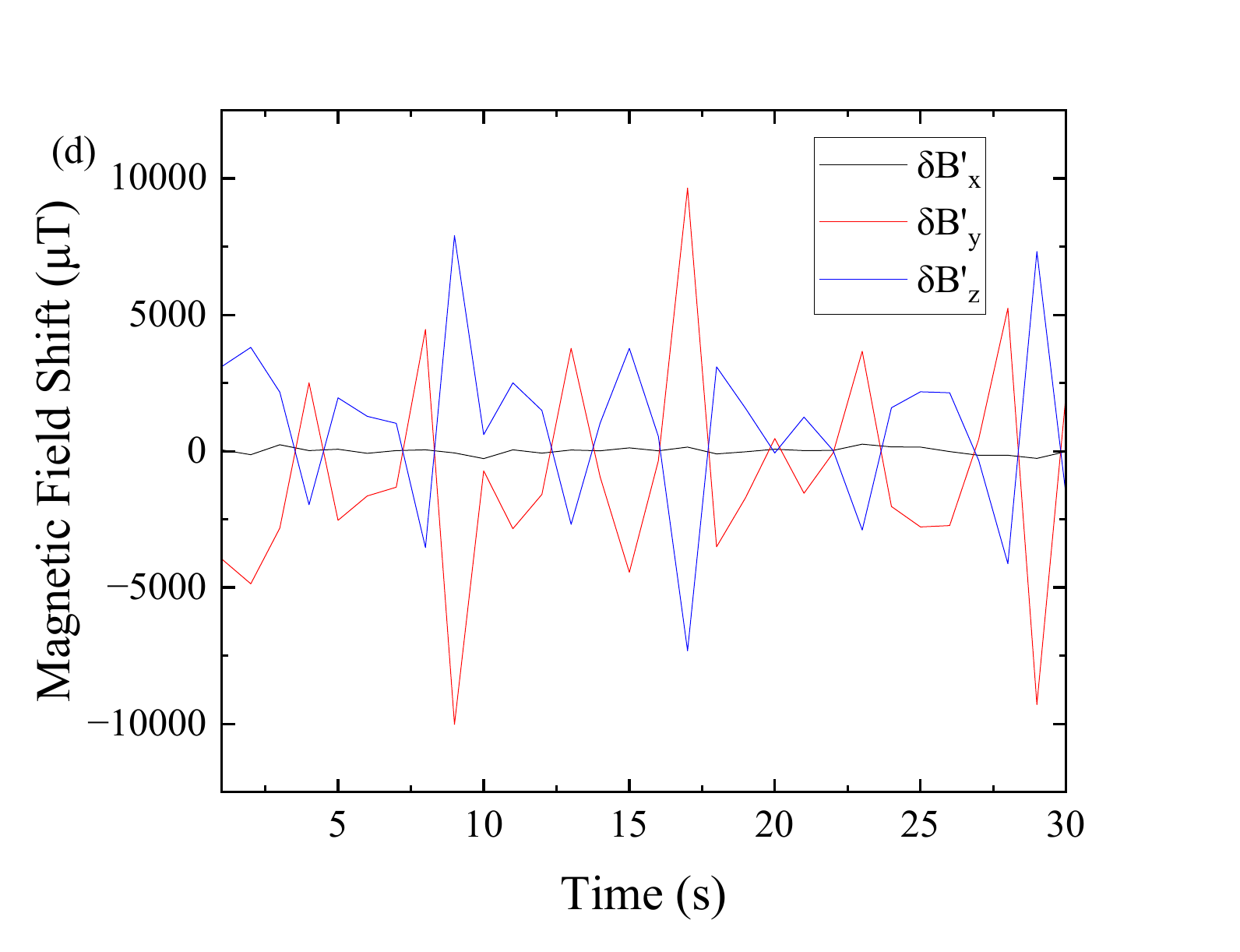} 
\caption{\small (a) High-field: Frequency shifts for four resonances (see Fig. \ref{fig:Near111ODMR}) as a result of a simulated test field with no added Gaussian noise. (b) Frequency shifts for four resonances as a result of a simulated test field with added Gaussian noise (standard deviation of 30 $\mu$T). (c) Magnetic field shifts calculated using an A-matrix from the no-added noise frequency shifts. (d) Magnetic field shifts used using the same A-matrix as (c) but from the added-noise frequency shifts. A bias field of 0.934 T was used for the simulation with a non-degenerate alignment.}
\label{fig: AppendixB-SyntheticData-HighField}
\end{figure*}

Note the link between the high-field A-matrix and the bias field direction, $n_{B}$ = (0.0353,0.5918,0.8053).
This is expected given that the four quantisation axes largely align to the single bias field direction, and the sensitivity to fields transverse to this direction collapses. The result is an ill-conditioned A-matrix with a condition number of 255.1, as opposed to 1.0 for the low-field case. The condition number is determined by dividing the maximum singular value of the matrix by the minimum singular value \cite{watkins2004fundamentals}.
The effects of this on vector field reconstruction are shown by the simulated test-field results in Figs. \ref{fig: AppendixB-SyntheticData-LowField} and \ref{fig: AppendixB-SyntheticData-HighField}. For a given level of Gaussian noise, it is apparent that the frequency measurement noise is more strongly amplified at high than low-field. This is expected for an ill-conditioned sensitivity matrix, as small variations in the input data are amplified to significant errors upon inversion in Eq. \ref{eq:MagFieldCalculation} \cite{watkins2004fundamentals}.

\begin{figure}[h!]
\hspace{-0.5cm}
\centering
\includegraphics[width=\columnwidth, trim={1.5cm 1cm 1.5cm 1.5cm}]{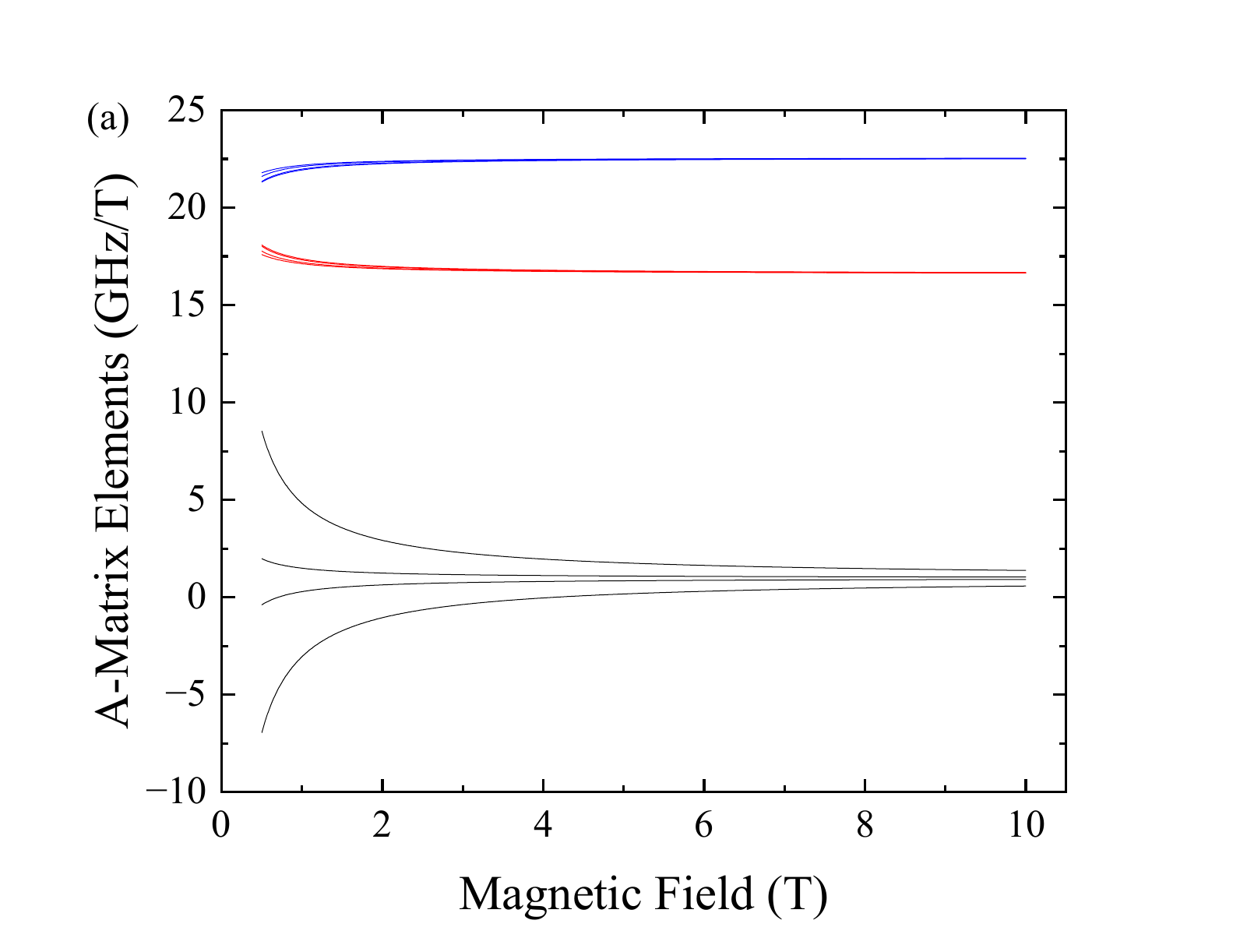} 
\includegraphics[width=\columnwidth, trim={1.5cm 1cm 1.5cm 1.5cm}]{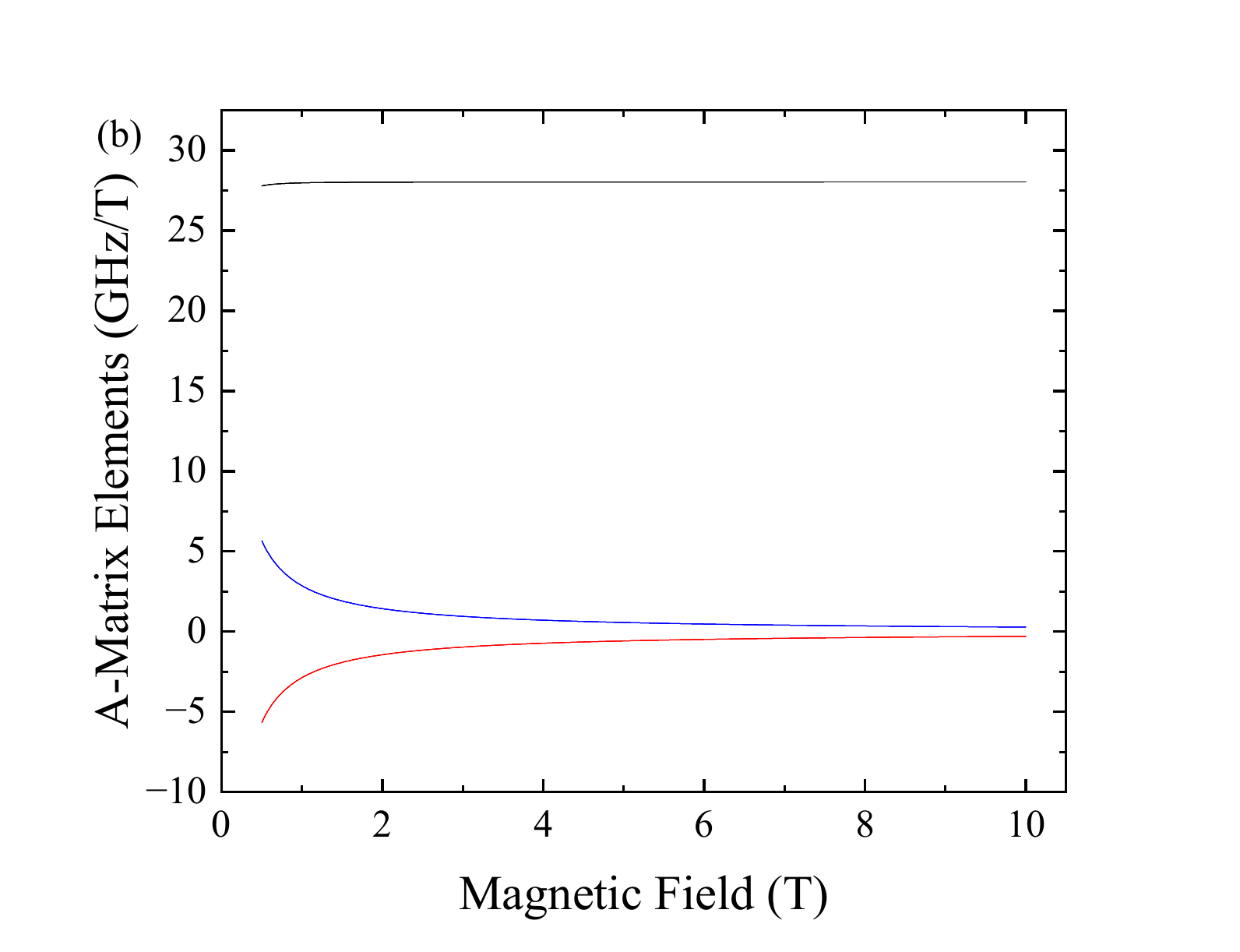} 
\caption{\small (a) and (b) show the simulated A-matrix elements as a function of magnetic field strength in the high-field regime for a non-degenerate and [100] bias field alignment respectively.}
\label{fig: AppendixB-AMatrixElements}
\end{figure}

These simulations suggest that vector performance would be significantly degraded at high-field in terms of accuracy and noise, on top of the reduction in single-axis magnetic-field sensitivity.  Figures \ref{fig: AppendixB-AMatrixElements}a and \ref{fig: AppendixB-AMatrixElements}b show the A-matrix elements for the non-degenerate and [100] bias field alignments. The latter clearly demonstrates the collapse in sensitivity to fields transverse to the bias field for each orientation. 

\begin{figure}[h!]
\hspace{-0.5cm}
\centering
\includegraphics[width=\columnwidth, trim={1.5cm 1cm 1.5cm 1.5cm}]{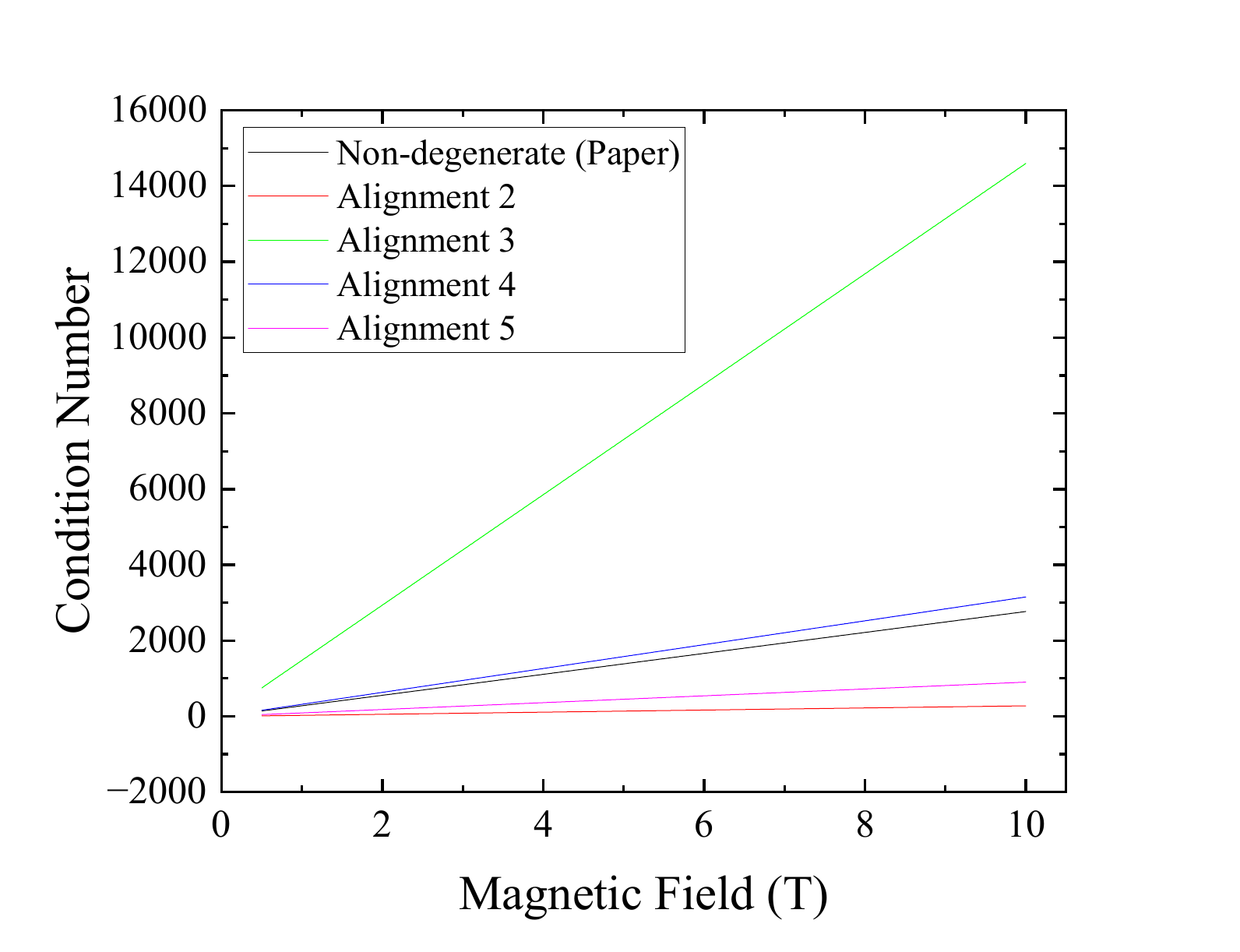} 
\caption{\small The simulated A-matrix condition number as a function of magnetic field strength for a range of arbitrary non-degenerate alignments, including the alignment used in the experimental measurements.}
\label{fig: AppendixB-CondNumberBiasFieldAlignments}
\end{figure}

These effects would be even more notable at 10 T, as the vector response is suppressed as D/$\gamma_e$$\textrm{B}$. 
While the elements of the A-matrix depend on the chosen Cartesian coordinate frame, the condition number is invariant. 
However, the condition number does vary with bias field alignment. Figure \ref{fig: AppendixB-CondNumberBiasFieldAlignments} shows the change in condition number as a function of magnetic field for a range of randomly selected non-degenerate bias field alignments. This suggests that some degree of optimisation would be possible. Improvements in the signal-to-noise ratio of the initial frequency shift measurements would also help to provide more stable vector performance at high-field. Alternative bias field alignments could also provide more uniform single-axis sensitivity across all eight resonances. 

\begin{figure}[h!]
\hspace{-0.5cm}
\centering
\includegraphics[width=\columnwidth, trim={1.5cm 1cm 1.5cm 1.5cm}]{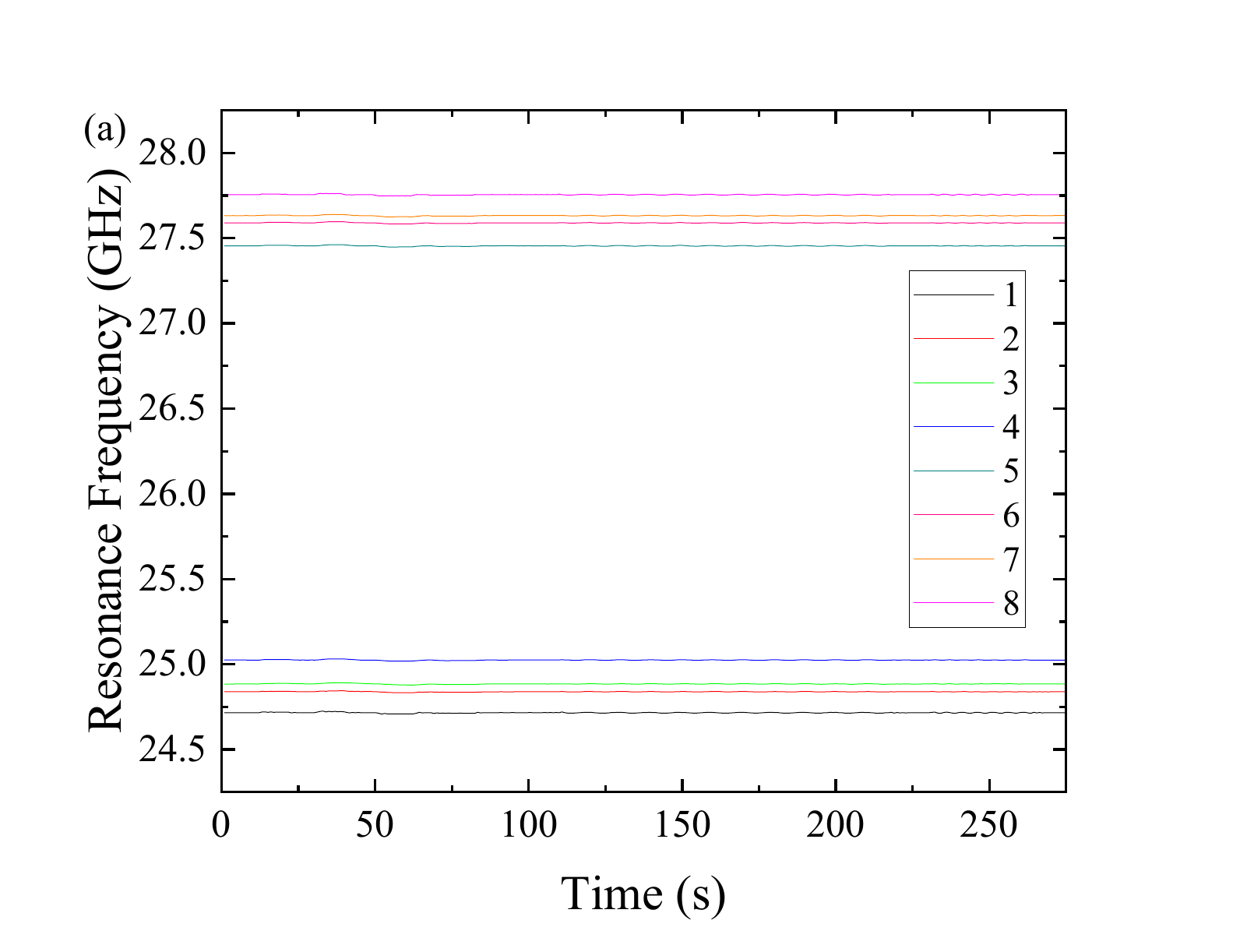} 
\includegraphics[width=\columnwidth, trim={1.5cm 1cm 1.5cm 1.5cm}]{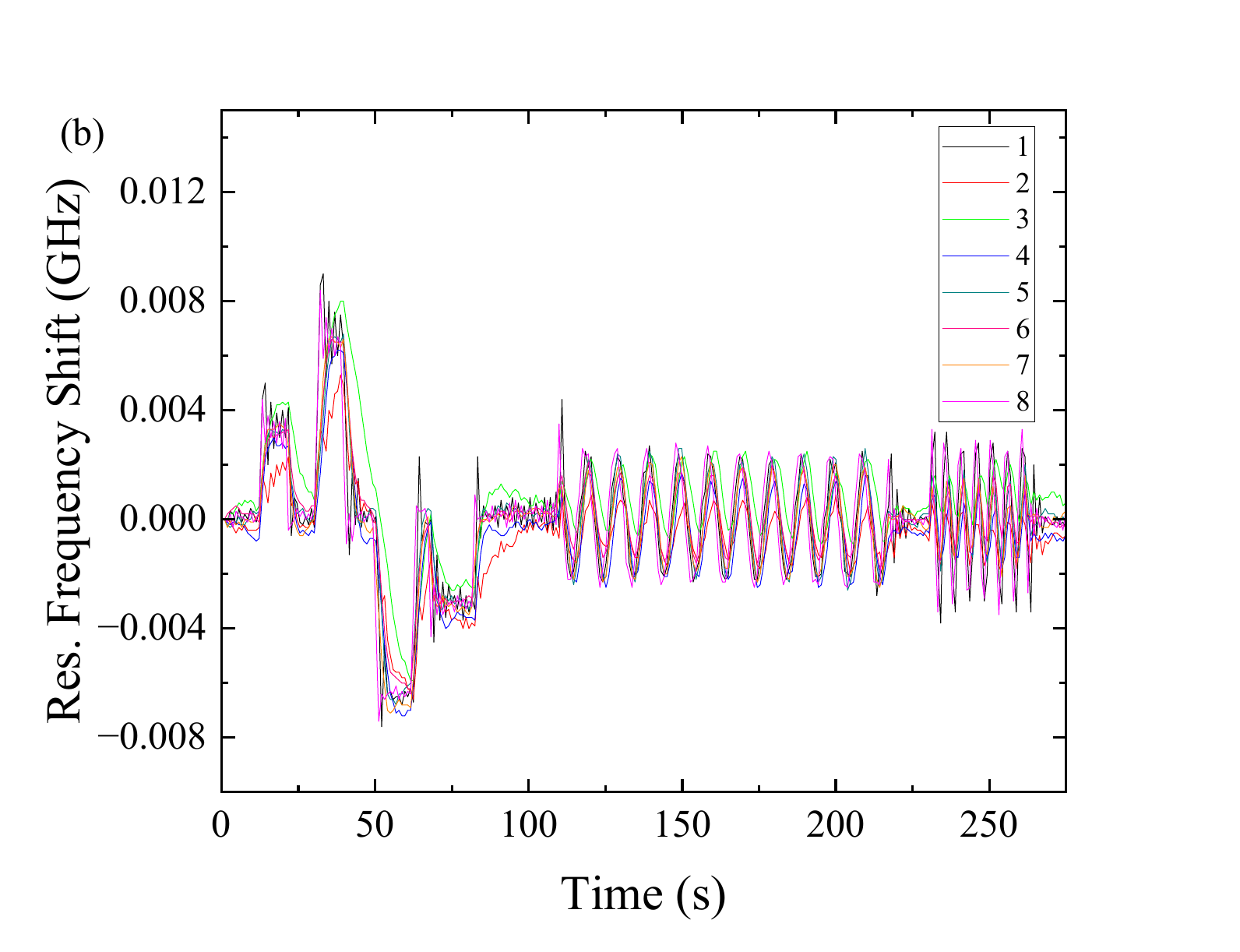} 
\caption{\small (a) Tracking all eight resonances by hopping the microwave frequency generated by a E8257D microwave source. (b) Subtracting off the initial resonance frequencies to observe the shift in frequency associated with the application of dc and ac-test fields.}
\label{fig: AppendixB-NondegenerateResonanceTracking}
\end{figure}

For real-time vector measurements, simultaneous measurement of each resonance frequency using a resonance tracking approach would be required \cite{clevenson2018robust}. A simpler approach is to hop the frequency of the microwave source sequentially across each resonance, as described in Ref. \cite{graham2025road}. This limits the frequency range and slew rate, but enables the use of a single microwave source. For a non-degenerate bias field alignment, Fig. \ref{fig: AppendixB-NondegenerateResonanceTracking} shows the resonance tracking of all eight resonances experimentally implemented. Improvements to the stability and speed of this tracking and the connected signal-to-noise ratio would be required to successfully reconstruct the vector field. 

\begin{figure*}[t]
\hspace{-0.5cm}
\centering
\includegraphics[width=\columnwidth, trim={1.5cm 1cm 1.5cm 1.5cm}]{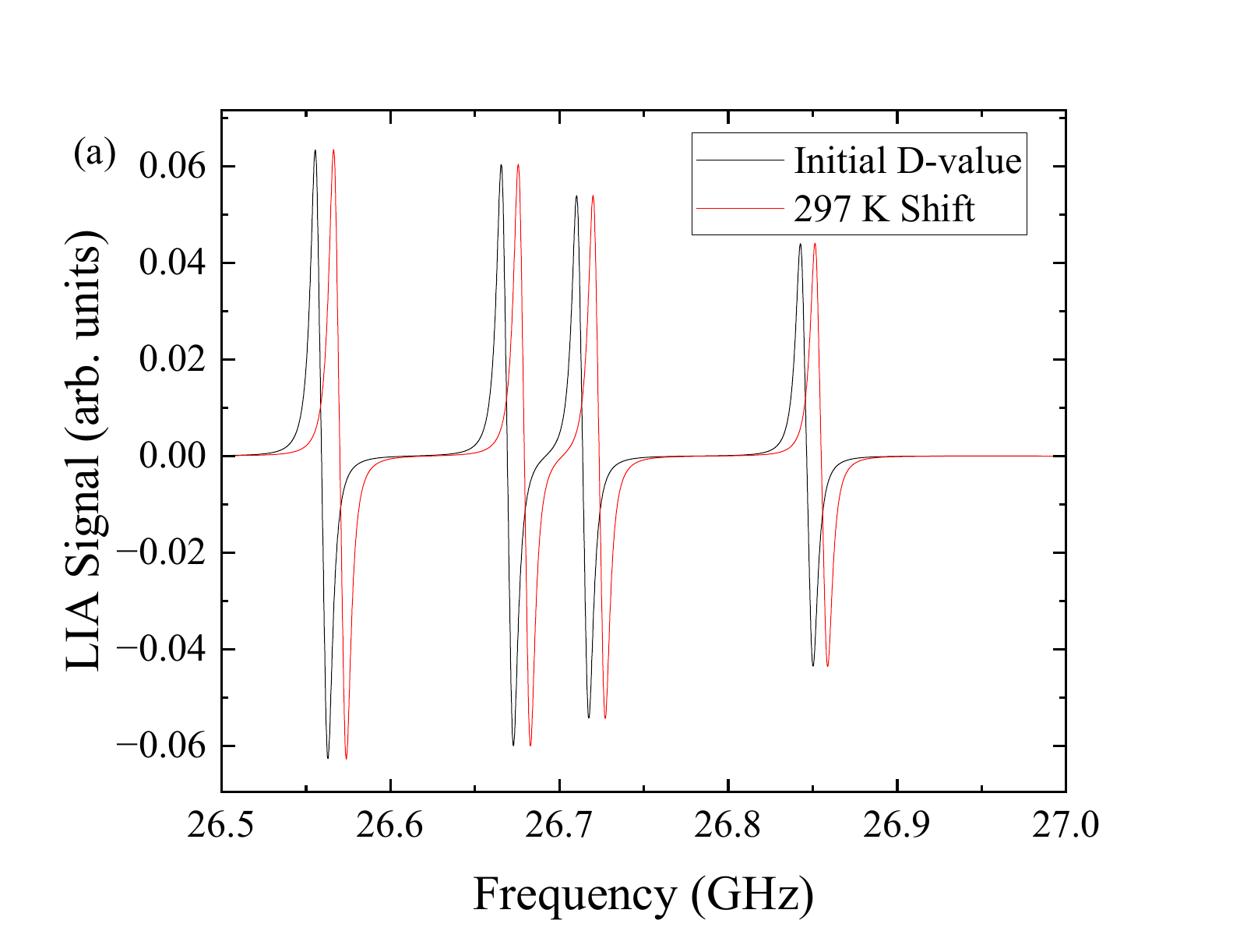} 
\includegraphics[width=\columnwidth, trim={1.5cm 1cm 1.5cm 1.5cm}]{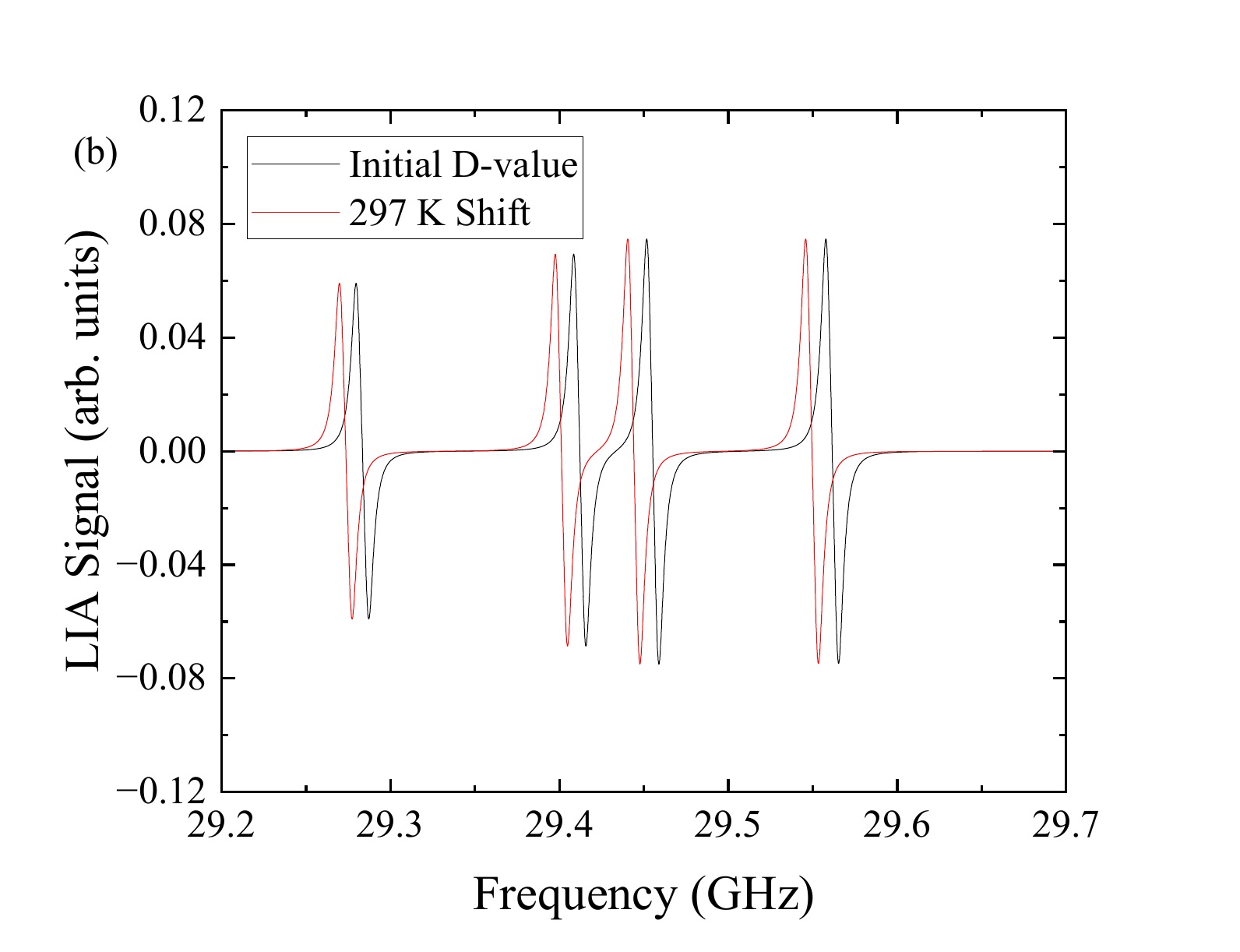} 
\includegraphics[width=\columnwidth, trim={1.5cm 1cm 1.5cm 1.5cm}]{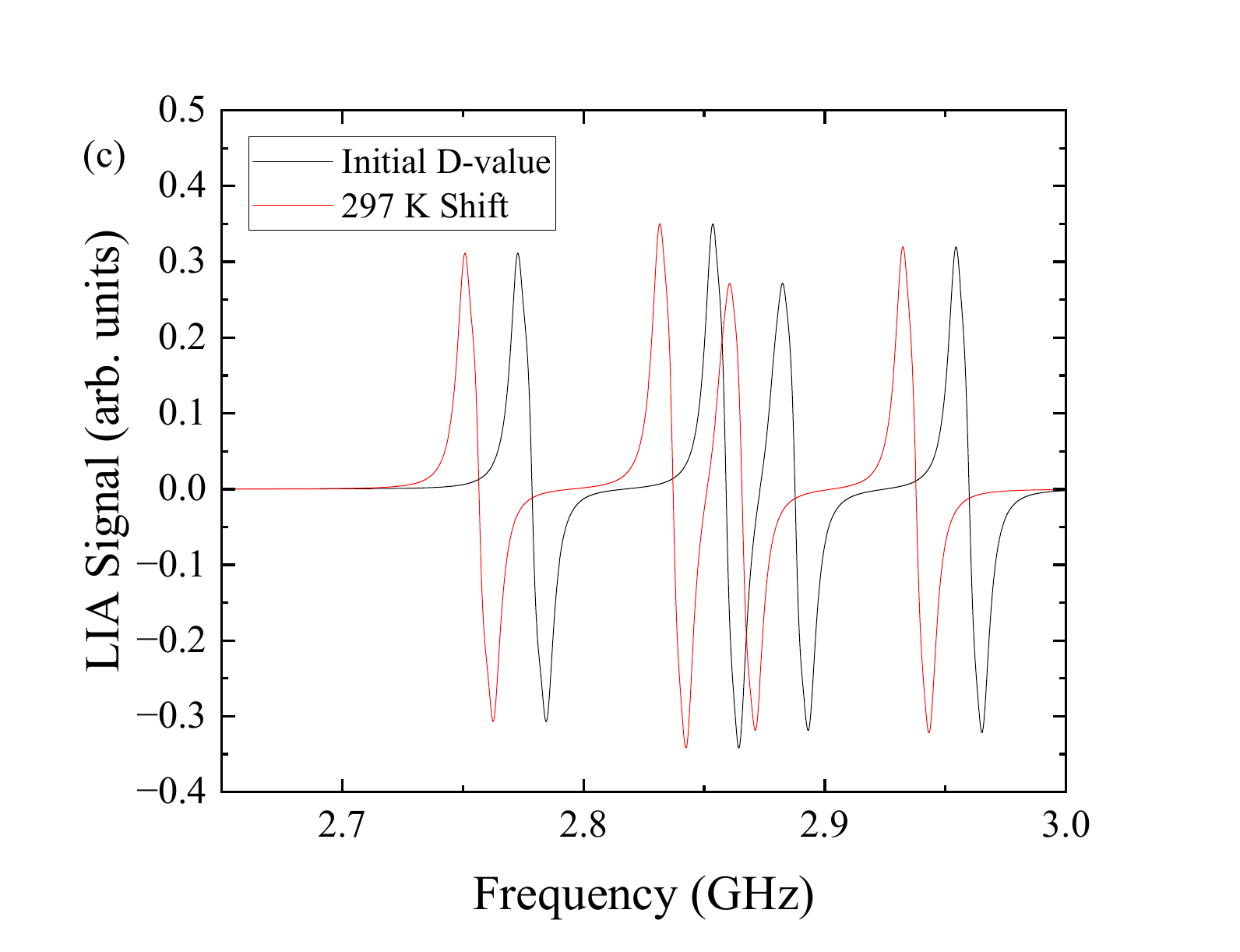} 
\includegraphics[width=\columnwidth, trim={1.5cm 1cm 1.5cm 1.5cm}]{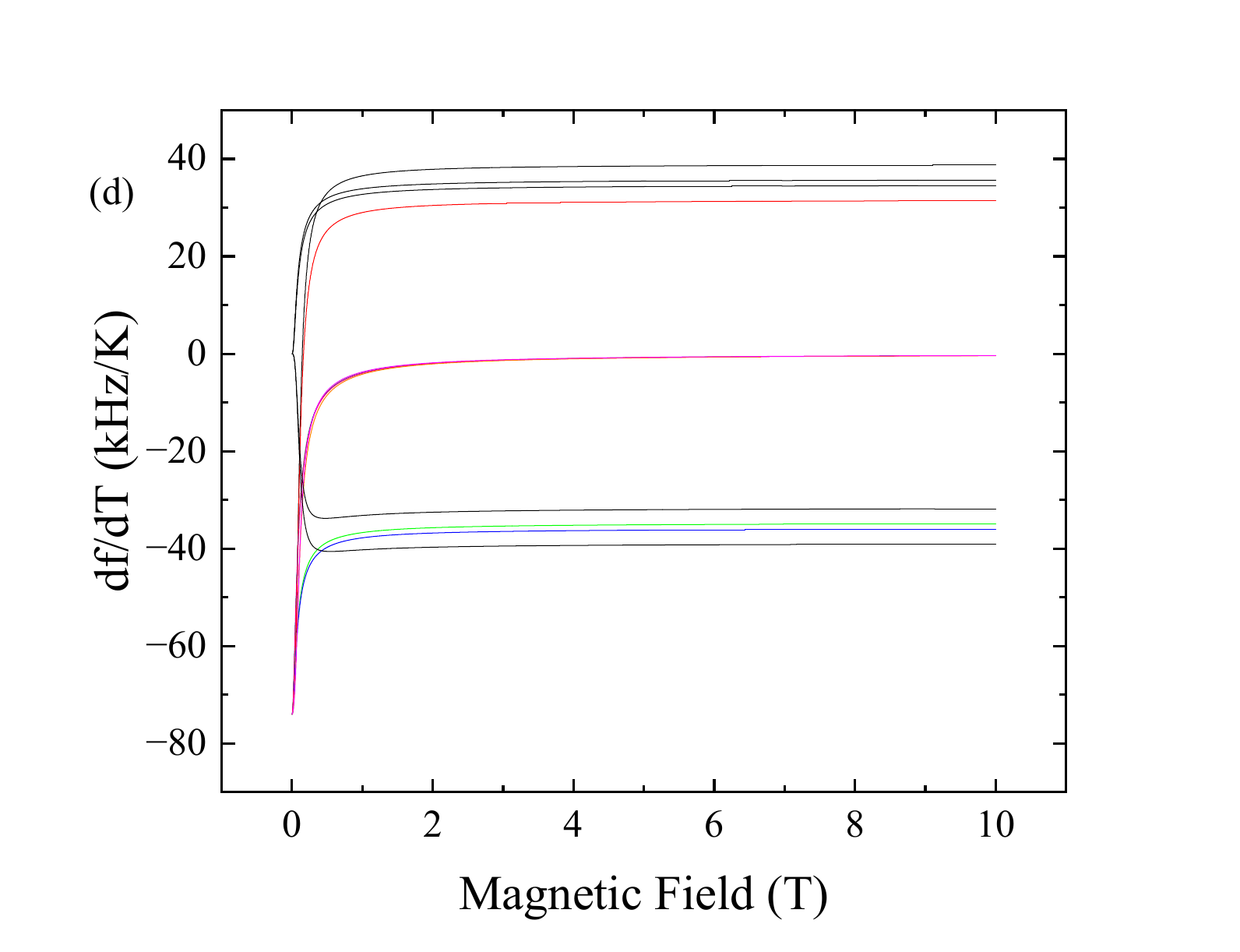} 
\caption{\small (a) and (b) show the lower (1 to 4) and upper (4 to 8) resonances of a simulated demodulated ODMR spectrum at a field strength of 1 T for a non-degenerate field alignment. (c) The corresponding simulated ODMR spectrum at 1 mT. Note this alignment is not non-degenerate at 1 mT. An exaggerated temperature shift of 297 K (D = 2.872 GHz to D = 2.85 GHz) was used for clarity. (d) Simulation of the shift in frequency of each resonance (including the DQ resonances) for a 13 K step-change in temperature. The pink lines are the DQ transition at high field, but not at low field. A $n_B$ = (0.0353, 0.5918, 0.8053) bias field alignment was used.
}
\label{fig: AppendixB-TemperatureShiftHighField}
\end{figure*}

It would also be necessary to account for temperature effects, beyond reductions in sensitivity, which are different in the high-field regime. At high fields, changes in temperature and thus in the ZFS constant, D, result in resonances shifting in opposite directions, as shown in Fig. \ref{fig: AppendixB-TemperatureShiftHighField}. As can be seen from Eq. \ref{eq:HighFieldPT}, at high-field taking the difference ($f_+ - f_-$) will give a term dependent on D and $\theta$, eliminating the Zeeman term. The total field magnitude can be determined from the sum $f_+ + f_-$ = 2$\gamma_e\textrm{B}$, neglecting higher order terms.
Figure \ref{fig: AppendixB-TemperatureShiftHighField}d shows the d$f_{\pm}$/dT (shift in frequency of each resonance for a given step change in temperature) as a function of the magnetic field strength from 0 to 10 T. As can be seen at low field an approximately -74 kHz/K shift is observed for each resonance, but for this bias field alignment at high field they shift as 30 to 38 kHz/K and -30 to -36 kHz/K respectively for the lower and upper resonances. Using Eq. \ref{eq:HighFieldPT}, it is possible to define an effective D-value at high-field, $\textrm{D}_{eff}$ = $\frac{D}{2}(3cos^2(\theta)-1)$ and 

\begin{equation}
\label{eq:TemperatureShifts}
\frac{df_{\pm}}{dT} \approx \pm \frac{dD_{eff}}{dT} = \pm \frac{dD}{dT} (\frac{3cos^2{\theta}-1}{2}).
\end{equation}

Accordingly, the d$f_{\pm}$/dT results also depend on the NVC orientation with respect to the bias field. For a $\langle$111$\rangle$ bias field alignment the d$f_{\pm}$/dT magnitudes would remain the same as at low-field for the outer resonances. 
The DQ transitions have a weaker temperature dependence, at both low and high fields. This effect can also be anticipated from Eq. \ref{eq:HighFieldPT}.
Note that it is not possible to consistently label the resonances in Fig. \ref{fig: AppendixB-TemperatureShiftHighField}, since the low- and high field regimes are distinct, with a different appropriate eigenbasis. 

\FloatBarrier

\section*{Appendix C: Additional Sensitivity Measurements}

\begin{figure}[h!]
\centering
\includegraphics[width=\columnwidth, trim={1.5cm 1cm 1.5cm 1.5cm}]{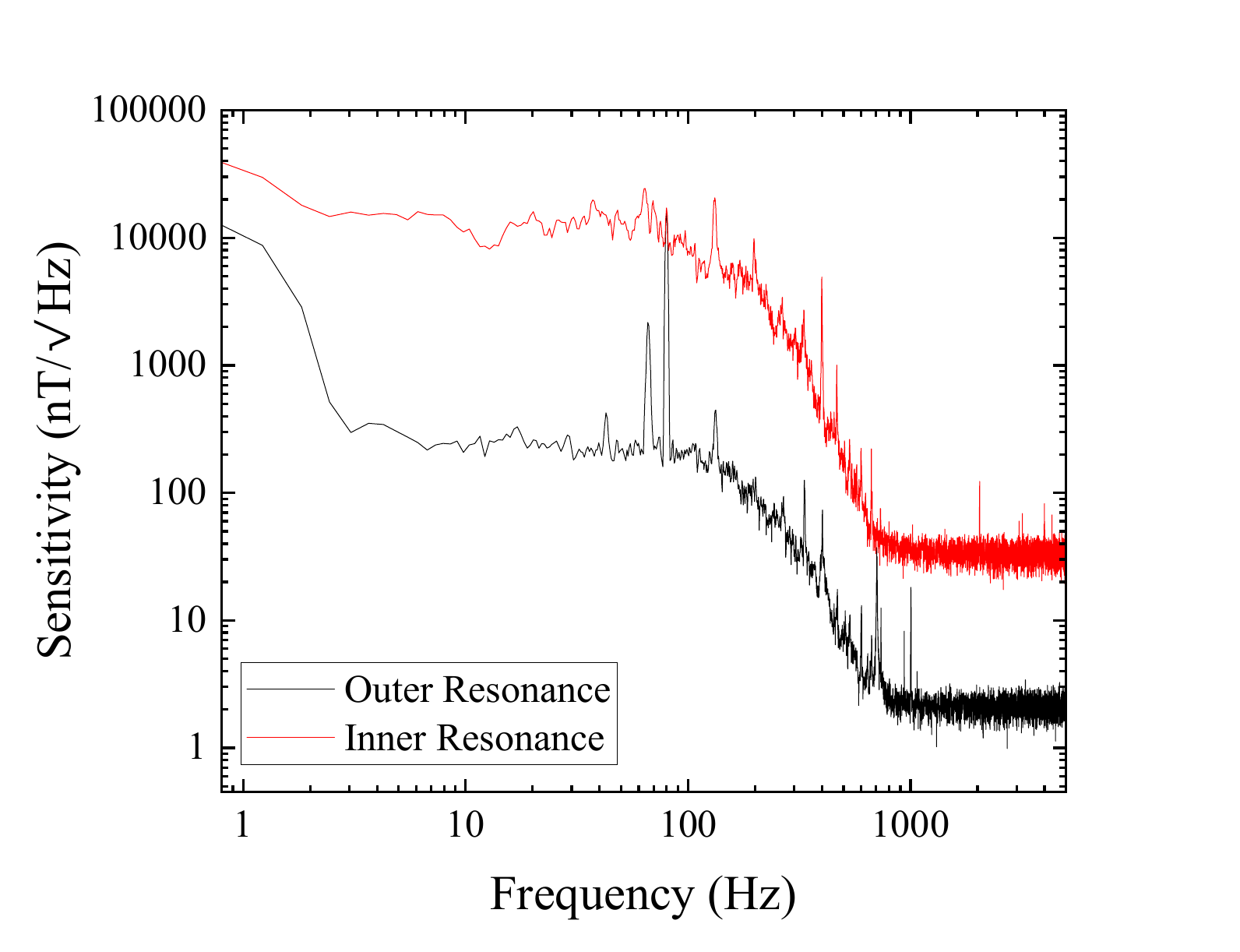}
\vspace{-5mm}
\caption{\small Sensitivity spectra taken on the inner (5,6,7 in Fig. \ref{fig:Near111ODMR}) and outer (8) resonances for a near-$\langle$111$\rangle$ alignment. An 80 Hz test field was applied parallel to the bias field.}
\label{fig: AppendixC-SensitivityNear111InnervsOuterResonances}
\end{figure}

Figure \ref{fig: AppendixC-SensitivityNear111InnervsOuterResonances} shows sensitivity measurements for a near-$\langle$111$\rangle$ alignment for the inner and outer resonances. This illustrates the benefits of aligning the bias field to one of the $\langle$111$\rangle$ directions, as well as the greater degradation in sensitivity that can occur for some non-degenerate alignments.

\FloatBarrier

\section*{Appendix D: Opto-Electronics and Sensor Head Design}

\begin{figure}[h!]
\hspace{-0.5cm}
\centering
\includegraphics[width=\columnwidth]{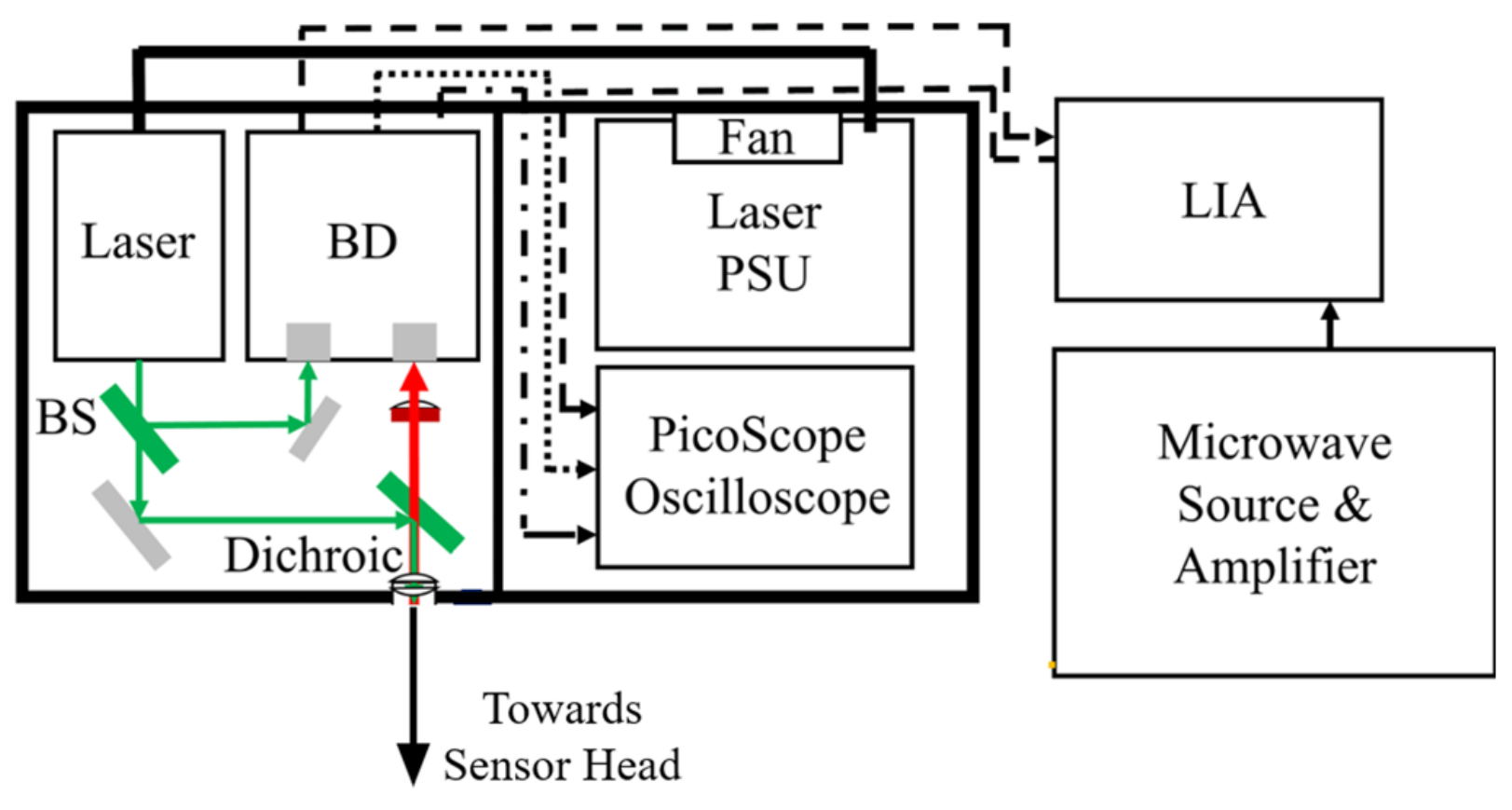} 
\caption{\small A schematic of the opto-electronics box.  PSU, power supply unit; BS, beam sampler; BD, balanced detector; LIA, lock-in amplifier.}
\label{fig: AppendixD-OptoelectronicsBox}
\end{figure}

\begin{figure}[h!]
\hspace{-0.5cm}
\centering
\includegraphics[width=\columnwidth]{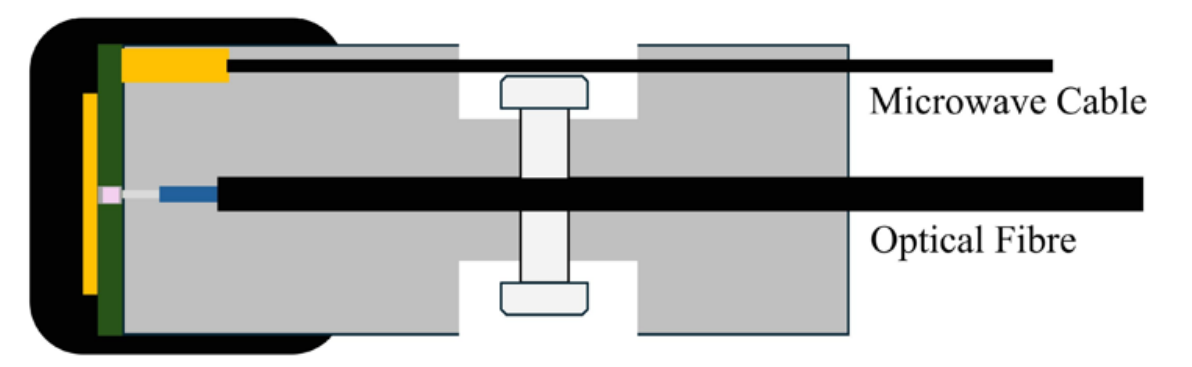} 
\caption{\small A schematic of the sensor head design.}
\label{fig: AppendixD-SensorHeadDesign}
\end{figure}

The opto-electronics box consisted of a (300 $\times$ 474 $\times$ 135)-mm box with a partition separating the optics from the electronics. Figure \ref{fig: AppendixD-OptoelectronicsBox} shows a schematic of this box. Further details are also found in Ref. \cite{graham2025road}. This opto-electronics box was placed on a portable trolley. The sensor head design used in this work is shown in Fig. \ref{fig: AppendixD-SensorHeadDesign}

\begin{figure}[t]
\centering
\includegraphics[width=\columnwidth]{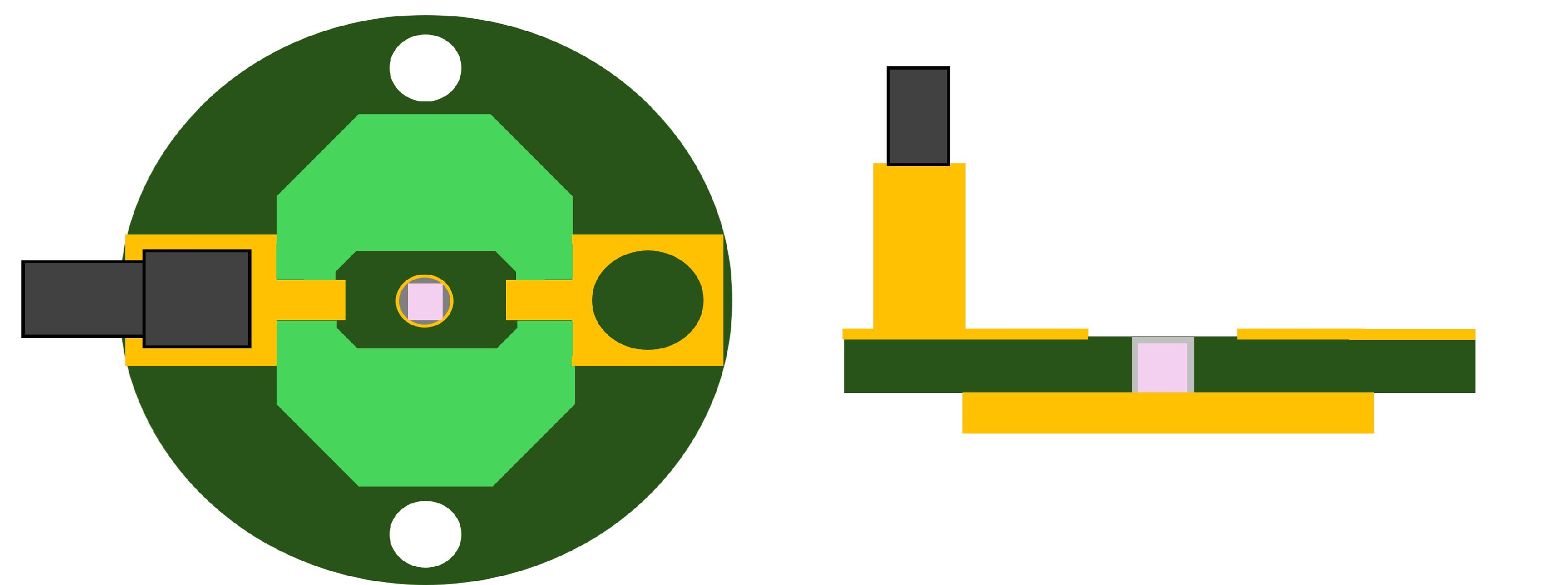}
\vspace{-5mm}
\caption{\small A schematic of the microwave transmission line viewed from the top and in profile.}
\label{fig: AppendixD-MicrowaveAntennaSchematic}
\end{figure}

Figure \ref{fig: AppendixD-MicrowaveAntennaSchematic} shows top and side-profile schematics of the microstrip design. The microstrip consists of a thin Cu conductor placed on a dielectric substrate, with a Cu ground plane below. There is also a top-surface ground plane producing a structure closer to a coplanar waveguide (GCPW). The dielectric substrate was made from FR4 and was covered in Honeywell TGP 8000PT thermal putty. The head was covered in metal tape for light-tightness. A mini-SMP launcher was soldered to one side to allow for a co-axial cable to be connected. On the other side it was terminated with a pair of 100 $\Omega$ RF resistors. The near-field $B_1$ drives the NVC ensemble.

\begin{figure}[h!]
\centering
\includegraphics[width=\columnwidth]{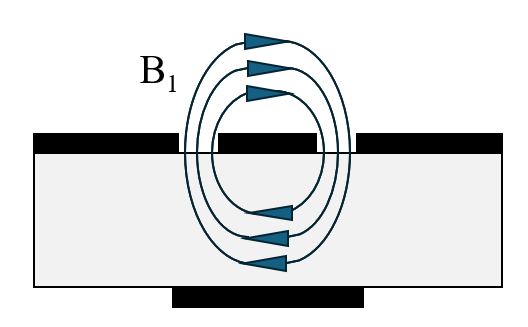}
\vspace{-5mm}
\caption{\small A schematic of the microwave transmission line in profile showing the $B_1$ field lines.}
\label{fig: AppendixD-B1FieldSchematic}
\end{figure}

The $B_1$ field for this microstrip design is shown in Fig. \ref{fig: AppendixD-B1FieldSchematic}. The ground plane acts to constrain the field lines into the dielectric between the surface and the ground plane. The field should become more concentrated in the dielectric as the microwave frequency is increased \cite{simons2004coplanar}. In order to place the diamond where $B_1$ was most concentrated, an aperture was drilled through the transmission line into the dielectric. 

\section*{Appendix E: ODMR vs Magnetic Field Strength}

\begin{figure*}[t]
\centering
\includegraphics[width=\columnwidth, trim={1.5cm 1cm 1.5cm 1.5cm}]{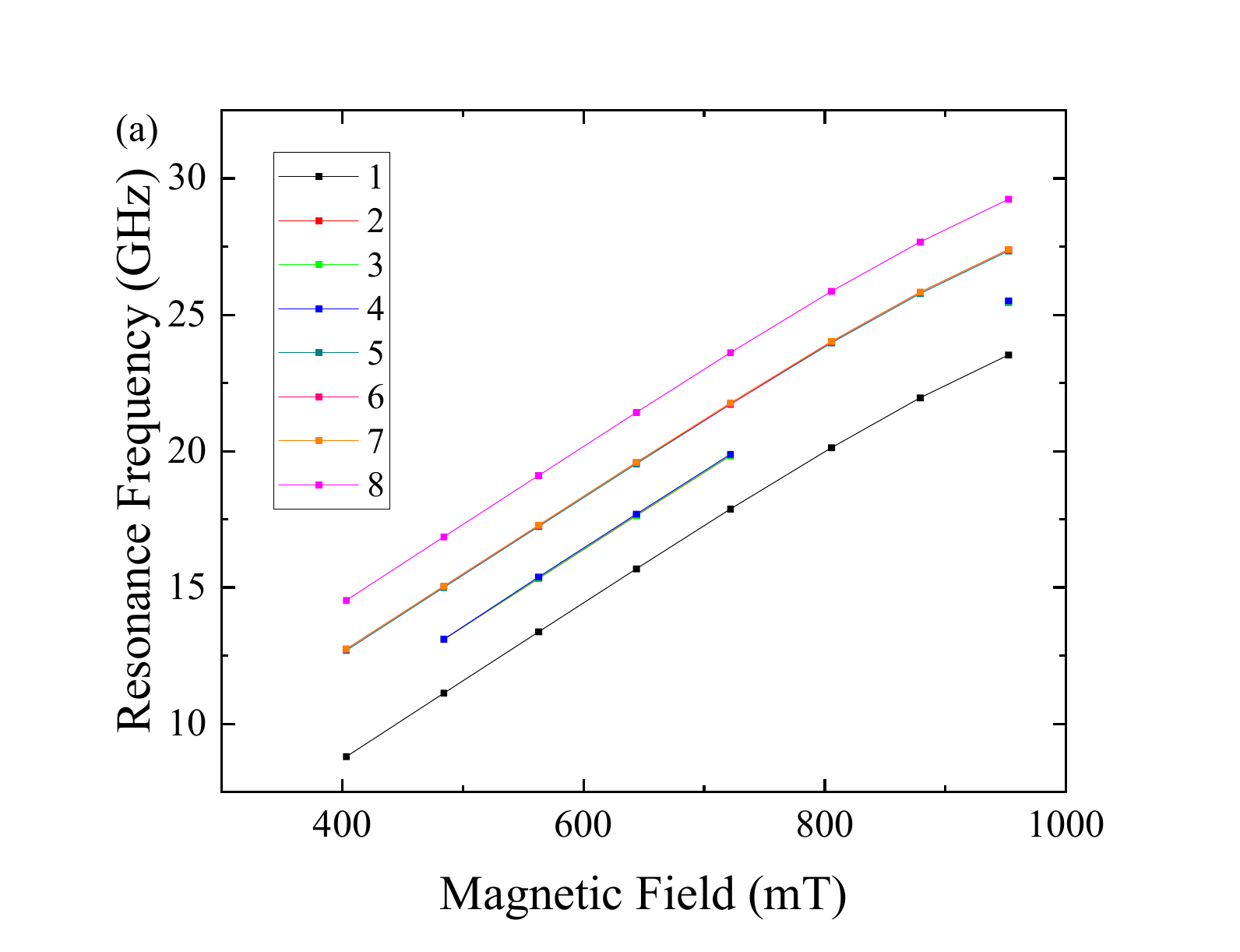}
\includegraphics[width=\columnwidth, trim={1.5cm 1cm 1.5cm 1.5cm}]{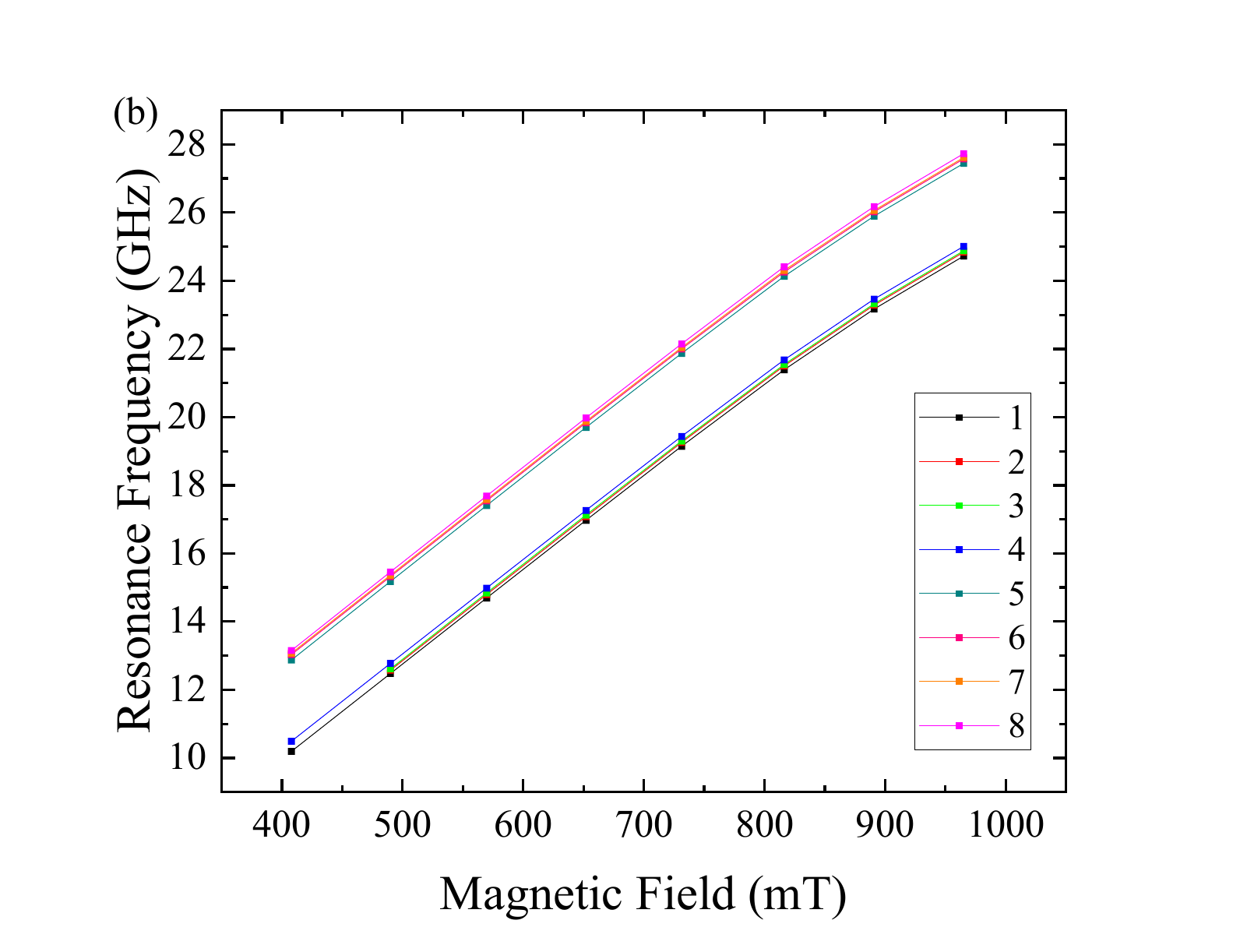}
\includegraphics[width=\columnwidth, trim={1.5cm 0.5cm 1.5cm 1.5cm}]{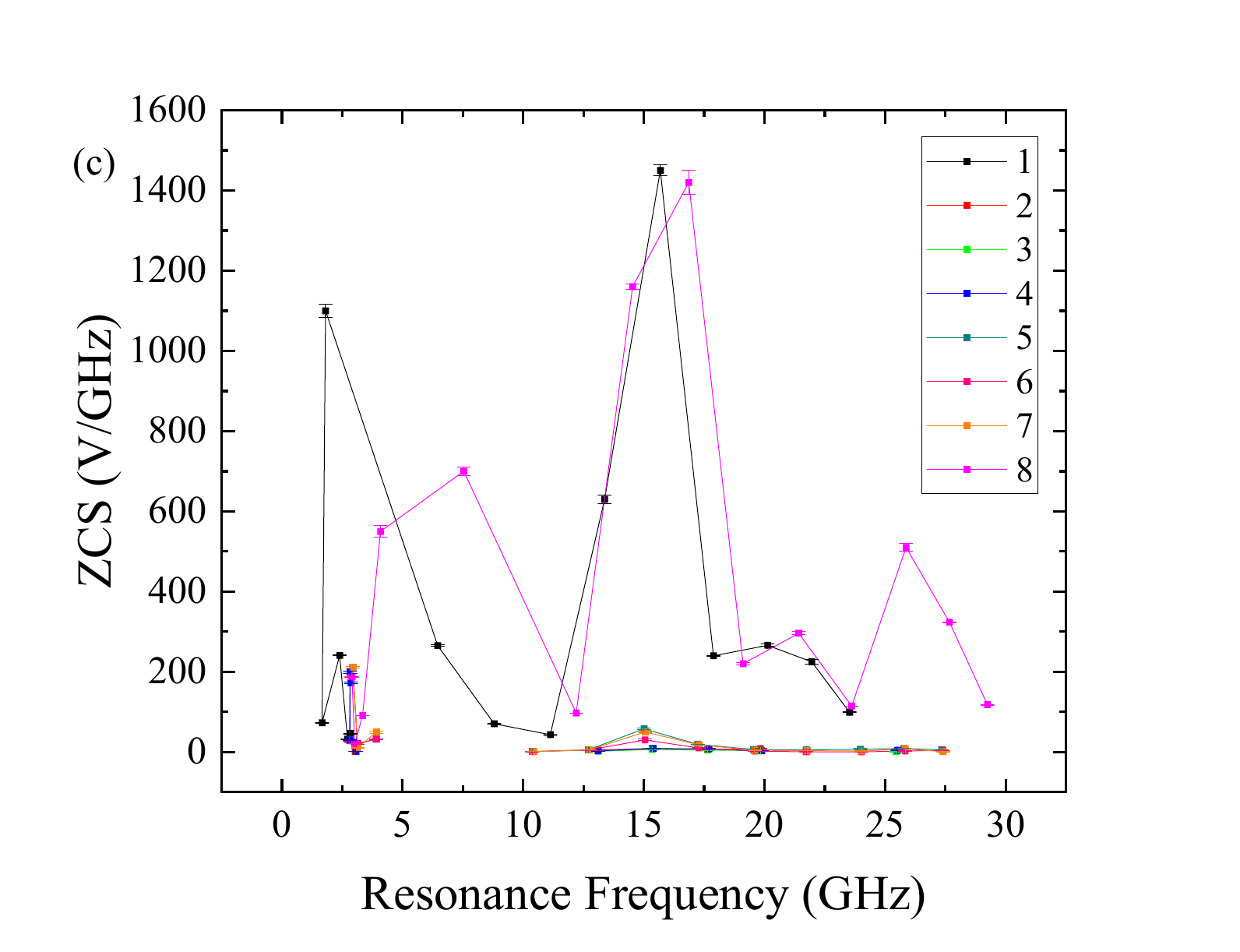}
\includegraphics[width=\columnwidth, trim={1.5cm 0.5cm 1.5cm 1.5cm}]{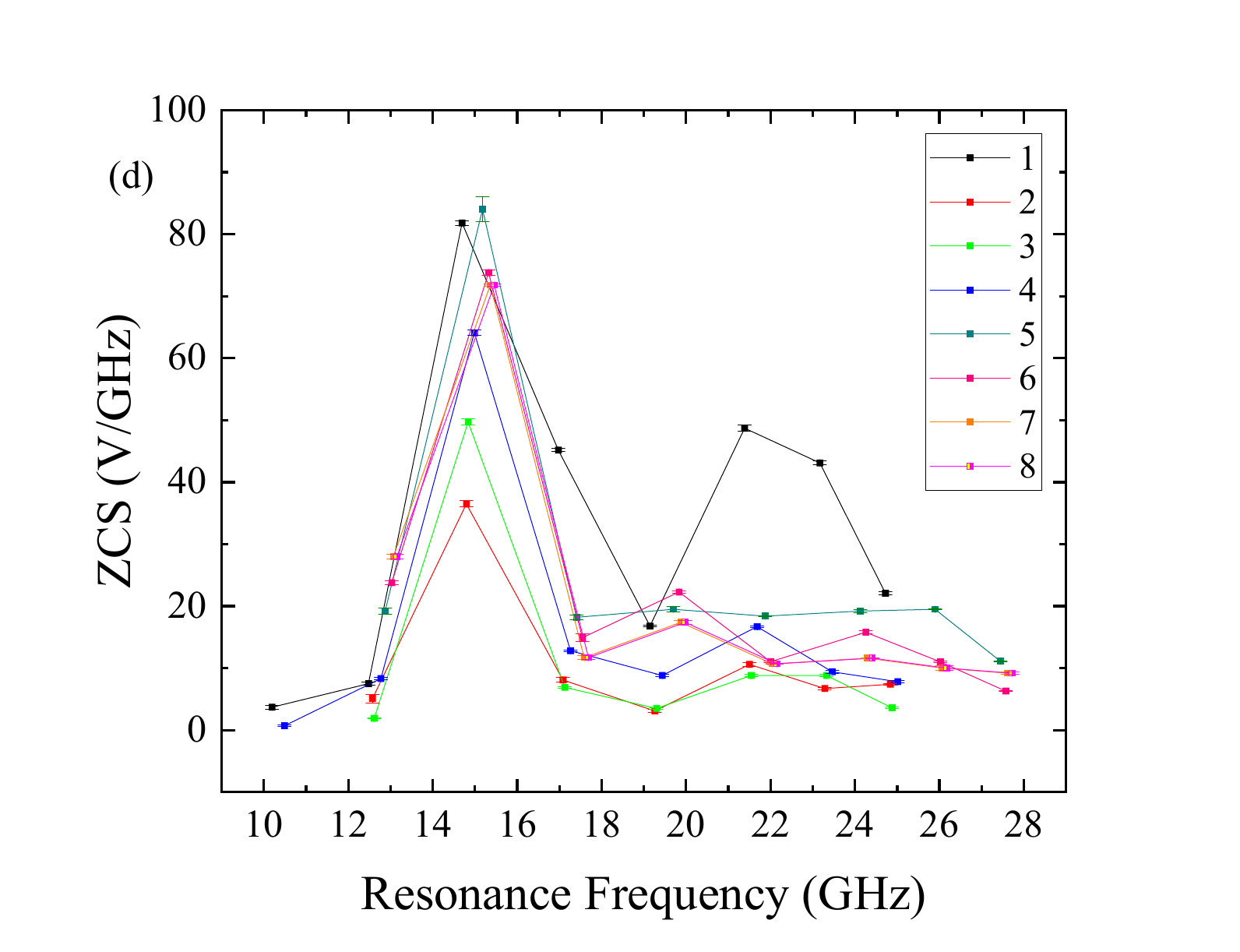}
\vspace{-5mm}
\caption{\small (a) and (b) show the resonance frequency measured for each resonance as a function of the magnetic field strength from 0.4 to 0.95 T for a near-$\langle$111$\rangle$ and non-degenerate alignment respectively. (c) and (d) show the corresponding ZCSs as a function of resonance frequency for a near-$\langle$111$\rangle$ and non-degenerate alignment respectively. For these measurements a microwave power of +20 dBm, a modulation frequency of 3.5 kHz and a modulation depth of 4 MHz were used. The missing data points for the near-$\langle$111$\rangle$ alignment are due to some of the inner-resonances being below the noise at certain magnetic field strengths (and thus microwave frequencies).}
\label{fig: AppendixE-ZCS-FreqvsMWFrequency}
\end{figure*}

ODMR spectra were also taken for intermediate field strengths between the low and high levels described in the main paper. Figure \ref{fig: AppendixE-ZCS-FreqvsMWFrequency} shows the measured resonance frequencies and ZCSs of the ODMR as a function of the magnetic field strength (and thus microwave frequency) for both the near-$\langle$111$\rangle$ and non-degenerate alignments. The apparent leveling in the resonance frequency towards 1 T can be attributed to the saturation of the magnet power supply at around 0.95 T (see appendix K).

It is likely that some of the decrease in the ZCSs as a function of the magnetic field strength can be attributed to the frequency performance of the microstrip design, as opposed to the NVC physics. This could also explain the drop in ZCS at approximately 15 GHz for both field alignments. ODMR measurements for a near-$\langle$100$\rangle$ are also shown in appendix I. A factor to consider is the orientation of the microwave field $B_1$ relative to the quantisation axis. This changes from being set by the NVC symmetry axes to depending upon the bias field orientation as the field strength is increased. Generally, the selection rules for the magnetic dipole transition are $\Delta$$m_s$ = $\pm$ 1 and $\Delta$$m_s$ = 0 for a $B_1$ field perpendicular and parallel to the quantisation axis respectively. For a misaligned magnetic field, the zero-field $m_s$ values are no longer good quantum numbers, and the selection rules no longer apply. The optimum $B_1$ field orientation differs between high and low field, being perpendicular to the bias field in the former case. 

Using the simulations described in appendix A, Figs. \ref{fig: AppendixE-Simulations-ZCSvsMagneticField}a and \ref{fig: AppendixE-Simulations-ZCSvsMagneticField}b show the resonance frequencies as a function of magnetic field. Note in the simulations the presence of a ninth resonance that appears to have a gyromagnetic ratio 2$\times$28 GHz/T in the high field regime. This can be attributed to the DQ transitions with $\Delta$$m_s$ = $\pm$ 2 \cite{jeong2017understanding}. The DQ transitions are generally forbidden, but become more probable as the misaligned field strength is increased. No DQ transitions are observed for the parallel NVC orientation for a $\langle$111$\rangle$ alignment. There are strictly four DQ transitions, but these tend towards a single frequency as the field is increased. The DQ transitions are insensitive to the orientation of the diamond, or equivalently to the magnetic field alignment, at high field \cite{jeong2017understanding}. 
The DQ transition is too weak and high in frequency to be observed in the experimental data. Figure \ref{fig: AppendixE-Simulations-ZCSvsMagneticField}c shows the simulated peak-to-peak for each resonance as a function of microwave frequency (and hence magnetic field) for a $\langle$111$\rangle$ alignment and Fig. \ref{fig: AppendixE-Simulations-ZCSvsMagneticField}d for the non-degenerate alignment used in the main paper. As can be seen, no peak is seen at 15 GHz, unlike the experimental data shown in Fig. \ref{fig: AppendixE-Simulations-ZCSvsMagneticField}. These simulations show the performance over a wider magnetic field range than could be accessed experimentally. Based on these simulations, the sensitivity should not be significantly degraded at fields up to 10 T, compared to 0.95 T, considering only NVC physics. In a future DEMO or commercial tokamak, fields are more likely to be on the order of 4 T at the plasma. A 4 T field would imply microwave frequencies $>$ 100 GHz. This would pose a challenge as rigid waveguides would be necessary instead of coaxial cables. It would also require an alternative microwave antenna design \cite{creely2020overview}.

\begin{figure*}[t]
\centering
\includegraphics[width=\columnwidth, trim={1.5cm 1cm 1.5cm 1.5cm}]{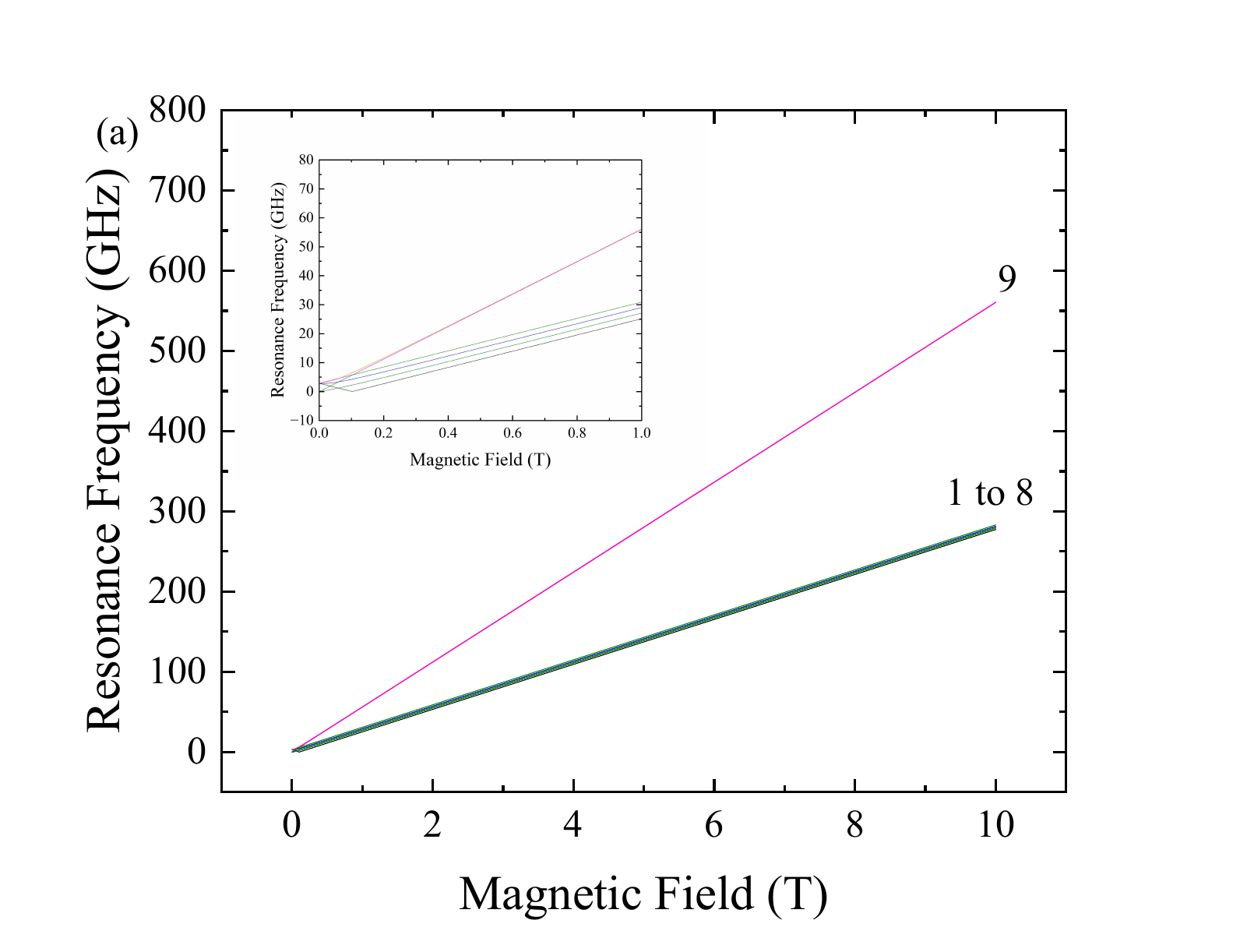} 
\includegraphics[width=\columnwidth, trim={1.5cm 1cm 1.5cm 1.5cm}]{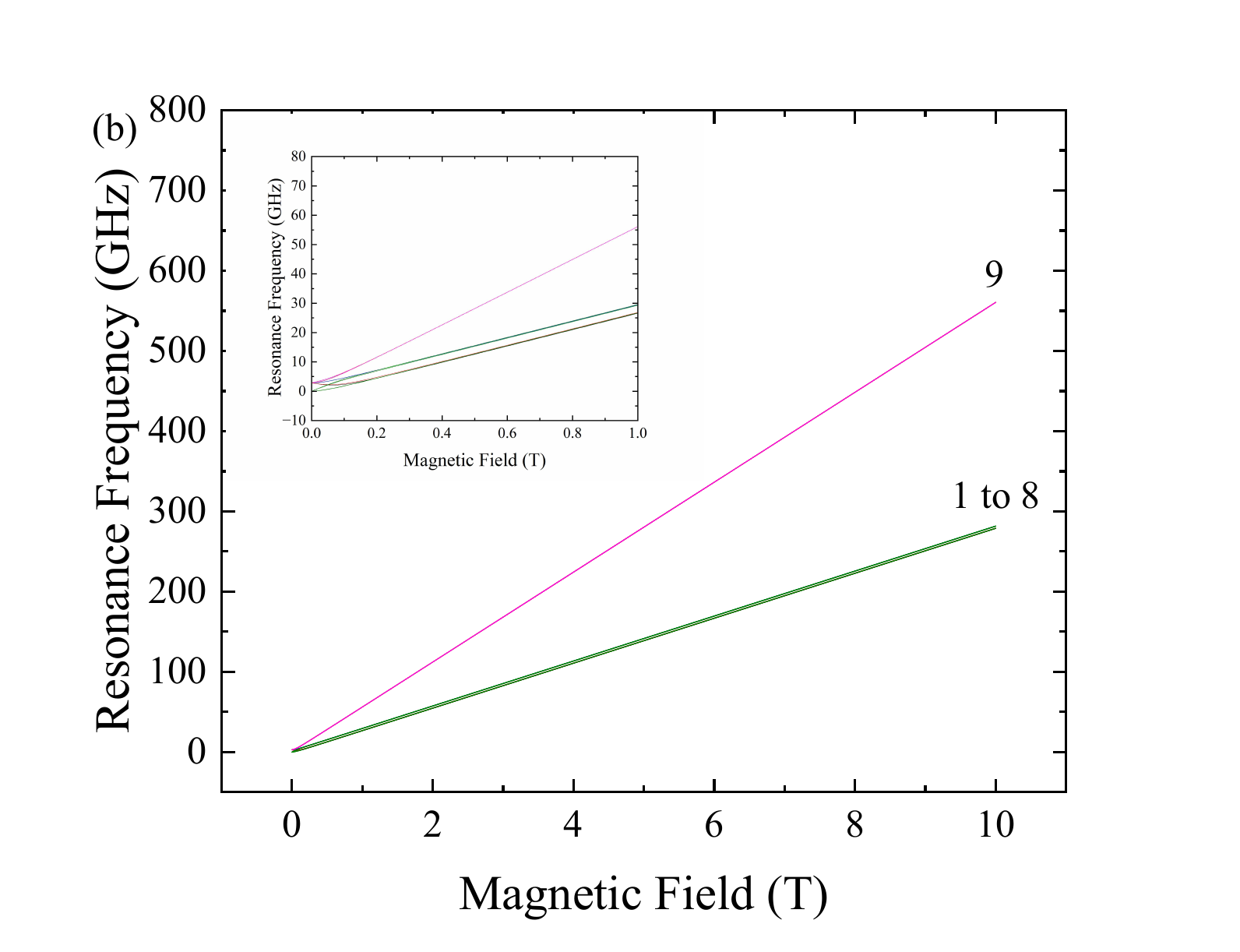} 
\includegraphics[width=\columnwidth, trim={1.5cm 1cm 1.5cm 1.5cm}]{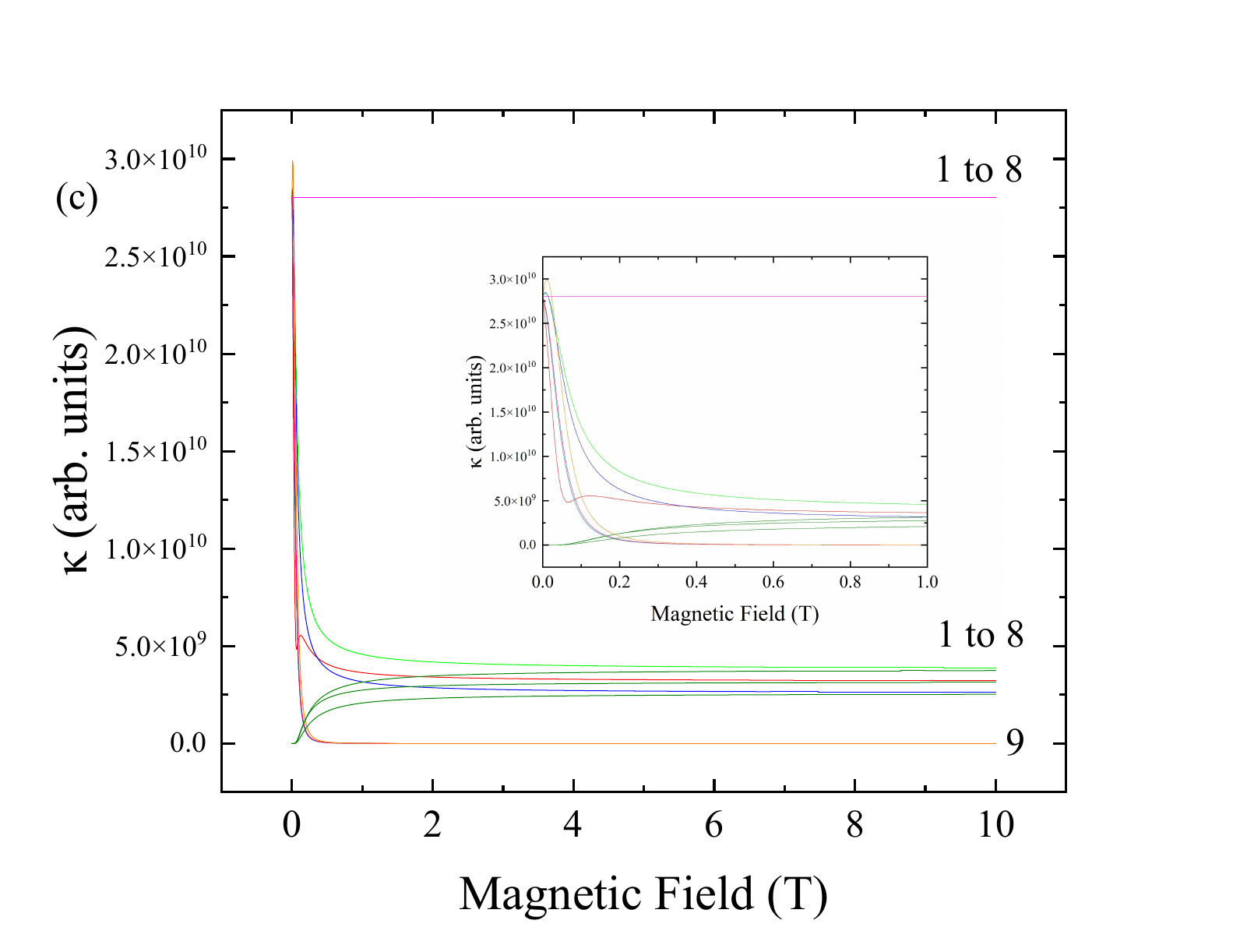} 
\includegraphics[width=\columnwidth, trim={1.5cm 1cm 1.5cm 1.5cm}]{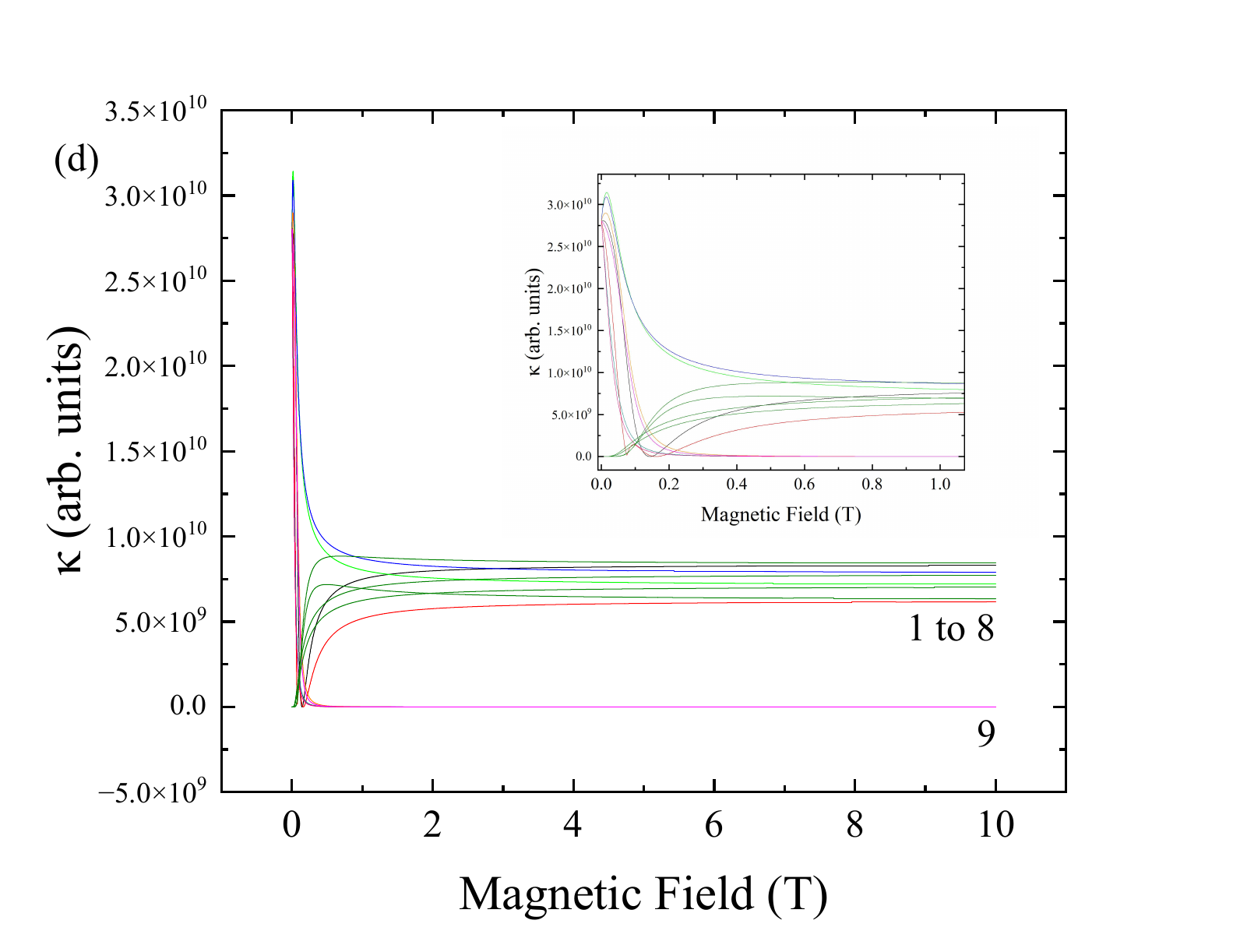} 
\caption{\small Simulations of the resonance frequencies for a (a) near-$\langle$111$\rangle$ alignment and (b) non-degenerate alignment from 0 to 10 T. The insets show from 0 to 1 T. Simulations of the $\kappa$ values corresponding to the resonance frequencies for (c) near-$\langle$111$\rangle$ and (d) non-degenerate alignment from 0 to 10 T. The insets show from 0 to 1 T. All generated using spin-Hamiltonian simulations described in appendix A. A linewidth of 4 MHz was used. Note that the appropriate eigenbasis changes between the low, intermediate and high field regimes with the low field $m_s$ quantum numbers no longer applying in the high field regime, for a non-degenerate alignment. A given line, therefore, can correspond to a single quantum transition in the low field, while being a DQ transition in the high field regime.}
\label{fig: AppendixE-Simulations-ZCSvsMagneticField}
\end{figure*}

\FloatBarrier

\section*{Appendix F: ODMR Hyperfine Structure}

\begin{figure}[h!]
\centering
\includegraphics[width=\columnwidth, trim={1.5cm 1cm 1.5cm 1.5cm}]{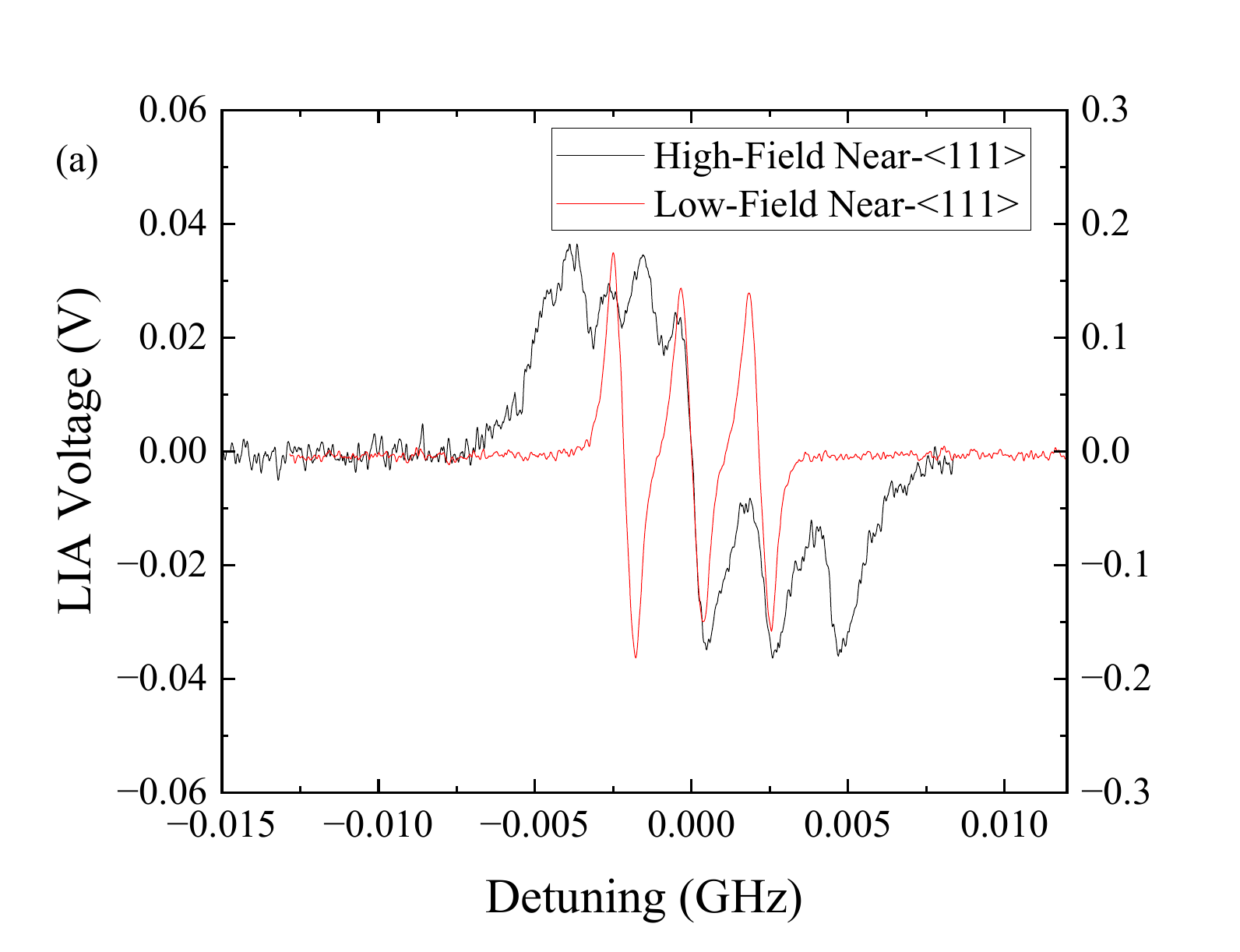} 
\includegraphics[width=\columnwidth, trim={1.5cm 1cm 1.5cm 1.5cm}]{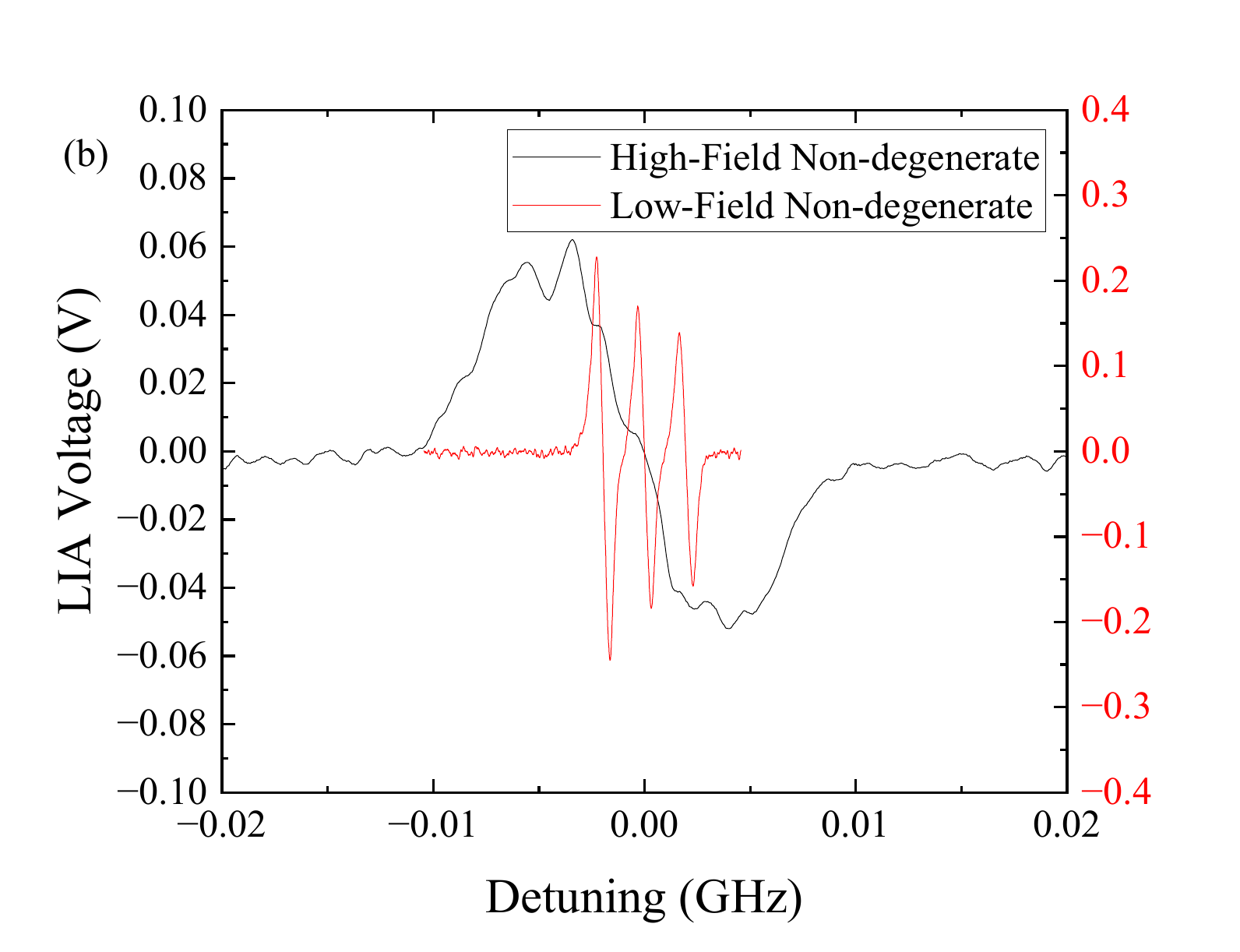}
\caption{\small (a) Demodulated ODMR spectra at low and high-field for a near-$\langle$111$\rangle$ alignment. (b) Demodulated ODMR spectra at low and high-field for a non-degenerate alignment. A modulation depth of 400 kHz was used for these measurements. Here detuning refers to the frequency deviation from the centre of each resonance feature.}
\label{fig: AppendixF-HyperfineStructureLowvsHigh}
\end{figure}

In the main paper, for high-field measurements, a modulation depth of 4 MHz was found to provide the highest sensitivity. However, this obscured the hyperfine structure of each ODMR resonance, and thus measurements were also taken with a lower modulation depth of 400 kHz. 
Figure \ref{fig: AppendixF-HyperfineStructureLowvsHigh}a and \ref{fig: AppendixF-HyperfineStructureLowvsHigh}b show the hyperfine structure at low (1 to 3 mT) and high field (0.95 T) for both the near-$\langle$111$\rangle$ and non-degenerate sensor head alignment.

\begin{figure}[h!]
\centering
\includegraphics[width=\columnwidth, trim={1.5cm 1cm 1.5cm 1.5cm}]{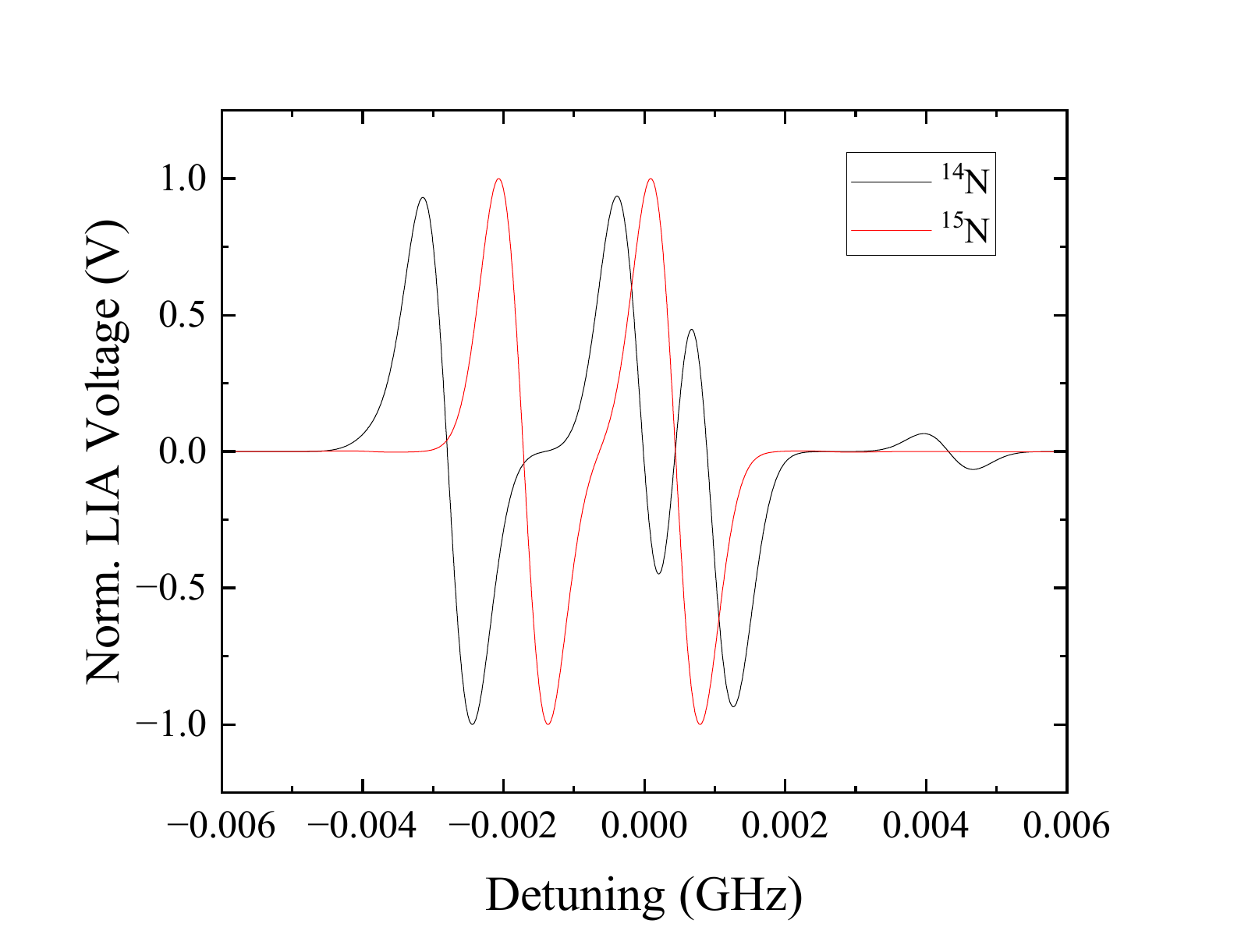} 
\caption{\small EasySpin simulations of normalised demodulated ODMR spectra at high-field (1 T), setting the nitrogen-atom to $^{15}\textrm{N}$ and $^{14}\textrm{N}$. Here detuning refers to the frequency deviation from the centre of each resonance feature.}
\label{fig: AppendixF-15vs14NHyperfineSimulations}
\end{figure}

As can be seen, for both alignments at low field only the three $m_I$ = -1, 0 and 1 hyperfine features are visible, with splittings of approximately 2.158 MHz. These features are due to the hyperfine interaction with the I = 1 $^{14}\textrm{N}$ nuclei of the NVC. 
At higher fields, at least for non-$\langle$111$\rangle$ alignments, the mixing of electron spin sub-levels by the transverse magnetic field components results in mixing of the product electron-nuclear spin sub-levels \cite{acosta2009diamonds, doherty2012theory, shin2014optically}. 
Under these conditions the $m_I$ values are no longer good quantum numbers and forbidden transitions involving nuclear spin flips are permitted. This results in additional resonances, with this structure becoming blurred out into a single broad peak as these features become comparable in amplitude to the nuclear spin conserving transitions \cite{felton2009hyperfine, doherty2012theory, schloss2018simultaneous}. This explains the difference in optimum modulation depth seen at high and low fields in appendix J. This is observed for both alignments, although no forbidden transitions would be expected for a perfect $\langle$111$\rangle$. 
Figure \ref{fig: AppendixF-15vs14NHyperfineSimulations} shows an EasySpin simulation of the expected hyperfine structure at 0.95 T for an NVC ensemble with both $^{14}\textrm{N}$ and $^{15}\textrm{N}$ nuclei. As can be seen, because the $^{15}\textrm{N}$ nuclei have a nuclear spin, I = 1/2, there is no quadrupole term in the Hamiltonian, and thus electron-nuclear product state mixing is not observed. 
This suggests that using $^{15}\textrm{N}$ purified diamonds could be useful to maintain contrast and thus sensitivity at high fields, in cases where the field is not parallel to a $\langle$111$\rangle$ alignment \cite{felton2009hyperfine}.

\FloatBarrier

\section*{Appendix G: Magnetic Field Spectra}

\begin{figure}[h!]
\centering
\includegraphics[width=\columnwidth, trim={1.5cm 1cm 1.5cm 1.5cm}]{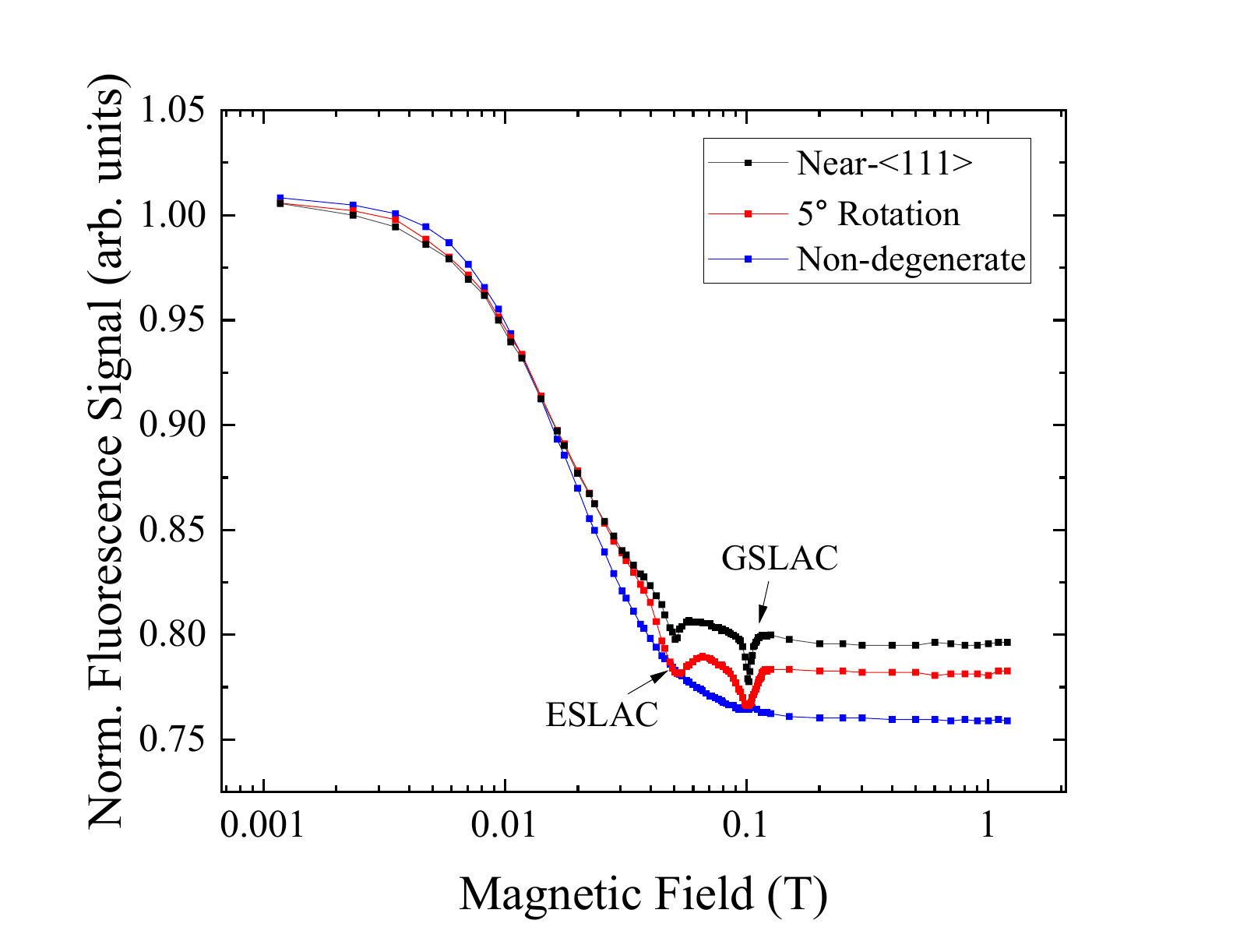} 
\caption{\small Magnetic spectra showing the change in fluorescence as a function of the magnetic field strength for various bias field alignments. 5$^{\circ}$ refers to a rotation about the sensor head's symmetry axis from the initial near-$\langle$111$\rangle$ alignment.}
\label{fig: AppendixG-MagneticSpectra}
\end{figure}

To confirm that the bias field was aligned close to a $\langle$111$\rangle$ direction, we measured the fluorescence signal as a function of the magnetic field. This is called a magnetic spectrum \cite{rogers2009time}. Figure \ref{fig: AppendixG-MagneticSpectra} shows this measurement taken for different orientations of the sensor head within the bias field, starting approximately along an $\langle$111$\rangle$, rotating away until the field was aligned such that eight, non-degenerate resonances were visible. As can be seen, the fluorescence level at high-field is higher for the near-$\langle$111$\rangle$ orientation than for the non-degenerate orientation. Some drop in fluorescence does occur due to the other three NVC orientations, however, for the parallel orientation the fluorescence level recovers following the excited-state and ground-state level anti-crossings (ESLAC and GSLAC) at 51.2 and 102.4 mT respectively. The presence of the GSLAC is a good indicator of a $\langle$111$\rangle$ alignment, although the presence of the ESLAC suggests that the bias field was not perfectly aligned along this direction \cite{rogers2009time}. 
The GSLAC and ESLAC peaks are observed to broaden as the alignment is rotated away from the initial near-$\langle$111$\rangle$.

\FloatBarrier

\section*{Appendix H: 1.2 T Measurements}

\begin{figure}[h!]
\centering
\includegraphics[width=\columnwidth, trim={1.5cm 1cm 1.5cm 1.5cm}]{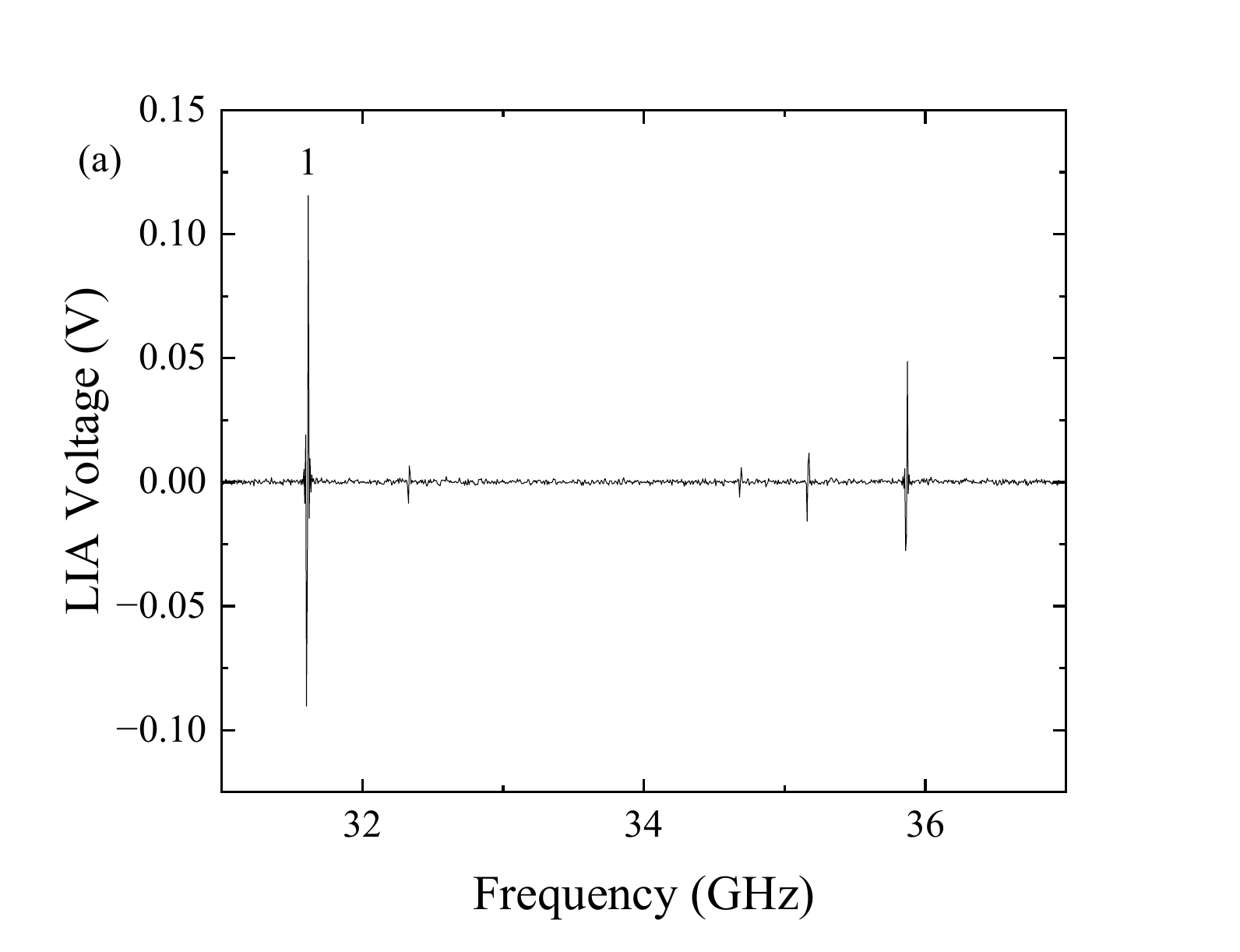} 
\includegraphics[width=\columnwidth, trim={1.5cm 1cm 1.5cm 1.5cm}]{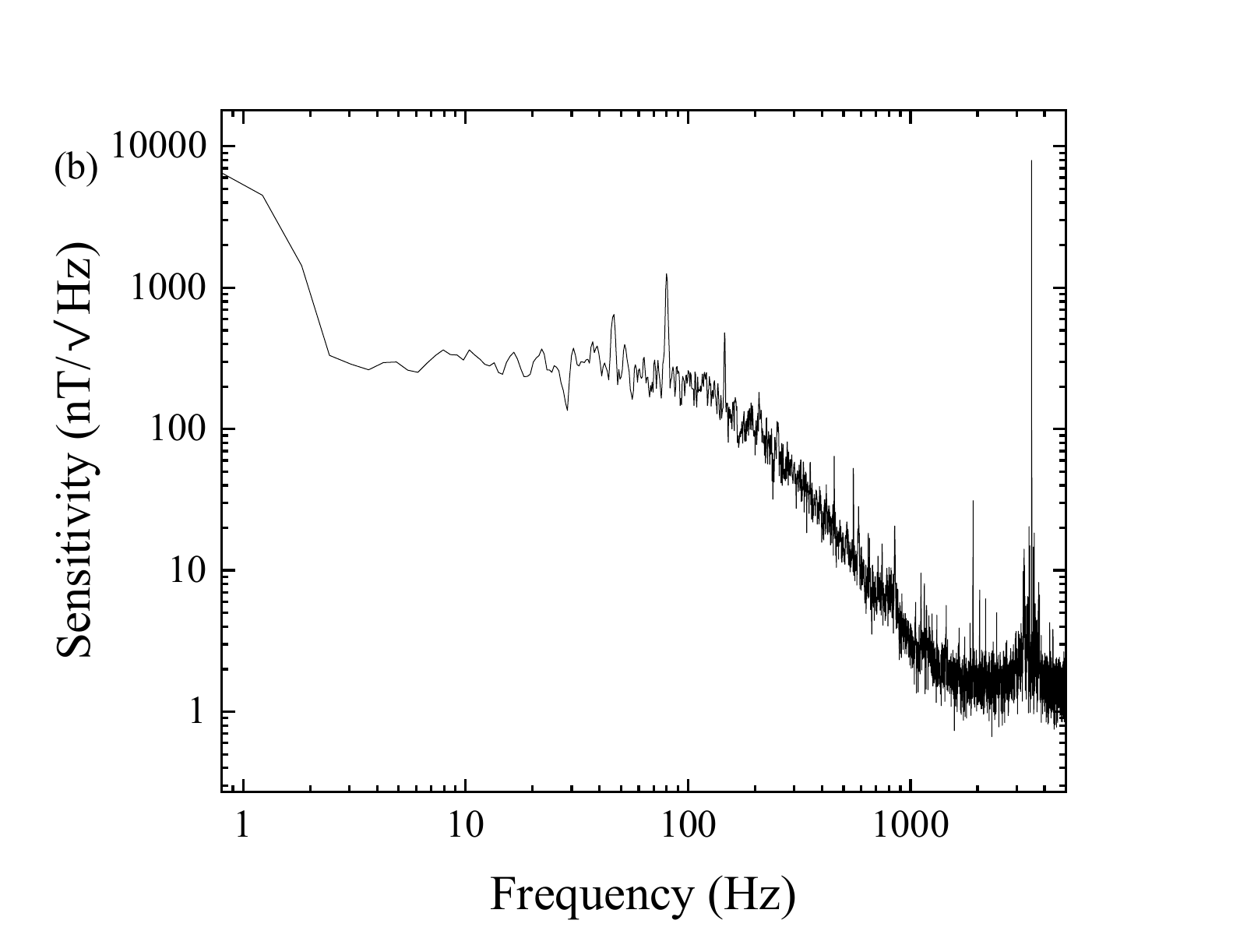}
\caption{\small (a) Demodulated ODMR spectrum at a magnetic field strength of approximately 1.2 T for a non-degenerate bias field alignment. (b) Sensitivity spectrum taken at a magnetic field strength of approximately 1.2 T for a non-degenerate bias field alignment. A LIA LPF of 150 Hz was used. An 80 Hz test field was applied parallel to the bias field.}
\label{fig: AppendixH-1.2TMeasurements}
\end{figure}

Figures \ref{fig: AppendixH-1.2TMeasurements}a and \ref{fig: AppendixH-1.2TMeasurements}b, respectively, show an ODMR spectrum and sensitivity measurement (with applied 80 Hz test field) at 1.2 T. This was for a non-degenerate alignment, different from that used for the main paper. A mean-sensitivity of (240 $\pm$ 60) nT/$\sqrt{\textrm{Hz}}$ was obtained for the leftmost resonance in an (10-150) Hz frequency range. 

\FloatBarrier

\section*{Appendix I: $\langle$100$\rangle$ Alignment}

\begin{figure}[t]
\centering
\includegraphics[width=\columnwidth, trim={1.5cm 1cm 1.5cm 1.5cm}]{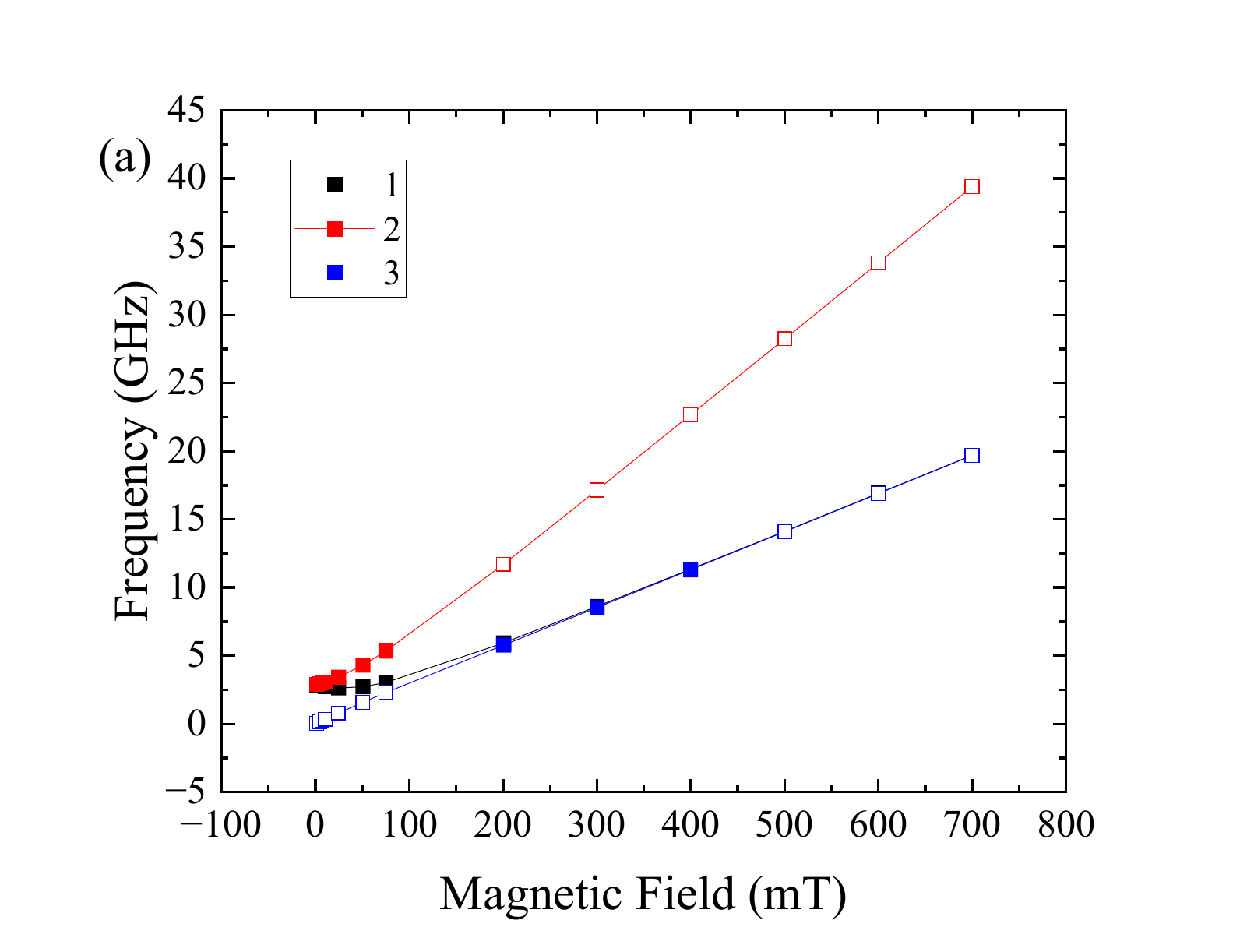}
\includegraphics[width=\columnwidth, trim={1.5cm 1cm 1.5cm 1.5cm}]{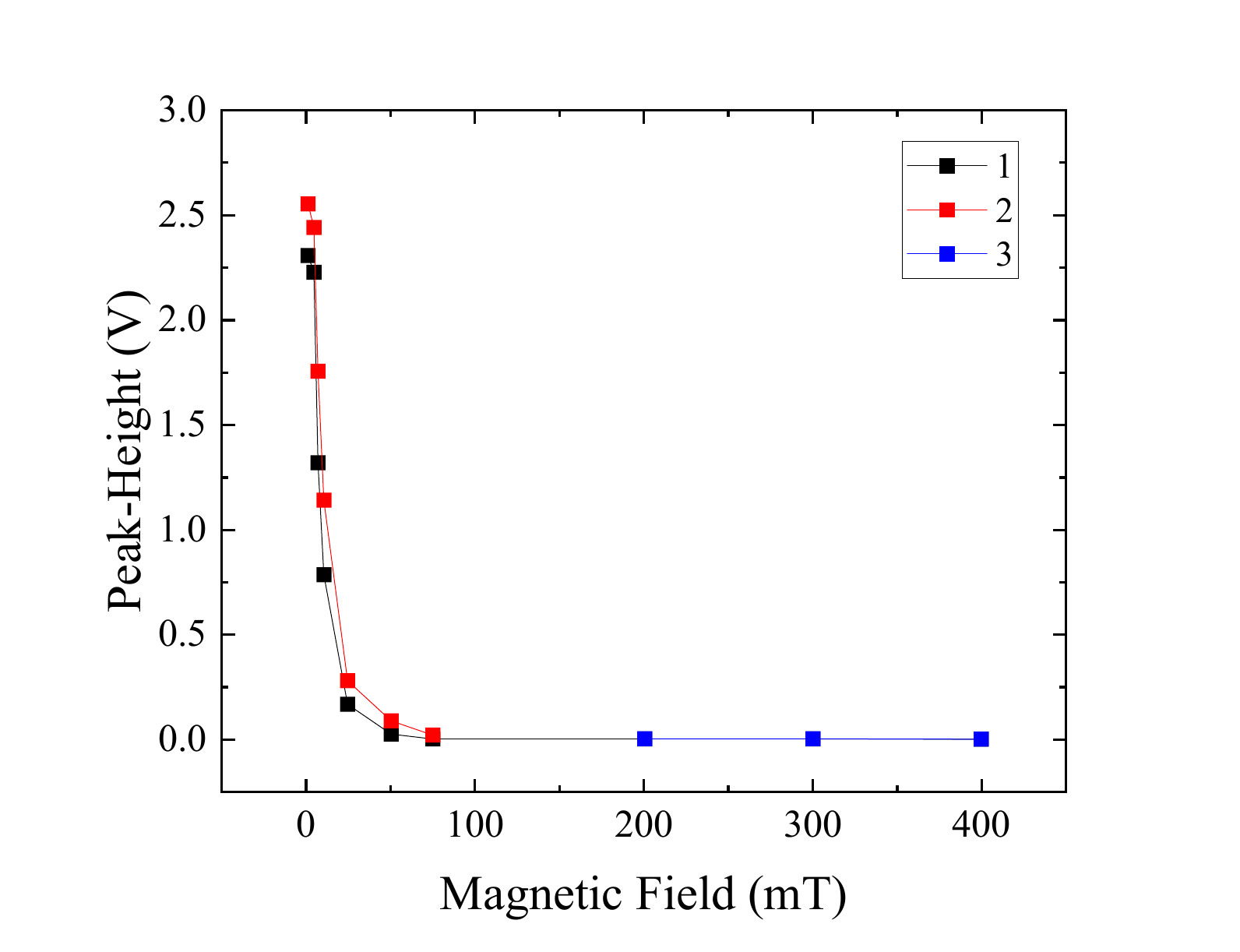}
\vspace{-5mm}
\caption{\small (a) Resonance frequencies and (b) corresponding peak heights measured as a function of the magnetic field strength for a near-$\langle$100$\rangle$ field alignment. A 2.003 kHz and 6 MHz modulation frequency and depth were used respectively. In (a) the missing resonance frequencies are indicated by open squares and these values were obtained using EasySpin simulations.}
\label{fig: AppendixI-100AlignmentODMR+ZCS}
\end{figure}

At high field, the orientation of the bias field becomes increasingly important for the sensitivity of the magnetometer. For a $\langle$100$\rangle$ bias field alignment no spin polarisation would be expected as this is the magic angle \cite{polenova2015magic}. These alignments are frequently used in low fields to increase sensitivity, as they enable simultaneous exploitation of the entire NVC population \cite{barry2024sensitive, graham2023fiber}. 
Figure \ref{fig: AppendixI-100AlignmentODMR+ZCS}a shows the resonance frequencies as a function of magnetic field for a $\langle$100$\rangle$ alignment. Resonance 2 can be attributed to the DQ transition in the high-field regime.
Figure \ref{fig: AppendixI-100AlignmentODMR+ZCS}b shows the resonance peak height (which is proportional for a fixed modulation depth and microwave power to the ZCS) corresponding to these resonance frequencies. No ODMR signal was visible above 400 mT. Given the degree of positioning control in the magnet bore, the diamond was not exactly aligned along a $\langle$100$\rangle$ alignment. For these measurements a modulation frequency of 2.003 kHz and a modulation depth of 6 MHz were used. 

\FloatBarrier

\section*{Appendix J: Parameter Optimisation}

Parameter optimisation was undertaken as described in Ref. \cite{patel2020subnanotesla}. ODMR spectra were taken as a function of the relevant microwave or FM (modulation frequency and depth) parameters. The ZCS was then determined by applying a linear fit to the derivative slope of the demodulated ODMR spectrum. A higher ZCS value indicates greater responsivity, and thus sensitivity. However, for modulation frequency measurements this does not account for potential variations in the noise level. Accordingly, a higher ZCS does not always suggest a superior sensitivity, and for these measurements corresponding noise measurements were taken to determine the sensitivity.

\begin{figure}[t]
\centering
\includegraphics[width=\columnwidth, trim={1.5cm 1cm 1.5cm 1.5cm}]{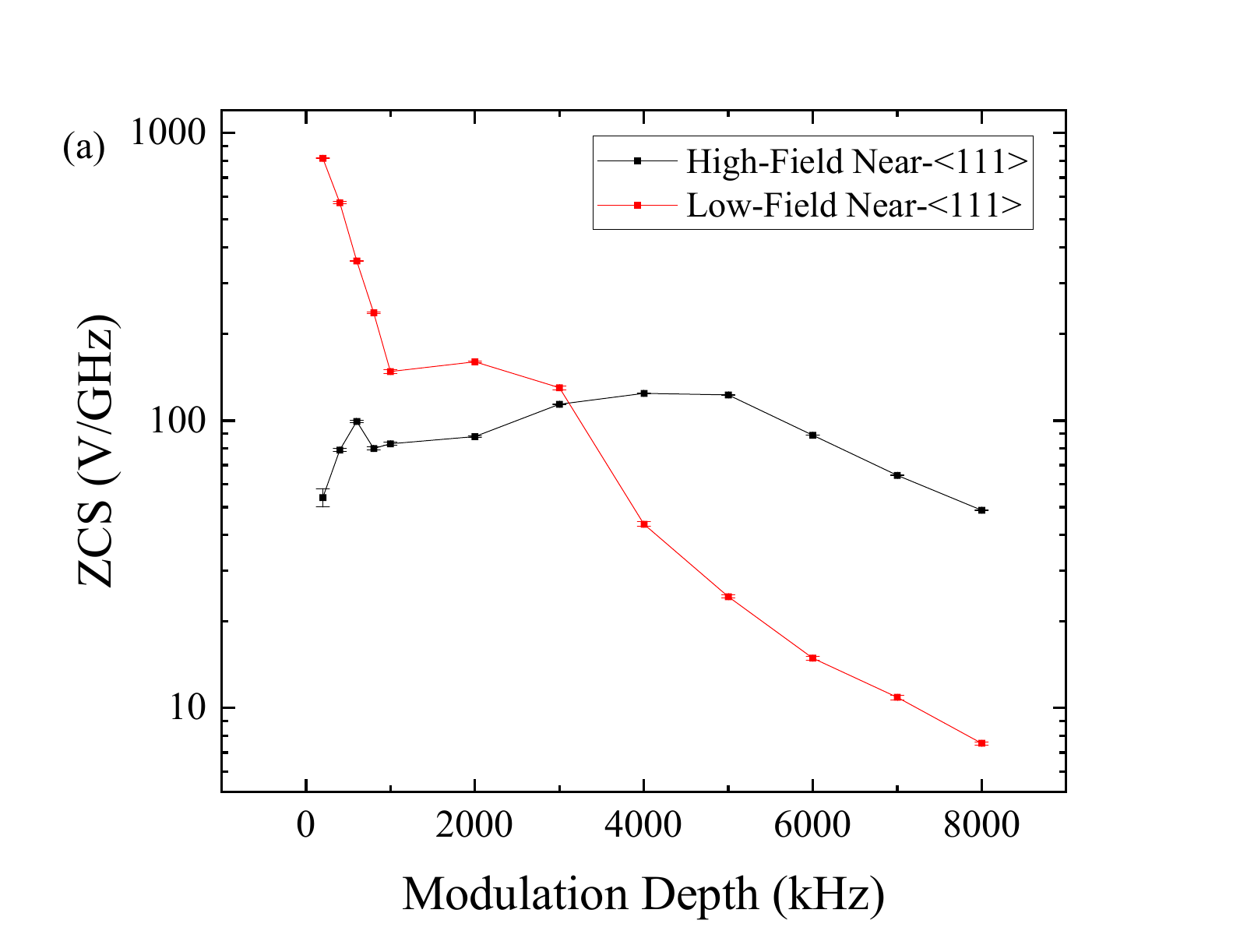}
\includegraphics[width=\columnwidth, trim={1.5cm 1cm 1.5cm 1.5cm}]{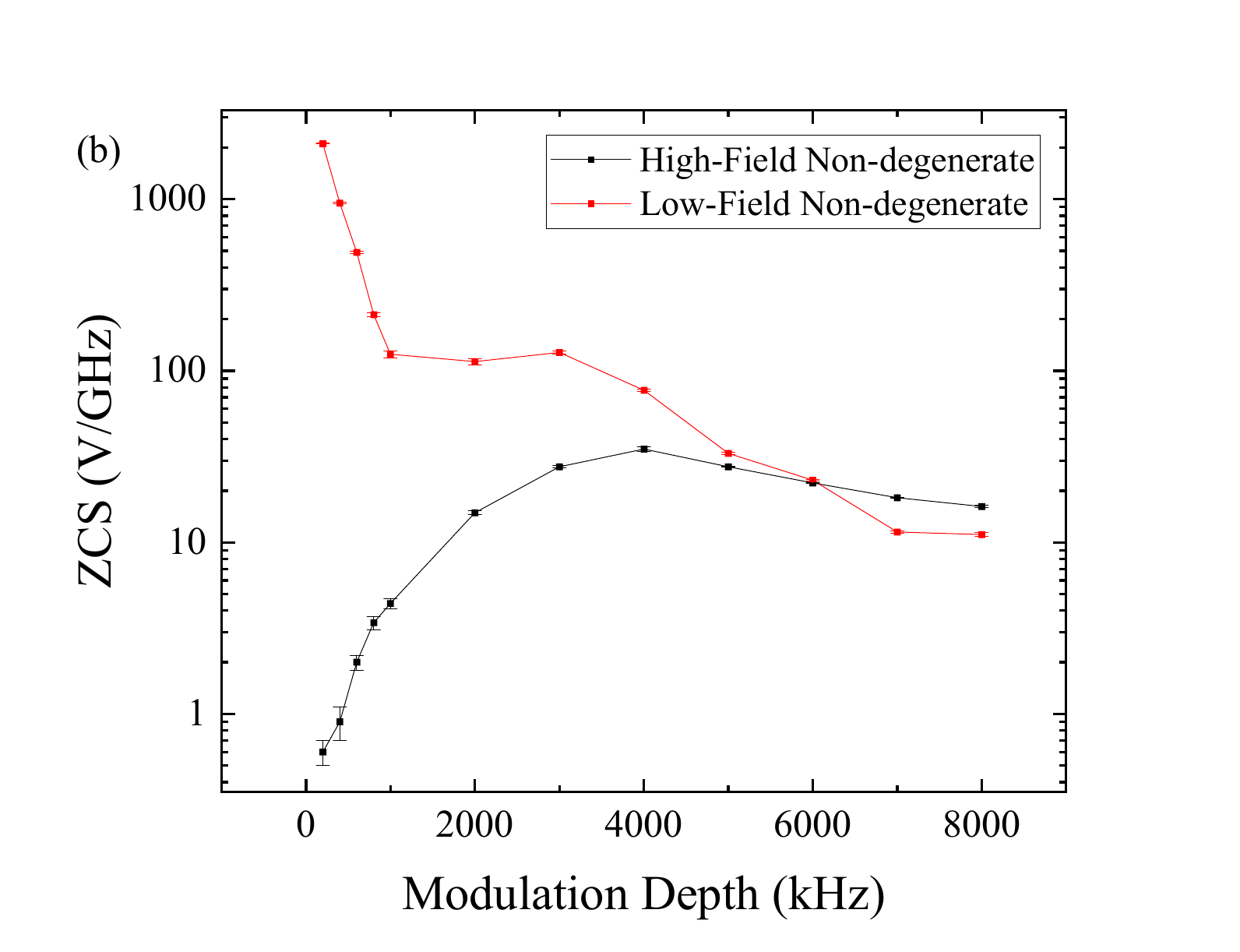}
\vspace{-5mm}
\caption{\small (a) The ZCS as a function of modulation depth at low and high-field for a near-$\langle$111$\rangle$ alignment. (b) The ZCS as a function of modulation depth at low and high-field for a non-degenerate alignment. A microwave power of +20 dBm and a modulation frequency of 3.5 kHz were used for these measurements.}
\label{fig: AppendixJ-ZCSvsModulationDepth}
\end{figure}

The ZCS was measured as a function of modulation depth at both low and high-fields, and for both the near-$\langle$111$\rangle$ and non-degenerate sensor head alignments. These measurements are shown in Figs. \ref{fig: AppendixJ-ZCSvsModulationDepth}a and \ref{fig: AppendixJ-ZCSvsModulationDepth}b. 

For both alignments, it is observed that the optimum modulation depth differs between low- and high-field. This can be attributed to the forbidden transitions that change the hyperfine structure of the ODMR. These are described further in appendix F. As a consequence, the relation $\Delta$$\nu$/2$\sqrt{3}$, where $\Delta$$\nu$ is the linewidth of the resonance, is not applicable \cite{barry2016optical, el2017optimised} and a higher modulation depth provides a higher sensitivity. This limits the possibility of improving sensitivity via techniques such as hyperfine excitation \cite{barry2016optical}. However, for a very good $\langle$111$\rangle$ alignment the hyperfine structure is largely unmodified, and so it might be anticipated that the optimum modulation depth would differ between a near-$\langle$111$\rangle$ and non-degenerate field alignment. However, this is not observed in the data in Fig. \ref{fig: AppendixJ-ZCSvsModulationDepth}.

\begin{figure*}[t]
\centering
\includegraphics[width=\columnwidth, trim={1.5cm 1cm 1.5cm 1.5cm}]{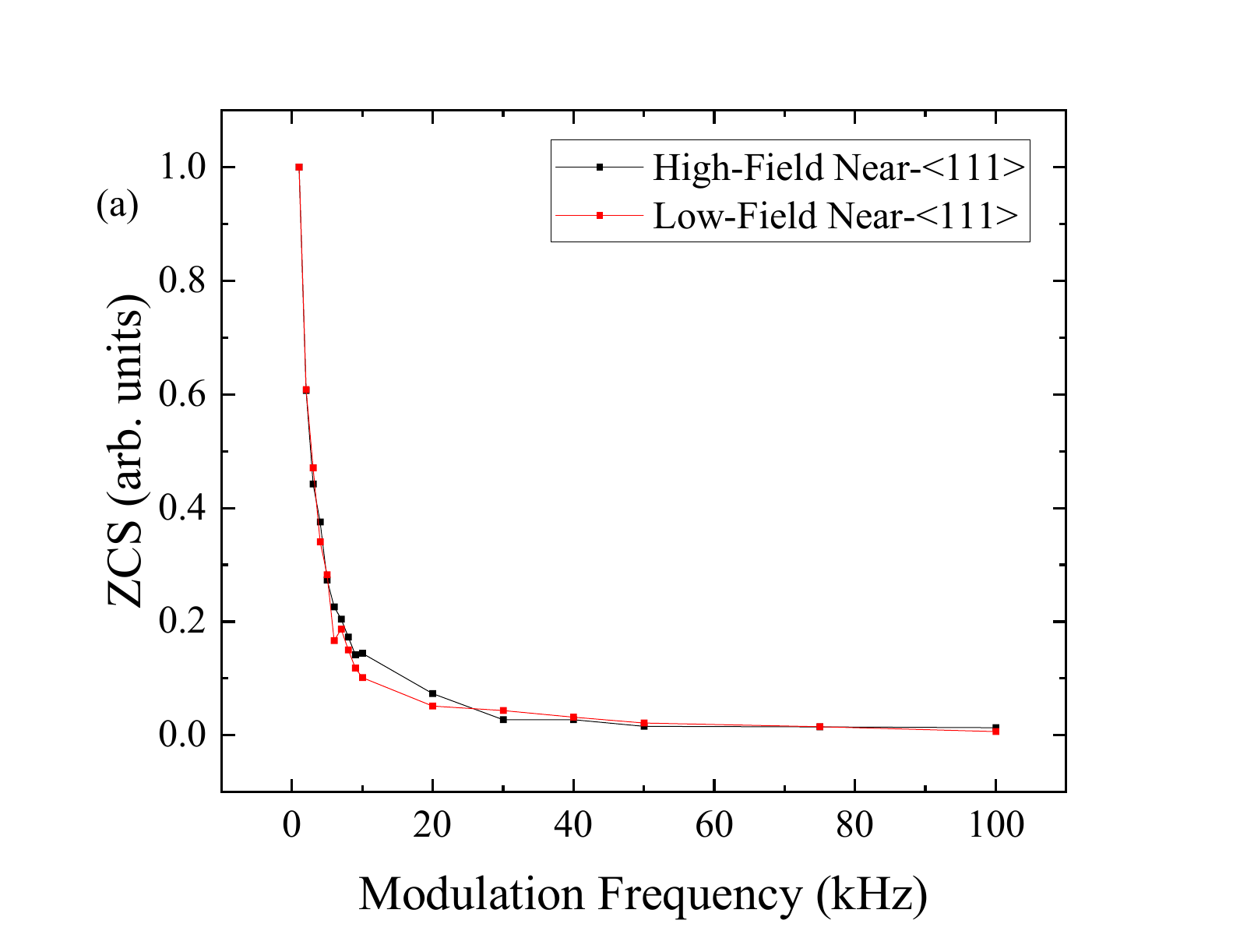}
\includegraphics[width=\columnwidth, trim={1.5cm 1cm 1.5cm 1.5cm}]{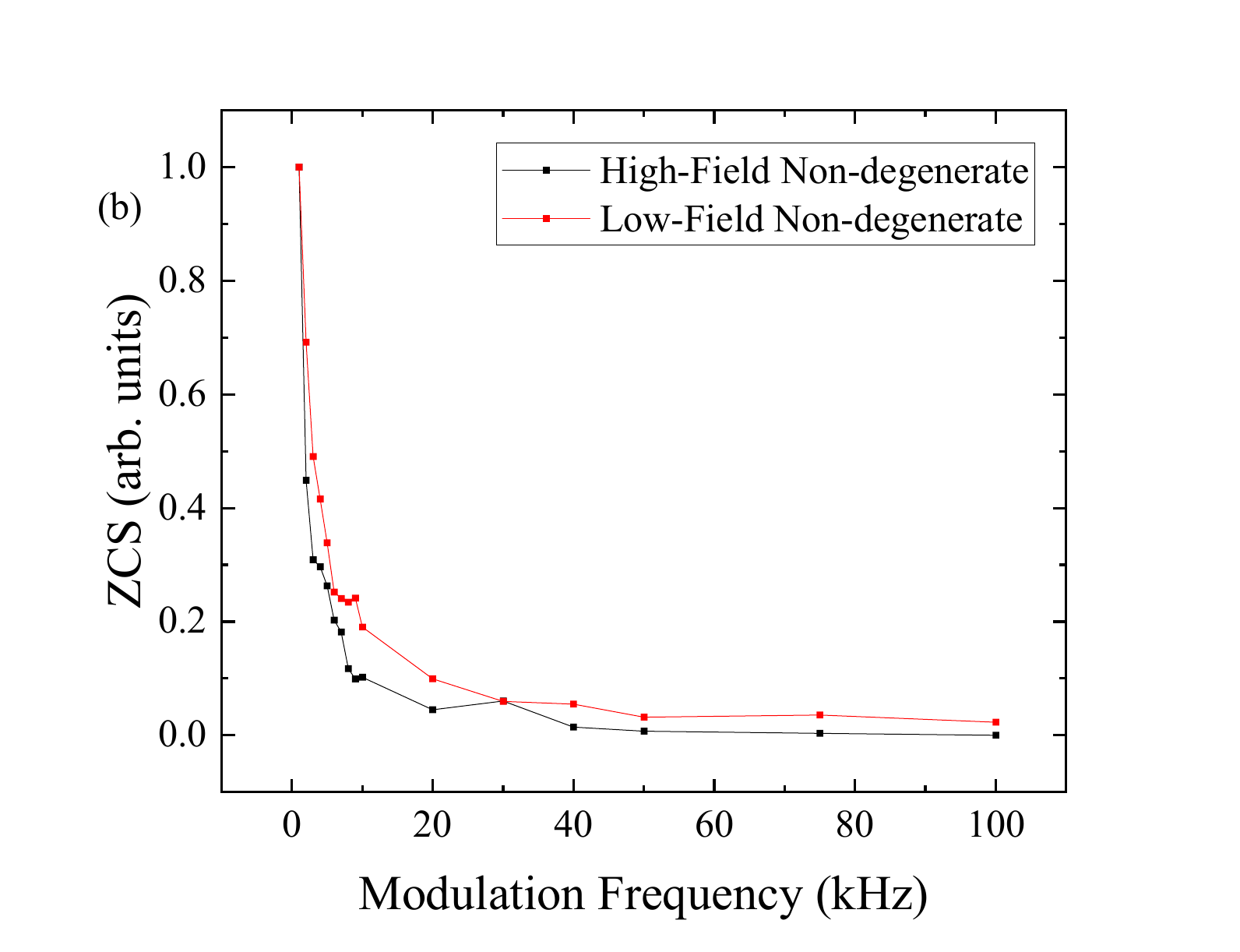}
\includegraphics[width=\columnwidth, trim={1.5cm 0.25cm 1.5cm 1.5cm}]{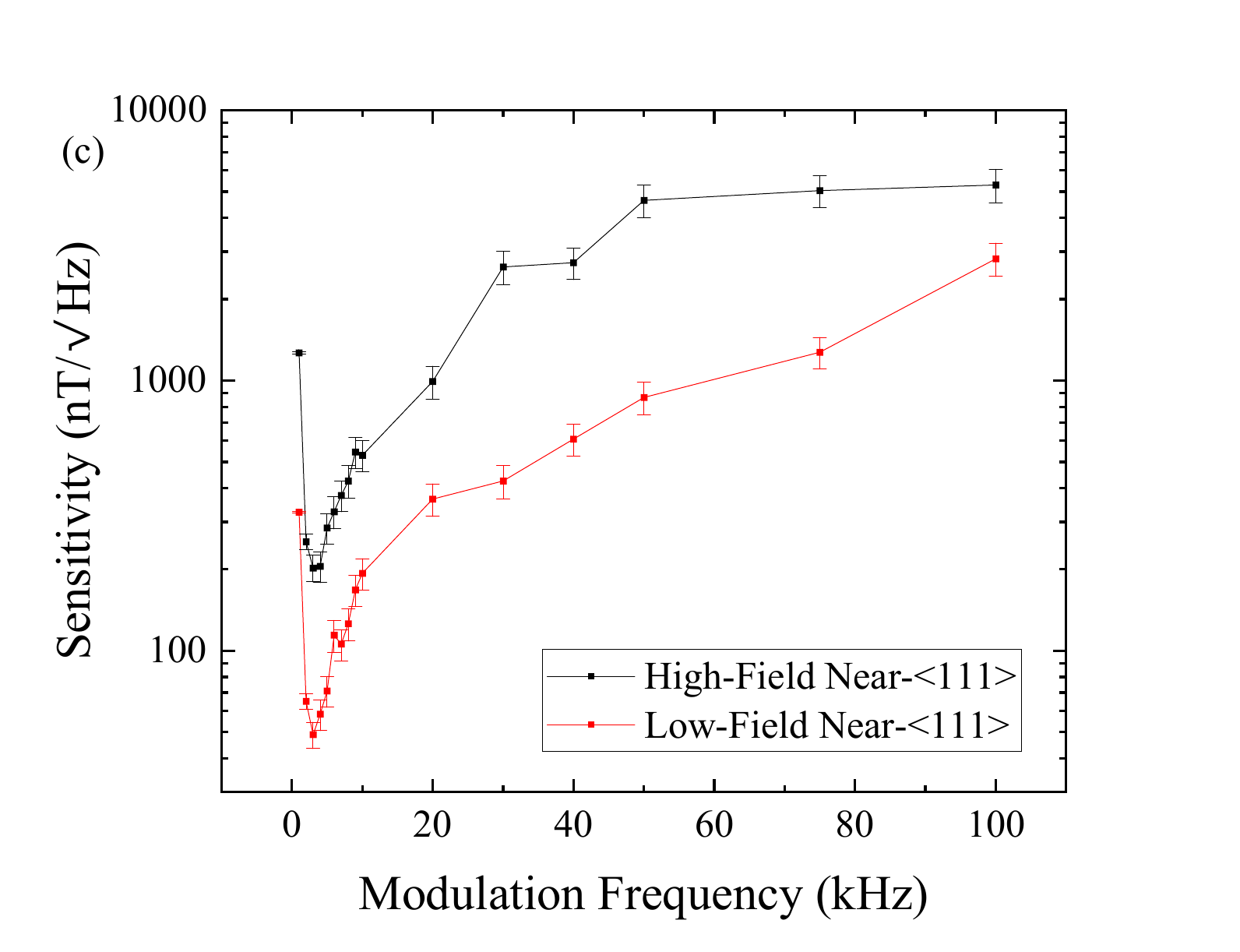}
\includegraphics[width=\columnwidth, trim={1.5cm 0.25cm 1.5cm 1.5cm}]{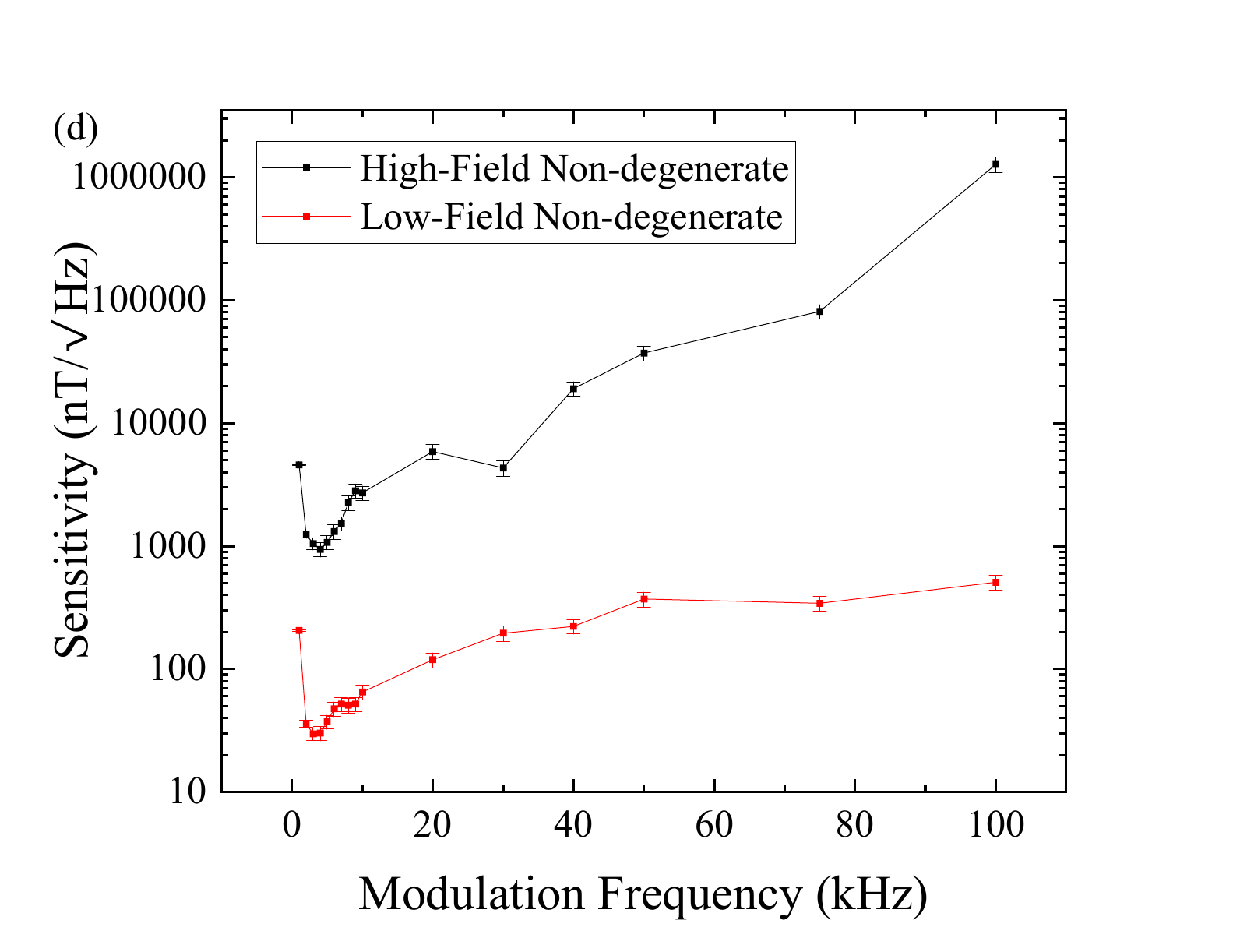}
\vspace{-5mm}
\caption{\small (a) and (b) The ZCS as a function of modulation frequency at low and high-field for a near-$\langle$111$\rangle$ alignment as well as for a non-degenerate alignment. (c) and (d) The sensitivity as a function of modulation frequency at low and high-field for a near-$\langle$111$\rangle$ alignment as well as for a non-degenerate alignment. A microwave power of +20 dBm and a modulation depth of 4 MHz were used for these measurements.}
\label{fig: AppendixJ-ZCSvsModulationFrequency}
\end{figure*}

The ZCS was also measured as a function of modulation frequency at high and low field for both alignments, as shown in Figs. \ref{fig: AppendixJ-ZCSvsModulationFrequency}a and \ref{fig: AppendixJ-ZCSvsModulationFrequency}b. The noise floors for each modulation frequency were also measured and the sensitivities calculated as shown in Figs. \ref{fig: AppendixJ-ZCSvsModulationFrequency}c and \ref{fig: AppendixJ-ZCSvsModulationFrequency}d. The ZCS was normalised for easier comparison. The ZCS was also determined, not from the demodulated X-channel of the LIA, but from R = $\sqrt{X^2+Y^2}$ to remove the dependence on the phase between the reference microwaves and the modulated fluorescence signal. The optimal phase would vary as the modulation frequency was adjusted. No clear differences are observed between the low- and high-field regimes.
Given the dependence of the contrast and thus ZCS on the optical re-polarisation rate, it might be predicted that mixing of the zero-field eigenstates could lead to some change in the optimum modulation frequency. The drop in ZCS with modulation frequency is more rapid for the non-degenerate high-field data than for the low-field data, though the difference is small accounting for errors. Given that the shelving time in the singlet states is on the order of 140 to 460 ns, depending strongly on temperature, changes in the excited state lifetimes would have minimal effect \cite{doherty2013nitrogen, kollarics2024terahertz}. Higher frequency ranges than those used in the main paper (10-150 Hz) could be achieved using a higher modulation frequency (and also LPF), at the expense of sensitivity. The degree of trade off required would depend upon the properties of the specific diamond sample.

\FloatBarrier

\section*{Appendix K: EPR Magnetic Field Calibration}

\begin{figure}[t]
\centering
\includegraphics[width=\columnwidth, trim={1.5cm 1cm 1.5cm 1.5cm}]{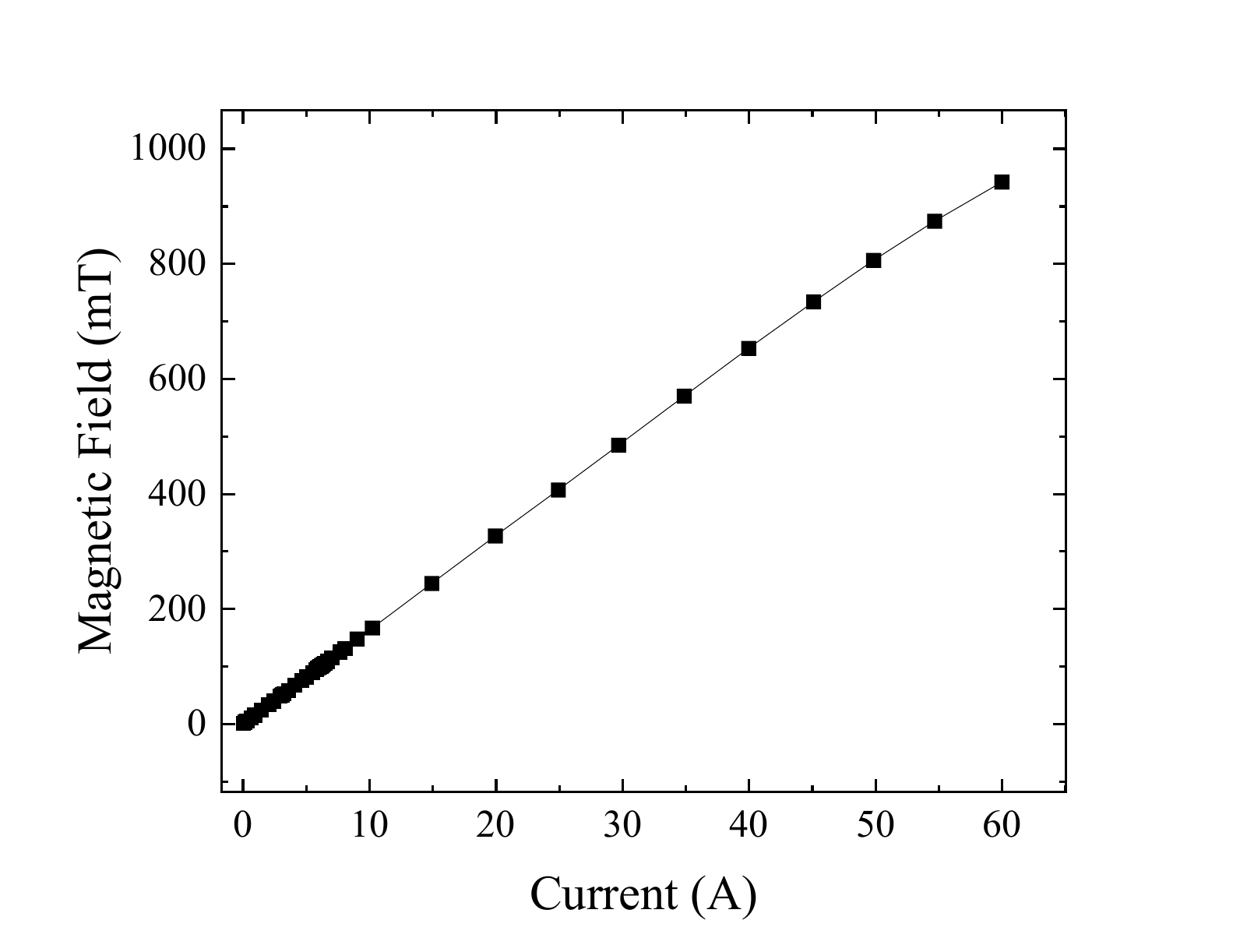}
\vspace{-5mm}
\caption{\small Calibration of the EPR magnetic field strength.}
\label{fig: AppendixK-EPRCalibration}
\end{figure}

The EPR magnets are controlled by a Bruker ER 082 power supply. The power supply was calibrated to determine the field strength from the current. Figure \ref{fig: AppendixK-EPRCalibration} shows this calibration. A Magnetics GM07 Hall probe was used to measure the magnetic field. For these measurements the EPR magnets were not controlled using the spectrometer's bridge, and a fault prevented them from reaching the full-rated magnetic field strength.
Once the bridge was restored, it was possible to increase the field to 1.2 T as shown in appendix H. 

\FloatBarrier

\section*{Appendix L: ODMR Road Map}

\begin{figure}[h!]
\centering
\includegraphics[width=\columnwidth, trim={1.5cm 1cm 1.5cm 1.5cm}]{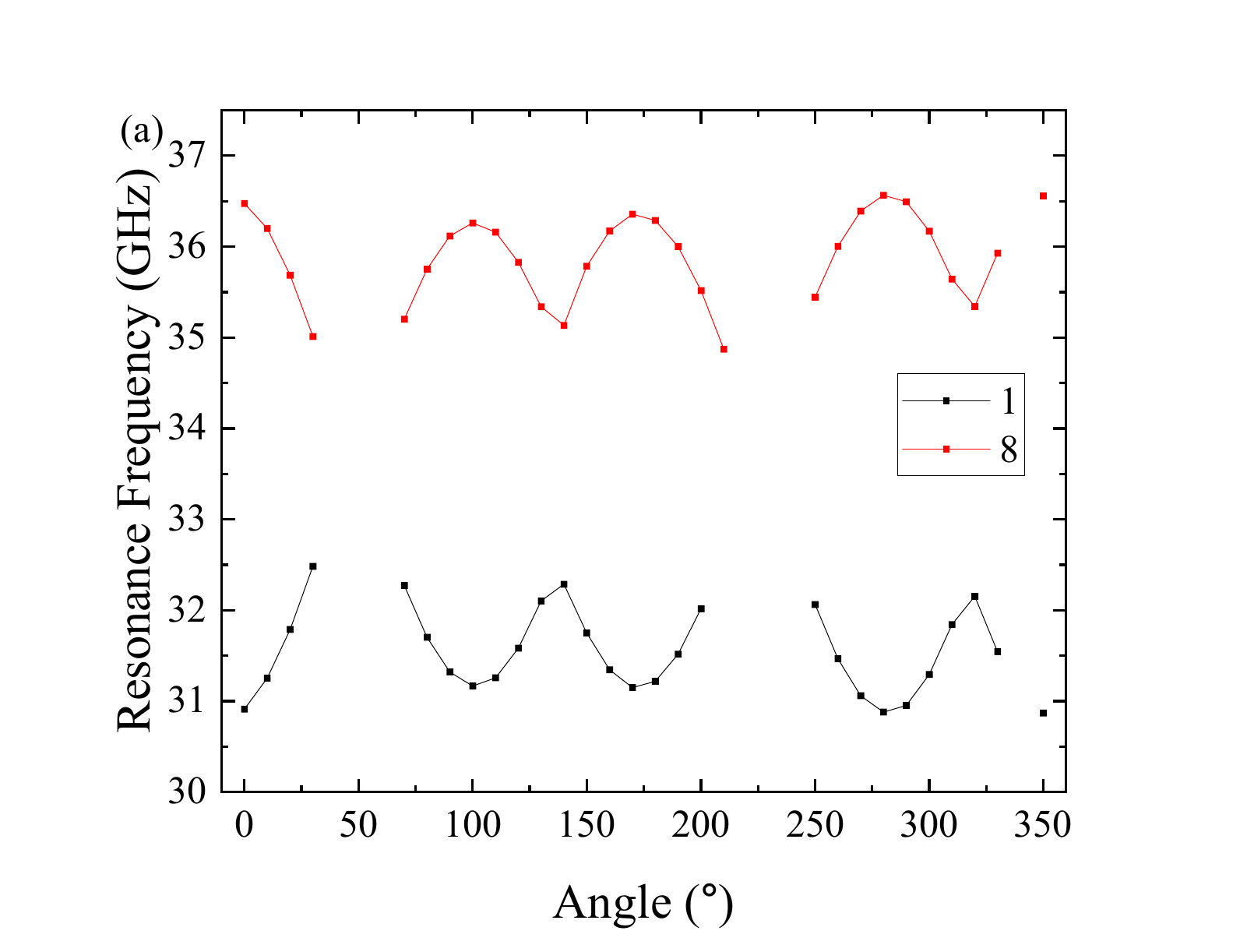} 
\includegraphics[width=\columnwidth, trim={1.5cm 1cm 1.5cm 1.5cm}]{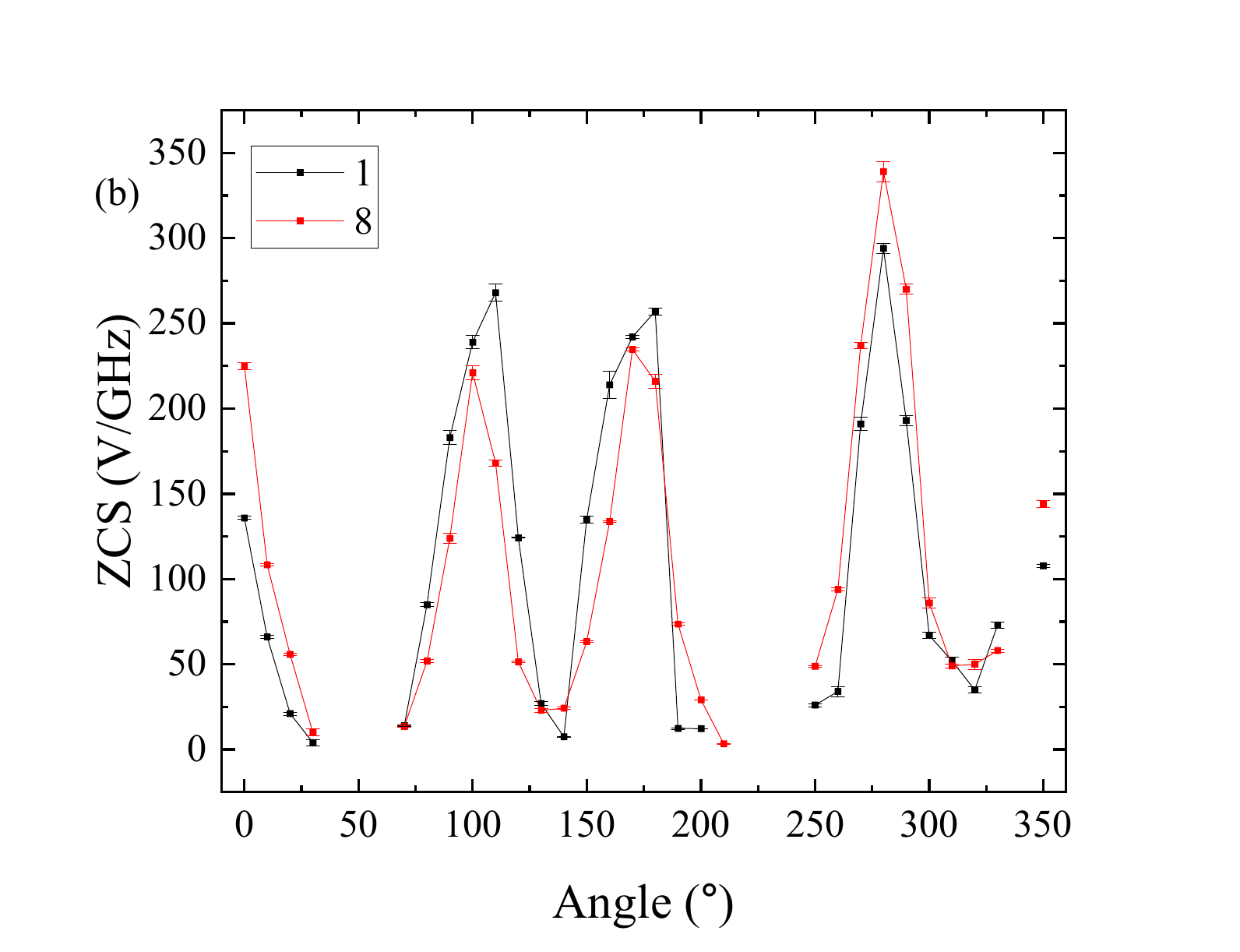} 
\caption{\small (a) and (b) Roadmap of the resonance frequencies and corresponding ZCSs as the sensor head is rotated about the sensor head symmetry axis by 350$^{\circ}$ in the EPR magnets field. This is for a near-$\langle$111$\rangle$ alignment.}
\label{fig: AppendixL-Roadmap}
\end{figure}

Figure \ref{fig: AppendixL-Roadmap}a shows a resonance frequency road map taken about a near-$\langle$111$\rangle$ alignment as the sensor head was rotated 350$^{\circ}$ about the symmetry axis of its cylindrical holder, with the angle measured using the attached goniometer. Figure \ref{fig: AppendixL-Roadmap}b shows the corresponding ZCSs for the two outermost resonances. The missing data points are due to the signal being too weak to resolve at those angles.  The road map further demonstrates the presence of vector information, with the resonance frequencies depending upon the field orientation with respect to each NVC symmetry axis (see the second term in Eq. \ref{eq:HighFieldPT}). However, as can be seen, there are near-dead zones for the sensor for certain field alignments. When placing diamond sensors within a tokamak it would be necessary to align them such that the total field was not along one of these (relative) dead zones.

\FloatBarrier

%


\end{document}